\documentclass{book}
\usepackage{graphicx}
\usepackage{afterpage}
\usepackage{fancyhdr}
\begin{document}
%
% macros
%
\def\expec#1{\big\langle{#1}\big\rangle}
\frontmatter
\begin{titlepage}
\title{An Introduction To Monte Carlo Simulations Of Surface Reactions}
\author{A. P. J. Jansen\footnote{e-mail: tgtatj@chem.tue.nl}
}
\end{titlepage}
\maketitle
\tableofcontents
\mainmatter
\chapter[Introduction]{Introduction}

If one is used to looking at chemical processes from an atomic point of
view, then the field of chemical kinetics is very complicated. Kinetics
is generally studied on meso- or macroscopic scales. Atomic scales are
of the order of \AA ngstr\o m and femtoseconds. Typical length scales in
laboratory experiments vary between micrometers to centimeters, and
typical time scales are often of the order of seconds or longer. This
means that there many orders of difference in length and time between
the individual reactions and the resulting kinetics.

The length gap is not always a problem. Many systems are homogeneous,
and the kinetics of a macroscopic system can be reduced to the kinetics
of a few reacting molecules. This is generally the case for reactions in
the gas phase and in solutions. For reactions on the surface of a
catalyst it is not clear when this is the case. It is certainly the case
that in the overwhelming number of studies on the kinetics in heterogeneous catalysis it is implicitly assumed that the adsorbates
are well-mixed, and that macroscopic rate equations can be
used. These equations have the form
\begin{eqnarray}
  {d\theta_{\rm A}\over dt}
  &=&-k_{\rm A}^{(1)}\theta_{\rm A}
     +\sum_{\rm B\ne A}k_{\rm B}^{(1)}\theta_{\rm B}\\
  & &-2k_{\rm A}^{(2)}\theta_{\rm A}^2
     -\sum_{\rm B\ne A}k_{\rm AB}^{(2)}\theta_{\rm A}\theta_{\rm B}
     +\sum_{\rm B,C\ne A}k_{\rm BC}^{(2)}\theta_{\rm B}\theta_{\rm C}
     +\ldots\nonumber
\end{eqnarray}
with $\theta_{\rm A}$ the so-called coverage of adsorbate A, which is
the number of A's per unit area of the surface on which the reactions
take place. The terms stand for the reactions ${\rm A}\to\ldots$, ${\rm
  B}\to{\rm A}$, $2{\rm A}\to\ldots$, ${\rm A}+{\rm B}\to\ldots$, and
${\rm B}+{\rm C}\to{\rm A}$, respectively. The $k$'s are rate constants.
If a position dependence $\theta=\theta({\bf r},t)$ is included we also
need a diffusion term. The result is called a reaction-diffusion
equation.  Simulations of reactions on surfaces and detailed studies in
surface science over the last few years have shown that the macroscopic
rate equations are only rarely correct. Moreover, there are systems that
show the formation of patterns with a characteristic length scale of
micro- to centimeters. For such systems it is not clear at all what the
relation is between the macroscopic kinetics and the individual
reactions.

Even more of a problem is the time gap. The typical atomic time scale is
given by the period of a molecular vibration. The fastest vibrations
have a reciprocal wavelength of up to $4000\,\hbox{cm}^{-1}$, and a
period of about $8.3\,$fs. Reactions in catalysis take place in seconds
or more. It is important to be aware of the origin of these fifteen
orders of magnitude difference. A reaction can be regarded as a movement
of the system from one local minimum on a potential-energy surface to
another. In such a move a so-called activation barrier has to be
overcome. Most of the time the system moves around one local
minimum. This movement is fast, in the order of femtoseconds, and
corresponds to a superposition of all possible vibrations. Every time
that the system moves in the direction of the activation barrier can be
regarded as an attempt to react. The probability that the reaction
actually succeeds can be estimated by calculating a Boltzmann factor
that gives the relative probability of finding the system at a local
minimum or on top of the activation barrier. This Boltzmann factor is
given by $\exp[-E_{\rm bar}/RT]$, where $E_{\rm bar}$ is the height of
the barrier, $R$ is the gas constant, and $T$ is the temperature. A
barrier of $E_{\rm bar}=100\,$kJ/mol at room temperature gives a
Boltzmann factor of about $10^{-18}$. Hence we see that the very large
difference in time scales is due to the very small probability that the
system overcomes activations barriers.

In Molecular Dynamics a reaction with a high activation barrier is
called a rare event, and various techniques have been developed to get a
reaction even when a standard simulation would never show it. These
techniques, however, work for one reacting molecule or two molecules
that react together, but not when one is interested in the combination
of thousands or more reacting molecules that one has when studying
kinetics. The purpose of this course is to show how one deals with such
a collection of reacting molecules. It turns out that one has to
sacrifices some of the detailed information that one has in Molecular
Dynamics simulations. One can still work on atomic length scales, but
one cannot work with the exact position of all atoms in a
system. Instead one only specifies near which minimum of the
potential-energy surface the system is. One does not work with the
atomic time scale. Instead one has the reactions as elementary events:
i.e., one specifies at which moment the system moves from one minimum of
the potential-energy surface to another. Moreover, because one doesn't
know where the atoms are exactly and how they are moving, one cannot
determine the times for the reactions exactly either. Instead one can
only give probabilities for the times of the reactions. It turns out,
however, that this information is more than sufficient for studying
kinetics.
\chapter[A Stochastic Model for the Description of Surface Reaction
Systems]{A Stochastic Model for the Description of Surface Reaction
  Systems}
\section{The lattice gas}

The size of the time step, and with this computational cost, in
simulations of the motion of atoms and molecules is determined by the
fast vibrations of chemical bonds.\cite{all87} Because the activation
energies of chemical reactions are generally much higher than the
thermal energies, chemical reactions take place on a time scale that is
many orders of magnitude larger. If one wants to study the kinetics on
surfaces, then one needs a method that does away with the fast motions.

The method that we present here does this by using the concept of {\it
  sites}. The forces working on an atom or a molecule that adsorbs on
the catalyst force it to well-defined positions on the
surface.\cite{zan88,tho97} These positions are called sites.
They correspond to minima on the potential-energy surface for the
adsorbate. Most of the time adsorbates stay very near these minima. Only
when they diffuse from one site to another or during a reaction they
will not be near a minima for a very short time. Instead of specifying
the precise positions, orientations, and configurations of the
adsorbates we will only specify for each sites its occupancy. A reaction
and a diffusion from one site to another will be modeled as a sudden
change in the occupancy of the sites. Because the elementary events are
now the reactions and the diffusion, the time that a system can be
simulated is no longer determined by fast motions of the adsorbates. By
taking a slightly larger length scale, we can simulate a much longer
time scale.

If the surface of the catalyst has two-dimensional translational
symmetry, or when it can be modeled as such, the sites form a regular
grid or a lattice. Our model is then a so-called {\it lattice-gas\/}
model. This chapter shows how this model can be used to describe a large
variety of problems in the kinetics of surface reactions.
\subsection{Definitions}

If the catalyst has two-dimensional translational symmetry then there
are two vectors, ${\bf a}_1$ and ${\bf a}_2$, with the property that
when the catalyst is translated over any of these vectors the result is
indistinguishable from the situation before the translation. It is said
that the system is {\it invariant\/} under translation over these
vectors. The vectors ${\bf a}_1$ and ${\bf a}_2$ are called {\it
  primitive vectors\/}. In fact the catalyst is invariant under
translational for any vector of the form
\begin{equation}
  {\bf x}=n_1{\bf a}_1+n_2{\bf a}_2
  \label{eq:ptrans}
\end{equation}
where $n_1$ and $n_2$ are integers. These vectors are the {\it lattice
  vectors\/}. The primitive vectors ${\bf a}_1$ and ${\bf a}_2$ are not
uniquely defined. For example a (111) surface of a fcc metal is
translationally invariant for ${\bf a}_1=a(1,0)$ and ${\bf
  a}_2=a(1/2,\sqrt{3}/2)$, where $a$ is the lattice spacing. But one can
just as well choose ${\bf a}_1=a(1,0)$ and ${\bf
  a}_2=a(-1/2,\sqrt{3}/2)$. The area defined by
\begin{equation}
  {\bf x}=x_1{\bf a}_1+x_2{\bf a}_2  
\end{equation}
with $x_1,x_2\in[0,1\rangle$ is called the {\it unit cell\/}. The whole
system is retained by tiling the plane with the contents of a unit cell.

Expression~(\ref{eq:ptrans}) defines a {\it simple lattice\/}, {\it
  Bravais lattice\/}, or {\it net\/}. Simple lattices have just one
lattice point, or grid point, per unit cell. It is also possible to have
more than one lattice point per unit cell. The lattice is then given by
all
\begin{equation}
  {\bf x}={\bf x}_0^{(i)}+n_1{\bf a}_1+n_2{\bf a}_2
\end{equation}
with $i=0,1,\ldots,N_{\rm sub}-1$. Each ${\bf x}_0^{(i)}$ is a different
vectors in the unit cell. The set ${\bf x}_0^{(i)}+n_1{\bf a}_1+n_2{\bf
  a}_2$ for a particular vector $i$ forms a {\it sublattice\/}, which is
itself a simple lattice. There are $N_{\rm sub}$ sublattices, and they
are all equivalent; they are only translated with respect to each
other. (For more information on lattices see for example
references~\cite{ash76} and \cite{zan88}).

We assign a {\it label\/} to each lattice point. The lattice points
correspond to the sites, and the labels specify properties of the sites.
The most common property that one wants to describe with the label is
the occupancy of the site. For example, the short-hand notation
$(n_1,n_2/s:{\rm A})$ can be interpreted as that the site at position
${\bf x}_0^{(s)}+n_1{\bf a}_1+n_2{\bf a}_2$ is occupied by a molecule A.
The labels can also be used the specify reactions. A reaction is nothing
but a change in the labels. An extension of the short-hand notation
$(n_1,n_2/s:{\rm A}\to{\rm B})$ indicates that during a reaction the
occupancy of the site at ${\bf x}_0^{(s)}+n_1{\bf a}_1+n_2{\bf a}_2$
changes from A to B. If more than one site is involved in a reaction
then the specification will consist of a set changes of the form
$(n_1,n_2/s:{\rm A}\to{\rm B})$.
\subsection{Examples}

\begin{figure}[ht]
\includegraphics[width=\hsize]{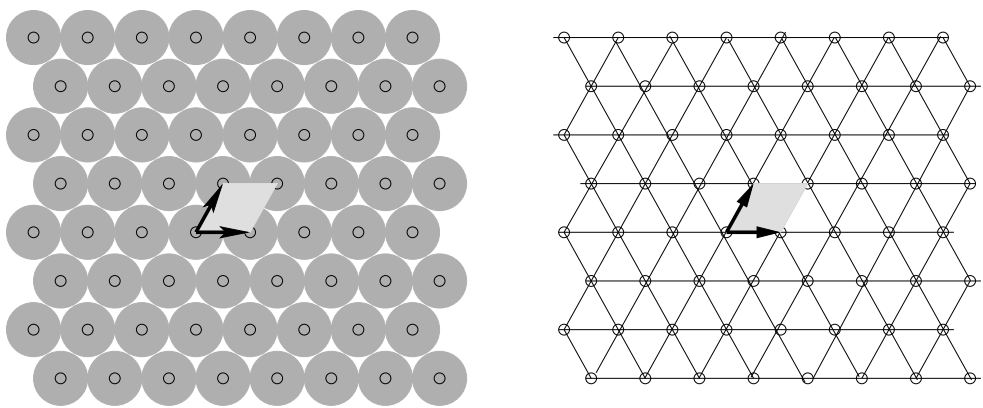}
\caption{On the left is shown the top layer of a Pt(111) surface with
  the primitive vectors, the unit cell, and the top sites. On the right
  is shown the lattice formed by the top sites. The lines are guides for
  the eyes.}
\label{fig:ptcoa}
\end{figure}

Figure~\ref{fig:ptcoa} shows a Pt(111) surface. CO prefers to adsorb on
this surface at the top sites.\cite{zan88} We can therefore model CO on
this surface with a simple lattice with the lattice points corresponding
to the top sites. We have ${\bf a}_1=a(1,0)$ and ${\bf
  a}_2=a(1/2,\sqrt{3}/2)$. As $N_{\rm sub}=1$ we choose ${\bf
  x}_0^{(0)}=(0,0)$ for simplicity. Each grid point has a label that we
choose to be equal to CO or $*$. The former indicates that the site is
occupied by a CO molecule, the latter that the site is vacant.

\begin{figure}[ht]
\includegraphics[width=\hsize]{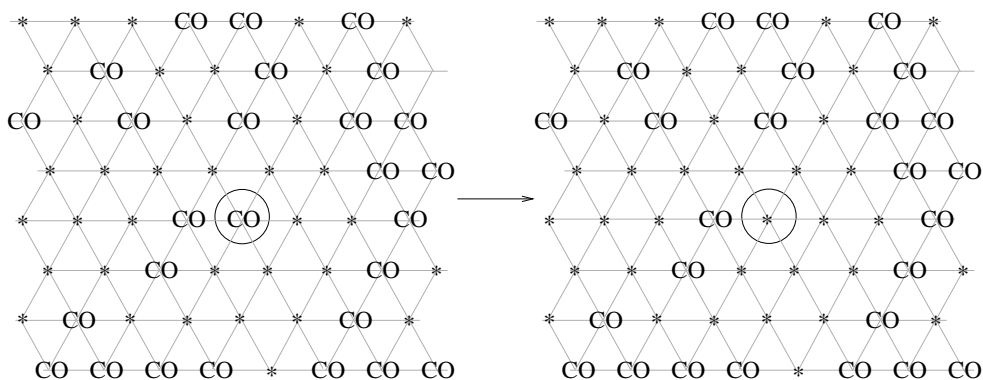}
\caption{Change of labels for CO desorption from a Pt(111) surface. The
  encircled CO molecule on the left desorbs and the label becomes $*$
  indicating a vacant site.}
\label{fig:lblch}
\end{figure}

Desorption of CO from Pt(111) can be written as $(0,0:{\rm CO}\to *)$,
where we have left out the index of the sublattice, because, as there is
only one, it is clear on which sublattice the reaction takes place (see
figure~\ref{fig:lblch}).  Desorption on other sites can be obtained by
translations over lattice vectors; i.e., $(0,0:{\rm CO}\to *)$ is
representative for $(n_1,n_2:{\rm CO}\to *)$ with $n_1$ and $n_2$
integers. Diffusion of CO can be modeled as hops from one site to a
neighboring site. We can write that as $\{(0,0:{\rm CO}\to *),(1,0:*\to
{\rm CO})\}$. Hops on other sites can again be obtained from these
descriptions by translations over lattice vectors, but also by rotations
that leave the surface is invariant.

\begin{figure}[ht]
\includegraphics[width=\hsize]{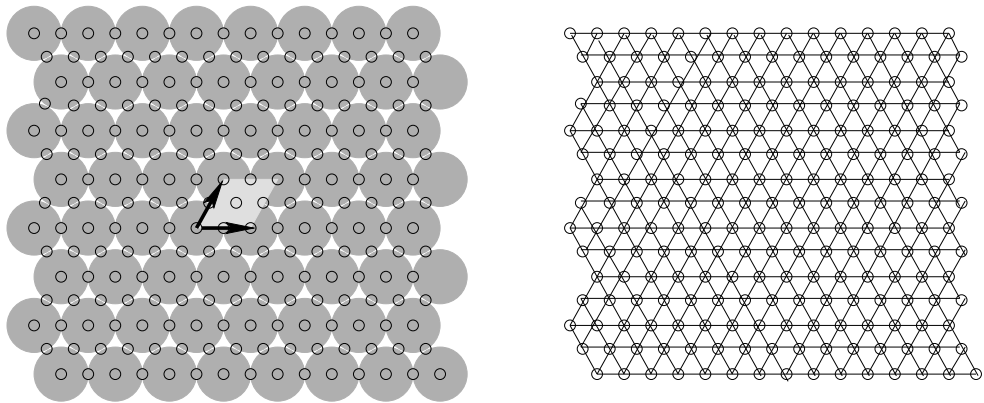}
\caption{On the left is shown the top layer of a Pt(111) surface with
  the primitive vectors, the unit cell, and the top and bridge sites. On
  the right is shown the lattice formed by the top and bridge sites
  sites. The lines are guides for the eyes.}
\label{fig:ptcob}
\end{figure}

At high coverages the repulsion between the CO molecules forces some of
them to bridge sites.\cite{pet02} Figure~\ref{fig:ptcob} shows the new
lattice. We have now for sublattices with ${\bf x}_0^{(0)}=(0,0)$, ${\bf
  x}_0^{(1)}=(1/2,0)$, ${\bf x}_0^{(2)}=(1/4,\sqrt{3}/4)$, ${\bf
  x}_0^{(3)}=(3/4,\sqrt{3}/4)$. The first one is for the top sites. The
others are for the three sublattices of bridge sites. The figure shows
that the lattice looks like a simple lattice. Indeed we can regard as
such, but only when we need not distinguish between top and bridge
sites.

\begin{figure}[ht]
\includegraphics[width=\hsize]{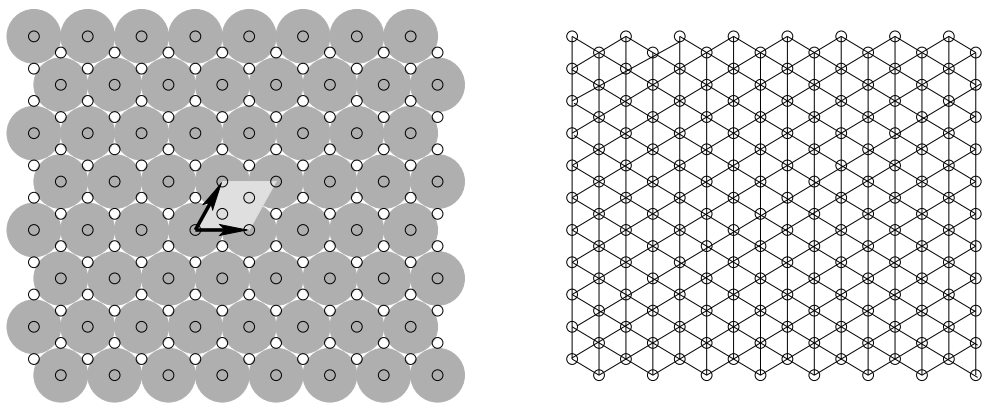}
\caption{On the left is shown the top layer of a Rh(111) surface with
the primitive vectors, the unit cell,
and the top and hollow sites. On the right is shown the lattice formed
by the top and hollow sites sites. The lines are guides for the eyes.}
\label{fig:rhno}
\end{figure}

NO on Rh(111) forms a $(2\times 2)$-3NO structure in which equal numbers
of NO molecules occupy top, fcc hollow, and hcp hollow
sites.\cite{kao89,har97a} Figure~\ref{fig:rhno} shows the sites that are
involved and the corresponding lattice. We now have three sublattices
with ${\bf x}_0^{(0)}=(0,0)$ (top sites), ${\bf
  x}_0^{(1)}=(1/2,\sqrt{3}/6)$ (fcc hollow sites), and ${\bf
  x}_0^{(2)}=(1,\sqrt{3}/3)$ (hcp hollow sites).  This is similar to the
case with high CO coverage on Pt(111).  $(0,0/0:{\rm NO})$ indicates
that there is an NO molecule at the top site $(0,0)$, and $(0,0/1:*)$
indicates that the fcc hollow site at $(1/2,\sqrt{3}/6)$ is vacant.

Note that also in this case the lattice resembles a simple lattice with
${\bf a}_1=a(1/2,\sqrt{3}/6)$ and ${\bf a}_2=a(\sqrt{3}/3,0)$. It is
indeed also possible to model the system with this simple lattice, but
one should note that then the difference between the top and hollow
sites is ignored. It is possible to use the simple lattice and at the
same time retaining the difference between the sites. The trick is to
use the labels not just for the occupancy, but also for indicating the
type of site. So instead of labels NO and $*$ indicating the occupancy,
we use NOt, NOf, NOh, $*$t, $*$f, and $*$h. The last letter indicates
the type of site (t stands for top, f for fcc hollow, and h for hcp
hollow) and the rest for the occupancy. Instead of $(0,0/0:{\rm NO})$
and $(0,0/1:*)$ we have $(0,0:{\rm NOt})$ and $(1,0:*{\rm f})$,
respectively. It depends very much on the reaction which way of
describing the system is more convenient and computationally more
efficient.

\begin{figure}[ht]
\includegraphics[width=\hsize]{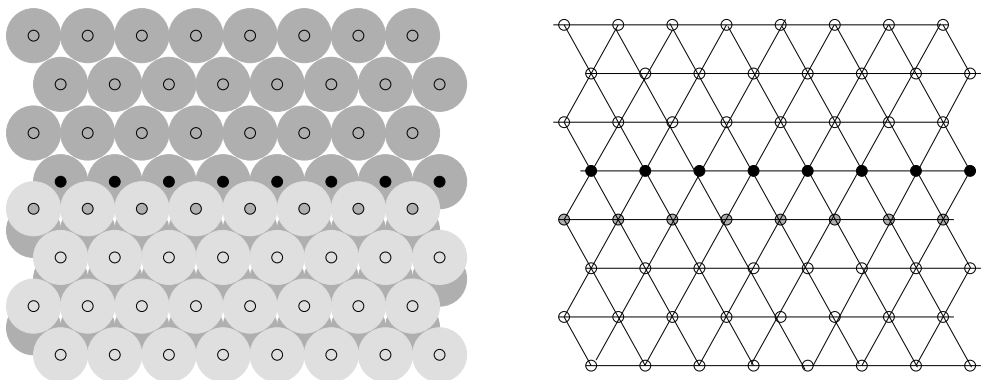}
\caption{A Ru(0001) surface with a step and the top sites on the
  left. On the right is shown the lattice. The open circles are top
  sites on the terraces. The small black circles are top sites at the
  bottom of the step, and the small dark grey circles are top sites at
  the top of the step. The lines are guides for the eyes.}
\label{fig:step}
\end{figure}

Using the label to specify other properties of the site than its
occupancy can be a very powerful tool. Figure~\ref{fig:step} shows how
to model a step.\cite{dah99,ham00a} If the terraces are small then it
might also be possible to work with a unit cell spanning the width of a
terrace, but when the terraces become large this will be inconvenient as
there will be many sublattices.

Site properties like the sublattice of which the site is part of and if
it is a step site or not are static properties. The occupancy of a site
is a dynamic property. There are also other properties of sites that are
dynamic. Bare Pt(100) reconstructs into a quasi-hexagonal
structure.\cite{imb95} CO oxidation on Pt(100) is substantially
influenced by this reconstruction because oxygen adsorbs much less
readily on the reconstructed surfaces than on the unreconstructed one.
This can lead to oscillation, chaos, and pattern formation.\cite{imb95,sli94}
It is possible to model the effect of the reconstruction on the CO
oxidation by using a label that specifies whether the surface is locally
reconstructed or not.\cite{gel98,kuz99a,kor99b}
\subsection{Shortcomings}

The lattice-gas model is simple yet very powerful, as it allows us to
model a large variety of systems and phenomena. Yet not everything can
be modeled with it. Let's take again CO oxidation on Pt(100). As stated
above this system shows reconstruction which can be modeled with a label
indicating that the surface is reconstructed or not. This way of
modeling has shown to be very successful,\cite{gel98,kuz99a,kor99b} but
it does neglect some aspects of the reconstruction. The reconstructed
and the unreconstructed surface have very different unit cells, and the
adsorption sites are also different.\cite{gru95,gru96} In fact, the unit
cell of the reconstructed surface is very large, and there are a large
number of adsorption sites with slightly different properties. These
aspects have been neglected in the kinetic simulations so far. As these
simulations have been quite successful, it seems that these aspects are
not very relevant in this case, but that need not be always be the case.
Catalytic partial oxidation (CPO) takes place at high temperature at
which the surface is so dynamic that all translational symmetry is lost.
In this case using a lattice to model the kinetics seems inappropriate.

The example of CO on Pt(111) has shown that at high coverage the
position at which the molecules adsorb change. The reason for this is
that these positions are not only determined by the interactions between
the adsorbates and the substrate, but also by the interactions between
the adsorbates themselves. At low coverages the former dominate, but at
high coverages the latter may be more important. This may lead to
adlayer structures that are incommensurate with the
substrate.\cite{zan88} Examples are formed by the nobles gases. These
are weakly physisorbed, whereas at high coverages the packing onto the
substrate is determined by the steric repulsion between them. At low and
high coverages different lattices are needed to describe the positions
of the adsorbates, but a single lattice describing both the low and the
high coverage sites is not possible. Simulations in which the coverages
change from low to high coverage and/or {\it vice versa\/} then cannot
be based on a lattice-gas model.
\section{The Master Equation}

\subsection{Definition}

Our treatment of Monte Carlo simulations of surface reactions differs in
one very fundamental aspect from that of other authors; the derivation
of the algorithms and a large part of the interpretation of the results
of the simulations are based on a Master Equation
\begin{equation}
  {dP_\alpha\over dt}=\sum_\beta
  \left[W_{\alpha\beta}P_\beta-W_{\beta\alpha}P_\alpha\right].
  \label{eq:MEdef}
\end{equation}
In this equation $t$ is time, $\alpha$ and $\beta$ are configurations of
the adlayer, $P_\alpha$ and $P_\beta$ are their probabilities, and
$W_{\alpha\beta}$ and $W_{\beta\alpha}$ are so-called transition
probabilities per unit time that specify the rate with which the adlayer
changes due to reactions. The Master Equation is a loss-gain
equation. The first term on the right stands for increases in $P_\alpha$
because of reactions that change other configurations into $\alpha$. The
second term stands for decreases because of reactions in $\alpha$. From
\begin{equation}
  {d\over dt}\sum_\alpha P_\alpha
  =\sum_\alpha{dP_\alpha\over dt}
  =\sum_{\alpha\beta}
  \left[W_{\alpha\beta}P_\beta-W_{\beta\alpha}P_\alpha\right]=0
\end{equation}
we see that the total probability is conserved. (The last equality can
be seen by swapping the summation indices in one of the terms.)

The Master Equation can be derived from first principles as will be
shown below, and hence forms a solid basis for all subsequent work.
There are other advantages as well. First, the derivation of the Master
Equation yields expressions for the transition probabilities that can be
computed with quantum chemical methods.\cite{lea96} This makes {\it
  ab-initio\/} kinetics for catalytic processes possible. Second, there
are many different algorithms for Monte Carlo simulations. Those that
are derived from the Master Equation all give necessarily results that
are statistically identical. Those that cannot be derived from the
Master Equation conflict with first principles and should be discarded.
Third, Monte Carlo is a way to solve the Master Equation, but it is not
the only one. The Master Equation can, for example, be used to derive
the normal macroscopic rate equation (see below). In general, it forms a
good basis to compare different theories of kinetic quantitatively, and
also to compare these theories with simulations.
\subsection{Derivation}
\label{sec:deriv}

The Master Equation can be derived by looking at the surface and its
adsorbates in phase space. This is, of course, a classical mechanics
concept, and one might wonder if it is correct to look at the reactions
on an atomic scale and use classical mechanics. The situation here is
the same as for the derivation of the rate equations for gas phase
reactions. The usual derivations there also use classical
mechanics.\cite{kec60,kec62,kec67,pec76,tru85} Although it is possible
to give a completely quantum mechanical derivation
formalism,\cite{mil74,mil75,vot93,ben94} the mathematical complexity
hides much of the important parts of the chemistry. Besides, it is
possible to replace the classical expressions that we will get by
semi-quantum mechanical ones, in exactly the same way as for gas phase
reactions.

A point in phase space completely specifies the positions and momenta of
all atoms in the system. In Molecular Dynamics simulations one uses
these positions and momenta at some starting point to compute them at
later times. One thus obtains a trajectory of the system in phase space.
We are not interested in that amount of detail, however. In fact as was
stated before too much detail is detrimental if one is interested in
simulating many reactions. The time interval that one can simulate a
system using Molecular Dynamics is typically of the order of
nanoseconds. Reactions in catalysis have a characteristic time that is
many orders of magnitude longer. To overcome this large difference we
need a method that removes the fast processes (vibrations) that
determine the time scale of Molecular Dynamics, and leaves us with the
slow processes (reactions).  This method looks as follows.

Instead of the precise position of each atom, we only want to know how
the different adsorbates are distributed over the sites of a surface. So
our physical model is a lattice. Each lattice point corresponds to one
site, and has a label that specifies which adsorbate is adsorbed. (A
vacant site is simply a special label.) A particular distribution of the
adsorbates over the sites, or, what is the same, a particular labeling
of the grid points, we call a {\it configuration\/}. As each point in
phase space is a precise specification of the position of each atom, we
also know which adsorbates are at which sites; i.e., we know the
corresponding configuration. Different points in phase space may,
however, correspond to the same configuration, which differ only in
slight variations of the positions of the atoms. This means that we can
partition phase space in many region, each of which corresponds to one
configuration. Reactions are then nothing but motion of the system in
phase space from one region to another.

\begin{figure}[ht]
\includegraphics[width=\hsize]{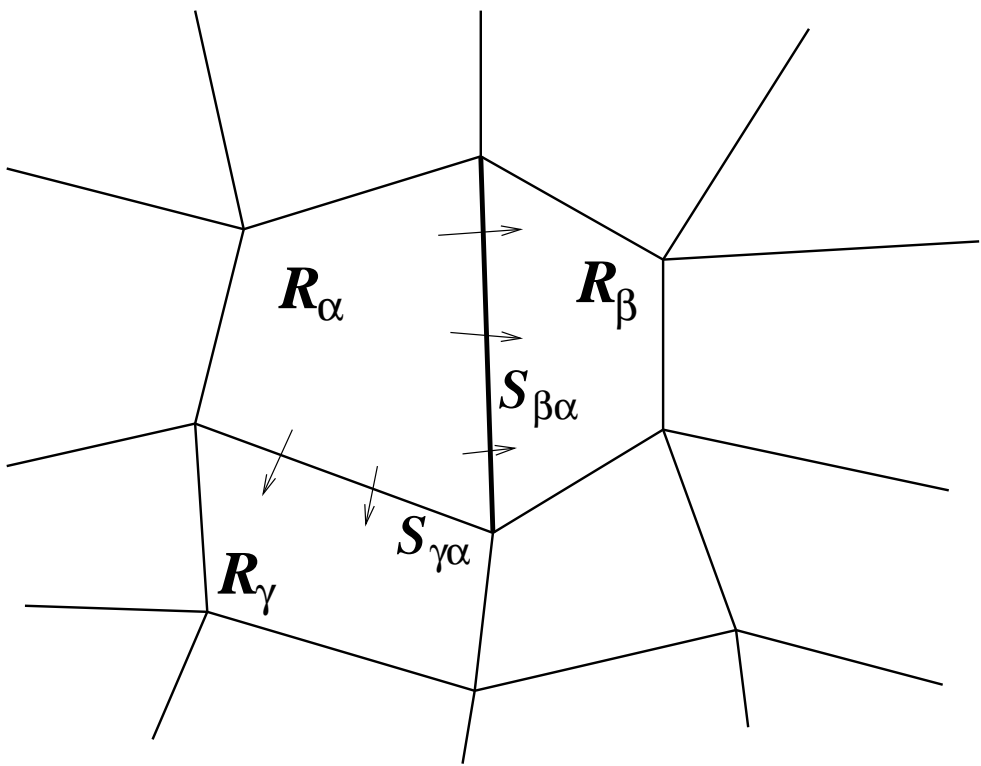}
\caption{Schematic drawing of the partitioning of configuration space into
  regions $R$, each of which corresponds to some particular
  configuration of the adlayer.  The reaction that changes $\alpha$ into
  $\beta$ corresponds to a flow from $R_\alpha$ to $R_\beta$. The
  transition probability $W_{\beta\alpha}$ for this reaction equals the
  flux through the surface $S_{\beta\alpha}$, separating $R_\alpha$ from
  $R_\beta$, divided by the probability to find the system in
  $R_\alpha$.
\label{fig:psreac}}
\end{figure}

Because it is not possible to reproduce an experiment with exactly the
same configuration, we are not only not interested in the precise
position of the atoms, we are not even interested in specific
configurations, but only in characteristic ones. Although there may be
differences on a microscopic scale, the behavior of a system on a
macroscopic, and often also on a mesoscopic, scale will be the same. So
we do not look at individual trajectories in phase space, but we average
over all possible trajectories. This means that we work with a phase
space density $\rho$ and a probability $P_\alpha$ of finding the system
in configuration $\alpha$. These are related via
\begin{equation}
  P_\alpha(t)=\int_{R_\alpha}\!\!{d{\bf q}\,d{\bf p}\over h^D}\,
  \rho({\bf q},{\bf p},t),
  \label{eq:prbdef}
\end{equation}
where ${\bf q}$ stands for all coordinates, ${\bf p}$ stands for all
momenta, $h$ is Planck's constant, $D$ is the number of degrees of
freedom, and the integration is over the region $R_\alpha$ in phase
space that corresponds to configuration $\alpha$ (see
figure~\ref{fig:psreac}). The denominator $h^D$ is not needed for a
purely classical description of the kinetics. However, it makes the
transition from a classical to a quantum mechanical description
easier.\cite{mcq76} 
The Master Equation tells us how these probabilities $P_\alpha$ change
in time. Differentiating equation~(\ref{eq:prbdef}) yields
\begin{equation}
  {dP_\alpha\over dt}=\int_{R_\alpha}\!\!{d{\bf q}\,d{\bf p}\over h^D}\,
  {\partial\rho\over\partial t}({\bf q},{\bf p},t).
\end{equation}
This can be transformed using the Liouville-equation\cite{bec85}
\begin{equation}
  {\partial\rho\over\partial t}
  =-\sum_{i=1}^D\left[{\partial\rho\over\partial q_i}
   {\partial H\over\partial p_i}
  -{\partial\rho\over\partial p_i}
   {\partial H\over\partial q_i}
  \right]
\end{equation}
into
\begin{equation}
  {dP_\alpha\over dt}=\int_{R_\alpha}\!\!{d{\bf q}\,d{\bf p}\over h^D}\,
  \sum_{i=1}^D\left[
  {\partial\rho\over\partial p_i}{\partial H\over\partial q_i}
  -{\partial\rho\over\partial q_i}{\partial H\over\partial p_i}
  \right],\label{eq:prate1}
\end{equation}
where $H$ is the system's classical Hamiltonian. To simplify the
mathematics, we will assume that the coordinates are Cartesian and the
Hamiltonian has the usual form
\begin{equation}
  H=\sum_{i=1}^D{p_i^2\over 2m_i}+V({\bf q}),
\end{equation}
where $m_i$ is the mass corresponding to coordinate $i$. We also assume
that the area $R_\alpha$ is defined by coordinates only, and that the
limits of integration for the momenta are $\pm\infty$. Although these
assumptions are hardly restrictive, we would like to mention
reference\cite{jan91} for a more general derivation. The assumptions
allow us to go from phase space to configuration space. (Not to be
confused with the configurations of the Master Equation.) The first term
of equation~(\ref{eq:prate1}) now becomes
\begin{eqnarray}
  \int_{R_\alpha}\!\!{d{\bf q}\,d{\bf p}\over h^D}\,
  \sum_{i=1}^D{\partial\rho\over\partial p_i}{\partial H\over\partial q_i}
  &&=\sum_{i=1}^D\int_{R_\alpha}\!\!d{\bf q}\,{\partial V\over\partial q_i}
  \int_{-\infty}^\infty\!{d{\bf p}\over h^D}\,
    {\partial\rho\over\partial p_i}\\
  &&=\sum_{i=1}^D\int_{R_\alpha}\!\!d{\bf q}\,{\partial V\over\partial q_i}
  \int_{-\infty}^\infty
  \!\!{dp_1\ldots dp_{i-1}dp_{i+1}\ldots dp_D\over h^D}\nonumber\\
  &&\qquad\times\big[\rho(p_i=\infty)-\rho(p_i=-\infty)\big]=0,\nonumber
\end{eqnarray}
because $\rho$ has to go to zero for any of its variables going to
$\pm\infty$ to be integrable. The second term becomes
\begin{equation}
  -\int_{R_\alpha}\!\!{d{\bf q}\,d{\bf p}\over h^D}\,
  \sum_{i=1}^D{\partial\rho\over\partial q_i}{\partial H\over\partial p_i}
  =-\int_{R_\alpha}\!\!{d{\bf q}\,d{\bf p}\over h^D}\,
  \sum_{i=1}^D{\partial\over\partial q_i}\left({p_i\over m_i}\rho\right).
\end{equation}
This particular form suggest using the divergence theorem for the
integration over the coordinates.\cite{kre93} The final result is then
\begin{equation}
  {dP_\alpha\over dt}=
  -\int_{S_\alpha}\!\!dS\int_{-\infty}^\infty\!{d{\bf p}\over h^D}\,
  \sum_{i=1}^Dn_i{p_i\over m_i}\rho,
  \label{eq:prate2}
\end{equation}
where the first integration is a surface integral over the surface of
$R_\alpha$, and $n_i$ are the components of the outward pointing normal
of that surface. Both the area $R_\alpha$ and the surface $S_\alpha$ are
now regarded as parts of the configuration space of the system. As
$p_i/m_i=\dot q_i$, we see that the summation in the last expression is
the flux through $S_\alpha$ in the direction of the outward pointing
normal (see figure~\ref{fig:psreac}).

The final step is now to decompose this flux in two ways. First, we
split the surface $S_\alpha$ into sections $S_\alpha=\cup_\beta
S_{\beta\alpha}$, where $S_{\beta\alpha}$ is the surface separating
$R_\alpha$ from $R_\beta$. Second, we distinguish between an outward
flux, $\sum_in_ip_i/m_i>0$, and an inward flux,
$\sum_in_ip_i/m_i<0$. Equation~({\ref{eq:prate2}) can then be rewritten as
\begin{eqnarray}
  {dP_\alpha\over dt}
  =&&\sum_\beta\int_{S_{\alpha\beta}}\!\!dS\,
  \int_{-\infty}^\infty\!{d{\bf p}\over h^D}\,
  \left(\sum_{i=1}^Dn_i{p_i\over m_i}\right)
  \Theta\left(\sum_{i=1}^Dn_i{p_i\over m_i}\right)\rho\label{eq:prate3}\\
  -&&\sum_\beta\int_{S_{\beta\alpha}}\!\!dS\,
  \int_{-\infty}^\infty\!{d{\bf p}\over h^D}\,
  \left(\sum_{i=1}^Dn_i{p_i\over m_i}\right)
  \Theta\left(\sum_{i=1}^Dn_i{p_i\over m_i}\right)\rho,\nonumber
\end{eqnarray}
where in the first term $S_{\alpha\beta}$ ($=S_{\beta\alpha}$) is
regarded as part of the surface of $R_\beta$, and the $n_i$ are
components of the outward pointing normal of $R_\beta$. The function
$\Theta$ is the Heaviside step function.\cite{zem87}
Equation~(\ref{eq:prate3}) can be cast in the form of the Master Equation
\begin{equation}
    {dP_\alpha\over dt}=\sum_\beta\left[
    W_{\alpha\beta}P_\beta-W_{\beta\alpha}P_\alpha\right],
\end{equation}
if we define the transition probabilities as
\begin{equation}
  W_{\alpha\beta}=\left[
  \int_{S_{\alpha\beta}}\!\!dS\,
  \int_{-\infty}^\infty\!{d{\bf p}\over h^D}\,
  \left(\sum_{i=1}^Dn_i{p_i\over m_i}\right)
  \Theta\left(\sum_{i=1}^Dn_i{p_i\over m_i}\right)\rho\right]
  /
  \left[
  \int_{R_\alpha}\!\!d{\bf q}\,
  \int_{-\infty}^\infty\!{d{\bf p}\over h^D}\,\rho\right].
  \label{eq:Wdef}
\end{equation}

The expression for the transition probabilities can be cast in a more
familiar form by using a few additional assumptions. We assume that
$\rho$ can locally be approximated by a Boltzmann-distribution
\begin{equation}
  \rho=N\exp\left[-{H\over k_BT}\right],
\end{equation}
where $T$ is the temperature, $k_B$ is the Boltzmann-constant, and $N$
is a normalizing constant. We also assume that we can define
$S_{\alpha\beta}$ and the coordinates in such a way that $n_i=0$, except
for one coordinate $i$, called the reaction coordinate, for which
$n_i=1$. The integral of the momentum corresponding to the reaction
coordinate can then be done and the result is
\begin{equation}
  W_{\alpha\beta}={k_BT\over h}{Q^\ddagger\over Q},
  \label{eq:Wts}
\end{equation}
with
\begin{eqnarray}
  Q^\ddagger&&\equiv\int_{S_{\alpha\beta}}\!\!dS\,
  \int_{-\infty}^\infty\!{dp_1\ldots dp_{i-1}dp_{i+1}\ldots dp_D\over
  h^{D-1}}
  \exp\left[-{H\over k_BT}\right],\label{eq:pfun}\\
  Q&&\equiv\int_{R_\alpha}\!\!d{\bf q}\,
  \int_{-\infty}^\infty\!{d{\bf p}\over h^D}\,
  \exp\left[-{H\over k_BT}\right].
  \label{eq:ppfun}
\end{eqnarray}
We see that this is an expression that is formally identical to the
Transition-State Theory (TST) expression for rate constants.\cite{san95}
There are differences in the definition of the partition functions $Q$
and $Q^\ddagger$, but even these can be neglected as will be shown in
chapter~\ref{ch:transprob}.
\section{Working without a lattice}

Although the use of a lattice is very important in the theory above, one
should realize that it is really not needed from a theoretical point of
view. No reference was made to a lattice in the derivation of the Master
Equation, and indeed one can use the Master Equation also for reactive
systems that do no have translational or any other kind of symmetry.

The idea is to look at the potential-energy surface (PES) of a
system,\cite{mez87} and associate each ``configuration'' $\alpha$ with a
minimum of the PES. The region $R_\alpha$ consists of the points in
phase space around the minimum (see figure~\ref{fig:catchpes}). (As
before the momenta can have any value.) The precise position of the
surfaces $S_{\beta\alpha}$ are hard to determine. In Variational
Transition-State Theory (VTST) they are chosen to minimize the
flux,\cite{kec60,kec62,kec67,pec76,tru85} but a more pragmatic approach
would be to put $S_{\beta\alpha}$ at the saddle point of the PES that
separates minimum $\alpha$ from $\beta$. The derivation in
section~\ref{sec:deriv} does not change, and we get a Master Equation
describing processes/reactions corresponding to transitions between the
minima of the PES. Again the fast motions in the system have been
removed.
\begin{figure}[ht]
\includegraphics[width=\hsize]{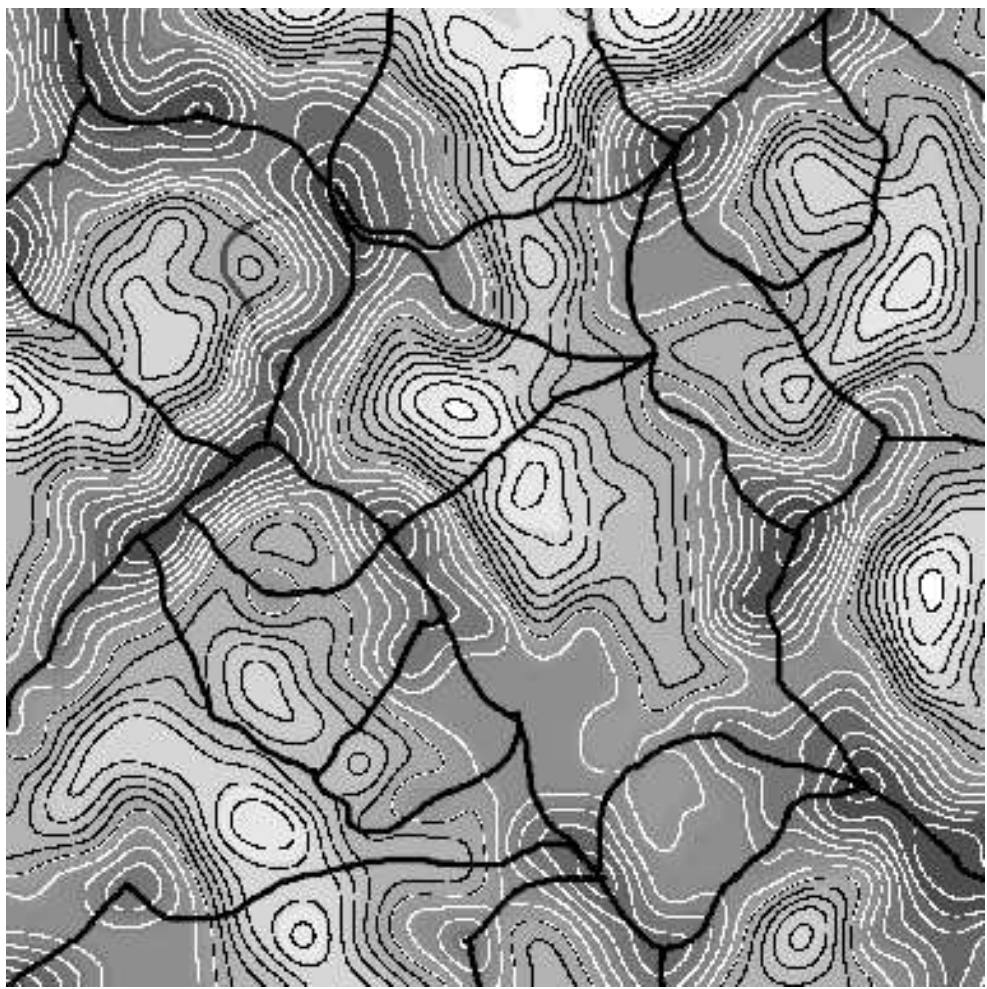}
\caption{A potential-energy surface and the regions $R_\alpha$ around the
  minima. The lighter areas and the black contour lines depict lower
  values. The darker areas and the white contour lines depict the higher
  values. The fat dark lines indicate the boundaries $S_{\beta\alpha}$.}
\label{fig:catchpes}
\end{figure}

The advantage of a system with translation symmetry has to do with the
number of different transition probabilities $W$. For the general case
based on minima of the PES there is a different transition probability
for each transition. For reactions on a surface the situation is
simpler, because the same reaction occurring at different sites
corresponds to different configuration changes but has the same
transition probability.
\chapter[How to Get the Transition Probabilities?]{How to Get the
  Transition Probabilities?}
\label{ch:transprob}

The Master Equation is only useful if one knows the transition
probabilities. There are basically two ways to get them. One way is to
calculation them. The other is to derive them from experimental data.
\section[Quantum chemical calculations of transition
probabilities]{Quantum chemical calculations of transition
  probabilities}

There are three difference between expressions~(\ref{eq:pfun})
and~(\ref{eq:ppfun}) for the partition function and those of
TST.\cite{san95} The first is the absence of an exponential factor of
the form $\exp(-E_{\rm bar}/k_{\rm B}T)$, the second is the boundaries
of the integrations, and the third is the absence of a reference to a
transition state. We deal with the boundaries first. Very often these
can simply be removed. Define ${\bf q}^{({\rm min})}$ as the point in
$R_\alpha$ at which $V$ is minimal, and approximate $V$ in $R_\alpha$ by
\begin{equation}
  V_{\rm harm}({\bf q})=V({\bf q}^{({\rm min})})
  +{1\over 2}\sum_{i,j}(q_i-q_i^{({\rm min})})
  {\partial^2V\over\partial q_i\partial q_j}({\bf q}^{({\rm min})})
  (q_j-q_j^{({\rm min})}).
\end{equation}
This is the harmonic approximation. A very common situation is the
following. $V_{\rm harm}$ differs from $V$ only appreciably where
$V-V({\bf q}^{({\rm min})})$ is large with respect to the thermal energy
$k_{\rm B}T$. Because of the Boltzmann-factor in the integrals we can
replace $V$ by $V_{\rm harm}$ in the integrals. The integration over
$R_\alpha$ can then also be extended to infinity. The reason for this is
that, in the region that has been added to the integral, the
Boltzmann-factor with $V_{\rm harm}$ is so small that the added part is
negligible. $Q$ thus becomes the normal expression for the classical
partition function.

Note that we do not really need to make the harmonic approximation.
Anharmonicities can be included. Instead of $V_{\rm harm}$ we can use
any approximation to $V$ that is accurate at ${\bf q}$ unless $V({\bf
  q}-V({\bf q}^{({\rm min})})\gg k_{\rm B}T$, and the approximation
should give negligible new contributions to the integrals when the
boundaries are extended beyond $R_\alpha$.

For $Q^\ddagger$ we can draw the same conclusion. We restrict ourselves
to $S_{\alpha\beta}$ and its extension defined by the coordinates used,
and ${\bf q}^{({\rm min})}$ is the point on $S_{\alpha\beta}$ where $V$
is minimal. The rest of the reasoning is then the same as for $Q$. This
also explains another difference with TST. The exponential factor is
obtained by taking the $V({\bf q}^{({\rm min})})$ out off the integrals
for $Q$ and $Q^\ddagger$. This immediately gives the exponential factor
with $E_{\rm bar}$ equal to the difference between the minima of $V$ on
$S_{\alpha\beta}$ and in $R_\alpha$.

There are two corrections to equation~(\ref{eq:Wdef}) that one might
want to make. The first has to do with dynamical
factors;\cite{gri81,tul81} i.e., trajectories leave $R_\alpha$, cross
the surface $S_{\beta\alpha}$, but then immediately return to
$R_\alpha$. Such a trajectory contributes to the transition probability
$W_{\beta\alpha}$, but is not really a reaction. We can correct for this
as in Variational Transition-State Theory (VTST) by shifting
$S_{\beta\alpha}$ along the surface normals.\cite{pec76,tru85} This is
related to the absence of any reference to any transition state so far.
Indeed, if the $S_{\beta\alpha}$ can be chosen more or less arbitrarily
provided the expression for $Q^\ddagger$ is corrected for the dynamical
factors. Using the VTST approach $S_{\beta\alpha}$ will be well-defined.
It turns out that with VTST the transition state (i.e., the saddle point
between the minima in $R_\alpha$ and $R_\beta$) is generally very close
to $S_{\beta\alpha}$, and taking $S_{\beta\alpha}$ so that it contains
the transition state is often a very good
approximation.\cite{pec76,tru85}

The second correction is for some quantum effects.
Equation~(\ref{eq:Wts}) indicates one way to include them. We can simply
replace the classical partition functions by their quantum mechanical
counterparts. (It is possible, of course, to do the integrals over the
momenta in equations~(\ref{eq:pfun}) and~(\ref{eq:ppfun}). The reason
why we did not do that was to retain the correspondence between
classical and quantum partition functions.)  This does not correct for
tunneling and interference effects, however. Inaccuracies due to
tunneling, interference, and dynamic effects are not specific for the
transition probabilities of the Master Equation. TST expressions have
them too. As these effects are often small, this means that in practice
one can use TST expressions to calculate the the transition
probabilities of the Master Equation using quantum chemical methods in
the same way as one calculates rate constants provided that the
partition functions get the dominant contribution from a region in the
integration range surrounding a minimum.

In the harmonic approximation we can write $Q$ as
\begin{equation}
  Q=e^{-V_{\rm min}/k_{\rm B}T}\prod_i q_v(\omega_i)
\end{equation}
with $V_{\rm min}$ the minimum of the potential energy in
$R_\alpha$. The vibrational partition function $q_v$ is given by
\begin{equation}
  q_v(\omega)={k_{\rm B}T\over \hbar\omega}
\end{equation}
classically, or
\begin{equation}
  q_v(\omega)
  ={e^{{1\over 2}\hbar\omega/k_{\rm B}T}\over
   1-e^{\hbar\omega/k_{\rm B}T}}
\end{equation}
quantum mechanically.\cite{mcq76,bec85} The frequencies $\omega_i$ are
the normal mode frequencies.\cite{wil55,gol81} Similarly we find for
$Q_\ddagger$
\begin{equation}
  Q^\ddagger=e^{-V_{\rm min}^\prime/k_{\rm B}T}\prod_j q_v(\omega_j^\prime)
\end{equation}
with $V_{\rm min}^\prime$ the minimum of the potential energy on
$S_{\beta\alpha}$ and $\omega_j^\prime$ the non-imaginary normal mode
frequencies at the transition state. Combining these results yields
\begin{equation}
  W_{\beta\alpha}={k_{\rm B}T\over h}
  {\prod_j q_v(\omega_j^\prime)\over
   \prod_i q_v(\omega_i)}
  e^{-E_{\rm bar}/k_{\rm B}T}
\label{eq:tpvib}
\end{equation}
with $E_{\rm bar}=V_{\rm min}^\prime-V_{\rm min}$.

It is very interesting to look in more detail at the case when it is not
correct to extend the boundaries of the integrals. When the substrate
has a closed-packed structure the potential-energy surface may be quite
flat parallel to the surface. If that is the case there may be
substantial contributions to the partition function up to the boundaries
of the integrals. It is then not possible to apply the reasoning above.
We will look at two examples; both dealing with simple desorption of an
atom. In the first example the potential is completely flat parallel to
the surface for all distances of the atom to the surface. In the second
example the potential only becomes flat at the transition state.

Because dealing with a phase space of many atoms is inconvenient, we
restrict ourselves to just one atom on the surface and calculate the
transition state for desorption for that single atom. This is the usual
approach; try to minimize the number of particles. The step to many
atoms on the surface is made by assuming that the transition probability
is independent on the number of atoms. This is correct if there are no
lateral interactions, as we are assuming here. The case with lateral
interactions will be discussed later.

Figure~\ref{fig:psatom} show regions in phase space corresponding to an
atom adsorbed on different sites. Crossing the upper horizontal plane
bounding a region constitutes a desorption. Crossing one of the vertical
planes bounding a region constitutes a diffusion to another site. The
integrals of each momentum in the expression for the partition functions
$Q$ and $Q^\ddagger$ become
\begin{equation}
  \int_{-\infty}^\infty\!\!{dp\over h}
  \exp\left[-{p^2\over 2mk_{\rm B}T}\right]
  ={1\over h}\sqrt{2\pi mk_{\rm B}T}.
\end{equation}
The integrals of the coordinates for $Q^\ddagger$ become
\begin{equation}
  \int_{S_{\alpha\beta}}\!\!dS
  \exp\left[-{V\over k_{\rm B}T}\right]
  =A_{\rm site}\exp\left[-{V(z_{\rm TS})\over k_{\rm B}T}\right],
\end{equation}
where $z_{\rm TS}$ is the value of the coordinate perpendicular to the
surface at the transition state for desorption, and $A_{\rm site}$ is
the area of the horizontal boundary plane of a region or the area of a
single site.
\begin{figure}[ht]
\includegraphics[width=\hsize]{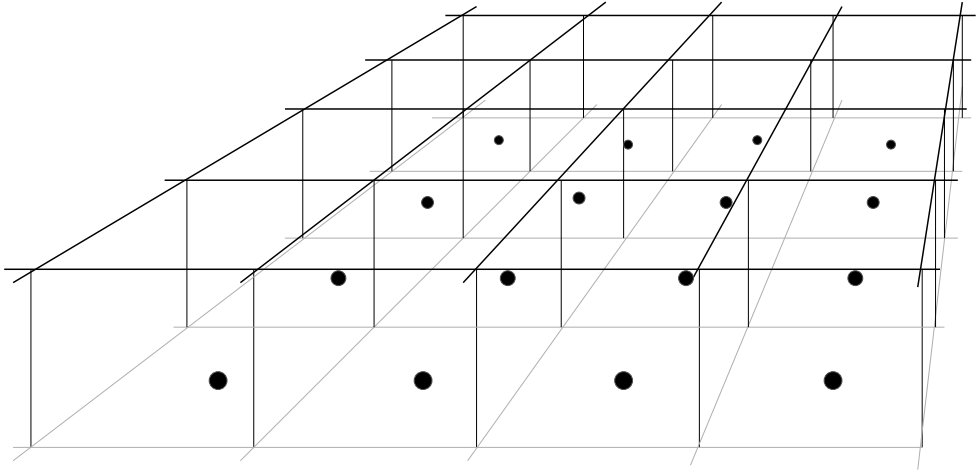}
\caption{The phase space (or rather the configuration space, as the
  momenta are not shown) of a single atom. The lattice indicates the
  adsorption site and the bounded regions are parts of phase space
  corresponding to the atom being adsorbed on particular sites. The
  upper horizontal boundary plane contains the transition state for
  desorption. The lower horizontal plane represents the surface.}
\label{fig:psatom}
\end{figure}

The integrals of the coordinates for $Q$ differ between our two
examples. If the potential has a well-defined minimum near the
adsorption site, then we can use the harmonic approximation (this is our
second example).
\begin{eqnarray}
  V_{\rm harm}&=&V(x^{({\rm min})},y^{({\rm min})},z^{({\rm min})})\\
  &+&{1\over 2}m\omega_\parallel^2(x-x^{({\rm min})})^2
  +{1\over 2}m\omega_\parallel^2(y-y^{({\rm min})})^2
  +{1\over 2}m\omega_\perp^2(z-z^{({\rm min})})^2.\nonumber
\end{eqnarray}
The integrals become
\begin{equation}
  \int_{R}\!\!d{\bf q}\exp\left[-{V\over k_{\rm B}T}\right]
  ={1\over \omega_\parallel^2\omega_\perp}
  \left({2\pi k_{\rm B}T\over m}\right)^{3/2}
  \exp\left[-{V(x^{({\rm min})},y^{({\rm min})},z^{({\rm min})})
           \over k_{\rm B}T}\right].
\end{equation}
If the potential is flat parallel to the surface (this is our first
example), then we can use the harmonic approximation only perpendicular
to the surface. We get the same expression as above but with
$\omega_\parallel=0$. The integrals now become
\begin{equation}
  \int_{R}\!\!d{\bf q}\exp\left[-{V\over k_{\rm B}T}\right]
  =A_{\rm site}{1\over \omega_\perp}
  \left({2\pi k_{\rm B}T\over m}\right)^{1/2}
  \exp\left[-{V(x^{({\rm min})},y^{({\rm min})},z^{({\rm min})})
           \over k_{\rm B}T}\right].
\end{equation}

All results can now be combined. For $Q^\ddagger$ the result is
\begin{equation}
  Q^\ddagger=q_f^2
  \exp\left[-{V(z_{\rm TS})\over k_{\rm B}T}\right]
\end{equation}
with
\begin{equation}
  q_f\equiv {1\over h}\sqrt{2\pi A_{\rm site}mk_{\rm B}T},
\end{equation}
which is the partition functions for one degree of freedom of a free
particle. For the first example the partition function $Q$ becomes
\begin{equation}
  Q=q_f^2q_v(\omega_\perp)
  \exp\left[-{V(x^{({\rm min})},y^{({\rm min})},z^{({\rm min})})
           \over k_{\rm B}T}\right]
\end{equation}
with
\begin{equation}
  q_v(\omega)\equiv{k_{\rm B}T\over\hbar\omega},
\end{equation}
which is the partition function for a one-dimensional harmonic
oscillator. The rate constant is then
\begin{equation}
  W_{\rm des}={\omega_\perp\over 2\pi}
  \exp\left[-{E_{\rm bar}\over k_{\rm B}T}\right]
\end{equation}
with $E_{\rm bar}\equiv V(z_{\rm TS})-V(x^{({\rm min})},y^{({\rm
    min})},z^{({\rm min})})$. (The derivation of
equation~(\ref{eq:tpvib}) is similar to what we show here.) Note that
this is a classical expression. The main quantum effect is included by
replacing the partition function $q_v$ by its quantum mechanical
counterpart.\cite{mcq76,bec85}
\begin{equation}
  q_v(\omega)
  ={e^{{1\over 2}\hbar\omega/k_{\rm B}T}\over
   1-e^{\hbar\omega/k_{\rm B}T}}
\end{equation}
If $k_{\rm B}T\ll\hbar\omega_\perp$ then $q_v=\exp[-{1\over
  2}\hbar\omega_\perp/k_{\rm B}T]$ is a good approximation so that
\begin{equation}
  W_{\rm des}={k_{\rm B}T\over h}
  \exp\left[-{E_{\rm bar}\over k_{\rm B}T}\right]
  \label{eq:Wdes1}
\end{equation}
with $E_{\rm bar}=V(z_{\rm TS})-[V(x^{({\rm min})},y^{({\rm
    min})},z^{({\rm min})})+{1\over 2}\hbar\omega_\perp]$. The last term
is the zero-point energy of the adsorbed atom.\cite{mes61}

For the second example we have
\begin{equation}
  Q=q_v(\omega_\perp)q_v^2(\omega_\parallel)
  \exp\left[-{V(x^{({\rm min})},y^{({\rm min})},z^{({\rm min})})
           \over k_{\rm B}T}\right]
\end{equation}
so that
\begin{equation}
  W_{\rm des}={2\pi A_{\rm site}\omega_\perp\omega_\parallel^2
     \over k_{\rm B}T}
  \exp\left[-{E_{\rm bar}\over k_{\rm B}T}\right]
\end{equation}
classically, and
\begin{equation}
  W_{\rm des}={2\pi A_{\rm site}m(k_{\rm B}T)^2\over h^3}
  \exp\left[-{E_{\rm bar}\over k_{\rm B}T}\right]
  \label{eq:Wdes2}
\end{equation}
quantum mechanically if $\omega_\perp,\omega_\parallel\gg k_{\rm
B}T/\hbar$ and with $E_{\rm bar}=V(z_{\rm TS})-[V(x^{({\rm
  min})},y^{({\rm min})},z^{({\rm min})})+\hbar\omega_\parallel+{1\over
2}\hbar\omega_\perp]$.

The expressions above show that the main properties that should be
determined in a quantum chemical calculation is the barrier $E_{\rm
  bar}$, and the vibrational frequencies $\omega_\perp$ and possible
$\omega_\parallel$. It is interesting to calculate the preexponential
factors in equations~(\ref{eq:Wdes1}) and~(\ref{eq:Wdes2}), assuming
that the vibrational excitation energies are large compared to the
thermal energies. This is probably correct for Xe desorption from
Pt(111). If there is little corrugation ($\omega_\parallel\approx 0$)
then we have to use equation~(\ref{eq:Wdes1}). The preexponential factor
in that expression at $T=100\,$K equals $k_{\rm B}T/h=2.1\cdot
10^{12}\,{\rm s}^{-1}$. If we assume that Xe adsorbs strongly to a
particular site then we are dealing with equation~(\ref{eq:Wdes2}). With
$m=2.2\cdot 10^{-25}\,$kg, $A=6.4\cdot 10^{-20}\,{\rm m}^2$ we get $2\pi
Am(k_{\rm B}T)^2/h^3=5.8\cdot 10^{14}\,{\rm s}^{-1}$.

The usual way to write a rate constant is $\nu\exp[-E{\rm act}/k_{\rm
  B}T]$. (We will see that rate constants and transition probabilities
are very similar. We will from now on often use the term rate constant
instead of transition probabilities.) The preexponential factor $\nu$
and the activation energy $E_{\rm act}$ is usually assumed to be
independent of temperature. This is done certainly in the experimental
literature when one determines these kinetic parameters from
measurements of rate constants as a function of temperature. Using this
form for the rate constant one may define the activation energy as
\begin{equation}
  E_{\rm act}\equiv -{d\ln W\over d(1/k_{\rm B}T)}
  =k_{\rm B}T^2{d\ln W\over dT},
\end{equation}
where $W$ is a transition probability, and the preexponential factor as
\begin{equation}
  \nu\equiv W\exp\left[{E_{\rm act}\over k_{\rm B}T}\right].
\end{equation}
With these definitions the kinetic parameters are often found not to be
temperature independent. Getting back to desorption from a surface in
which the corrugation of the potential is negligible we find from
equation~(\ref{eq:Wdes1}) that $E_{\rm act}=E_{\rm bar}+k_{\rm B}T$ and
$\nu=ek_{\rm B}T/h$. For desorption from a surface with corrugation we
find from equation~(\ref{eq:Wdes2}) that $E_{\rm act}=E_{\rm
  bar}+2k_{\rm B}T$ and $\nu=2\pi A_{\rm site}(ek_{\rm B}T)^2/h^3$. We
see that indeed the activation energy and the preexponential factor are
temperature dependent. The dependence for the activation energy is
small, because the thermal energy $k_{\rm B}T$ is general small compared
to the barrier height $E_{\rm bar}$.  The effect on the preexponential
factor seems larger, but one should remember that rate constants vary
over many orders of magnitude, and the effect of temperature on the
preexponential factor affects the order of magnitude of the
preexponential factor only a little. Moreover, there is a {\it
  compensation effect\/}.  Increasing the temperature increases the
preexponential factor, but also the activation energy, so the effect on
the rate constant is reduced.

The experimental determination of the activation energy and
preexponential factor does not use the expression above of course.
Experimentalists plot the logarithm of a rate constant versus the
reciprocal temperature and then fit a linear curve to it. This is
something we can do as well with equations~(\ref{eq:Wdes1}) and
\ref{eq:Wdes2}. The result will depend on the temperature interval on
which we do the fit, but we will see that the dependence is generally
small. If we take, for example, equation~(\ref{eq:Wdes1}) and plot
$\ln(hW_{\rm des}/E_{\rm bar})$ versus $\beta\equiv E_{\rm bar}/k_{\rm
  B}T$ we get figure~\ref{fig:kexp}. The function that is plotted in
this figure is $-\beta-\ln(\beta)$. Although this function is not
linear, we see that only if $\beta$ is small $\ln(\beta)$ is of similar
size as $\beta$ and deviations of non-linearity are noticeable. This
only occurs at such high temperature that $\beta<1$, whereas
experimentally one usually works at temperatures with $\beta\gg 1$.
\begin{figure}[ht]
\includegraphics[width=\hsize]{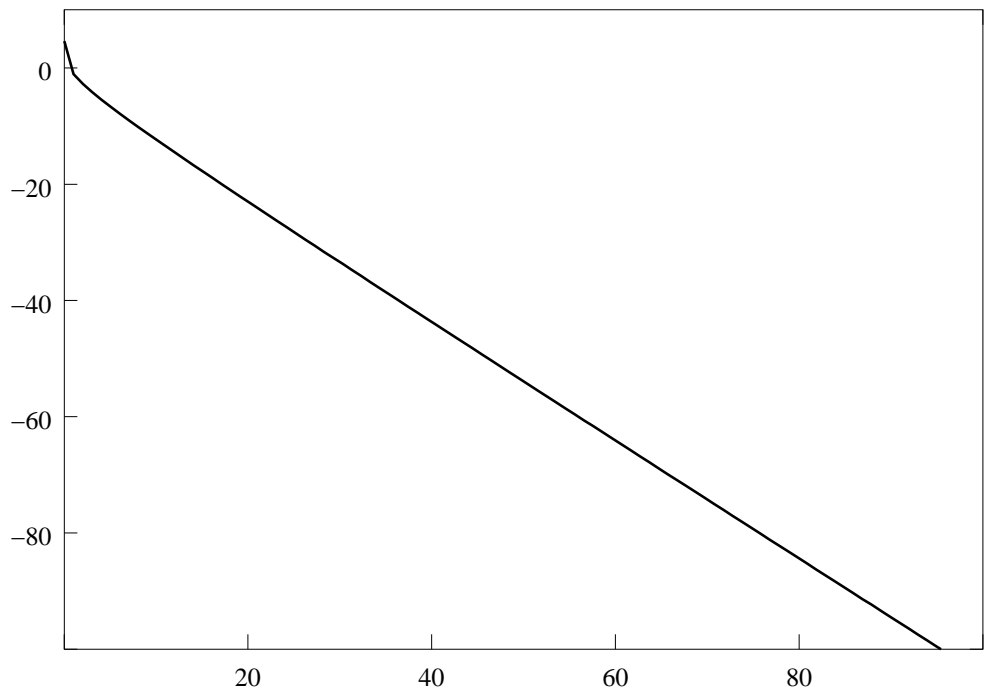}
\caption{The logarithm of the rate constant $\ln(Wh/E_{\rm bar})$
  according to equation~(\ref{eq:Wdes1}) plotted versus reciprocal
  temperature $E_{\rm bar}/k_{\rm B}T$. Although the expression is not
  linear, the deviation of linearity is only very small, and only
  visible at high temperatures (i.e., small values of $E_{\rm
  bar}/k_{\rm B}T$).}
\label{fig:kexp}
\end{figure}

If we fit $\nu\exp[-E{\rm act}/k_{\rm B}T]$ to equation~(\ref{eq:Wdes1})
on the interval $[T_{\rm low},T_{\rm high}]$ in the experimental way, we
have to minimize
\begin{equation}
  \int_{1/T_{\rm high}}^{1/T_{\rm low}}
  \!\!d\left({1\over T}\right)
  \left[
  \ln W_{\rm des}-\ln\nu+{E_{\rm act}\over k_{\rm B}T}
  \right]^2
\end{equation}
as a function of $\ln\nu$ and $E_{\rm act}$. The mathematics is
straightforward, and the result is
\begin{equation}
  E_{\rm act}=E_{\rm bar}
  \left[
  1+{3k_{\rm B}T_{\rm high}\over E_{\rm bar}}
  {f^2-1-2f\ln f\over (f-1)^3}
  \right],
\end{equation}
with $f=T_{\rm high}/T_{\rm low}$. The second term in square brackets is
small because the barrier height is generally much larger than the
thermal energy. The factor with the $f$'s decreases monotonically from 1
for $f=1$ to 0 for $f\to\infty$. For the preexponential factor we find
\begin{equation}
  \ln\nu=\ln{k_{\rm B}T_{\rm high}\over h}
  +\left[
  {5f^2+2f+5\over 2(f-1)^2}-{f(f^2+f+4)\over (f-1)^3}\ln f
  \right].
\end{equation}
The expression in square brackets with the $f$'s varies from 1 for $f=1$
to $-\infty$ for $f\to\infty$ and becomes 0 at $f\approx 5.85$. This
means that it is generally small compared to the first term in the
expression for $\ln\nu$.

If we fit $\nu\exp[-E{\rm act}/k_{\rm B}T]$ to equation~(\ref{eq:Wdes2})
on the interval $[T_{\rm low},T_{\rm high}]$ in the experimental way, we
get
\begin{equation}
  E_{\rm act}=E_{\rm bar}
  \left[
  1+{6k_{\rm B}T_{\rm high}\over E_{\rm bar}}
  {f^2-1-2f\ln f\over (f-1)^3}
  \right],
\end{equation}
and
\begin{equation}
  \ln\nu=\ln{2\pi A_{\rm site}m(k_{\rm B}T_{\rm high})^2\over h^3}
  +2\left[
  {5f^2+2f+5\over 2(f-1)^2}-{f(f^2+f+4)\over (f-1)^3}\ln f
  \right].
\end{equation}
We see that the result is very similar to the previous case and that
here too the choice of the temperature interval has only a marginal
effect.
\section{Transition probabilities from experiments}

One of the problems of calculating transition probabilities is the
accuracy. The method that is mostly used to calculate the energetics on
adsorbates on a transition metal surface is Density-Functional Theory
(DFT).\cite{hoh64,koh65,par89a} Estimates of the error made using DFT
for such systems are at least about $10\,$kJ/mol. An error of this size
in the activation energy means that at room temperature the transition
probability is off by about two orders of magnitude. How well a
preexponential factor can be calculated is not really known at all. This
does not mean that calculating transition probabilities is useless. The
errors in the energetics have less effect, if the temperature is higher,
but even more important is that one can calculate transition
probabilities for processes that are experimentally hardly or not
accessible. If, one the other hand, one can obtain transition
probabilities from an experiment, then the value that is obtained is
generally more reliable than one calculated.

In general, one has to deal with a system in which several reactions can
take place at the same time. The crude approach to obtain transition
probabilities from experiments is then to try to fit all transition
probabilities to the experiments at the same time. This is often not a
good idea. First of all such a procedure can be quite complicated. The
data that one gets from an experiment are seldom a linear function of
the transition probabilities. Consequently the fitting procedure
consists of minimizing a nonlinear function that stands for the
difference between experimental and the calculated or simulated data.
Such a function normally has many local minima, and it is very hard to
find the best set of transition probabilities. But this isn't even the
most important drawback. Although one may be able to do a very good fit
of the experimental data, this need not mean that the transition
probabilities are good; given enough fit parameters, one can fit
anything.

Deriving kinetic parameters from experiments does work well, when one
has an experiment of a single simple process that can be described by
just one or two parameters. The process should be simple in the sense
that one has an analytical expression with which one can derive
relatively easily the kinetic parameters given experimental data. The
analytical expression should be exact or at least a very good
approximation. If one has to deal with a reaction system that is
complicated and consists of many reactions, then one should try to get
experiments that measure just one of the reactions. For example, in CO
oxidation one has at least adsorption of CO, dissociative adsorption of
oxygen, and the formation of ${\rm CO}_2$. Instead of trying to fit rate
constants of these three reactions simultaneously, one should look at
experiments that show only one of these reactions. An experiment that
only measures sticking coefficients as a function of CO pressure can be
used to get the CO adsorption rate constant. The following sections show
a number of processes which can be used to get kinetic parameters, and
we show how to get the parameters.
\subsection{Relating macroscopic properties to microscopic processes}

The analytical expressions mentioned above should relate some property
that is measured to the transition probabilities. We will address first
the general relation. This relation is exact, but often not very useful.
In the next sections we will show situations were the general relation
can be simplified either exactly or with the use of some approximation.

If a system is in a well-defined configuration then a macroscopic
property can generally be computed easily. For example, the number of
molecules of a particular type in the adlayer can be obtained simply be
counting. If the property that we are interested in is denoted by $X$,
then its value when the system is in configuration $\alpha$ is given by
$X_\alpha$. As our description of the system uses probabilities for the
configurations, we have to look at the expectation value of $X$, which
is given by
\begin{equation}
  \expec{X}=\sum_\alpha P_\alpha X_\alpha.
\end{equation}
Kinetic experiment measure changes, so we have to look at
$d\expec{X}/dt$. This is given by
\begin{equation}
  {d\expec{X}\over dt}=\sum_\alpha {dP_\alpha\over dt} X_\alpha,
\end{equation}
because $X_\alpha$ is a property of a fixed configuration. We can remove
the derivative of the probability using the Master Equation.  This gives
us
\begin{eqnarray}
  {d\expec{X}\over dt}
  &=&\sum_{\alpha\beta}\left[W_{\alpha\beta}P_\beta
   -W_{\beta\alpha}P_\alpha\right] X_\alpha,\nonumber\\
  &=&\sum_{\alpha\beta}W_{\alpha\beta}P_\beta
   \left[X_\alpha-X_\beta\right].
\label{eq:changeX}
\end{eqnarray}
The second step is obtained by swapping the summation indices. The final
result can be regarded as the expectation value of the change of $X$ in
the reaction $\beta\to\alpha$ times the rate constant of that reaction.
This general equation forms the basis for deriving relations between
macroscopic properties and transition probabilities.
\subsection{Unimolecular desorption}

Suppose we have atoms or molecules that adsorb onto one particular type
of site. We assume that we have of surface of area $A$ with $S$
adsorption sites. If $N_\alpha$ is the number of atoms/molecules in
configuration $\alpha$ then
\begin{equation}
  {d\expec{N}\over dt}
  =\sum_{\alpha\beta}W_{\alpha\beta}P_\beta
   \left[N_\alpha-N_\beta\right].
\end{equation}
Diffusion does not change the number atoms/molecules, and it does not
matter in this case whether we include it or not. The only relevant
process that we look at is desorption. For the summation over $\alpha$
we have to distinguish between two types of terms; the ones where
$\alpha$ can originate from $\beta$ by a desorption, and the ones where
it cannot. The latter terms have $W_{\alpha\beta}=0$ and so they to not
contribute to the sum. The former do contribute and we have
$W_{\alpha\beta}=W_{\rm des}$, with $W_{\rm des}$ the transition
probability for desorption, and $N_\alpha-N_\beta=-1$. So all these
non-zero terms contribute equally to the sum for a given configuration
$\beta$. Moreover, the number of these terms is equally to the number of
atoms/molecules in $\beta$ that can desorb, because each desorbing
atom/molecule yields a different $\alpha$. So
\begin{equation}
  {d\expec{N}\over dt}
  =-W_{\rm des}\sum_\beta P_\beta N_\beta
  =-W_{\rm des}\expec{N}.
\end{equation}
This is an exact expression. Dividing by the number of sites $S$ gives
the rate equation for the coverage $\theta=\expec{N}/S$.
\begin{equation}
  {d\theta\over dt}=-W_{\rm des}\theta.
\end{equation}
If we compare this to the macroscopic rate equation $d\theta/dt=-k_{\rm
  des}\theta$ with $k_{\rm des}$ the macroscopic rate constant, we see
that $k_{\rm des}=W_{\rm des}$.

For isothermal desorption $k_{\rm des}$ does not depend on time and the
solution to the rate equation is
\begin{equation}
  \theta(t)=\theta(0)\exp[-k_{\rm des}t],
\end{equation}
where $\theta(0)$ is the coverage at time $t=0$. Kinetic experiments often
measure rates, and for the desorption rate we have
\begin{equation}
  {d\theta\over dt}(t)=-k_{\rm des}\theta(0)\exp[-k_{\rm des}t].
\end{equation}
We can now obtain the rate constant by measuring, for example, the rate
of desorption as a function of time and plotting minus the logarithm of
the rate as a function of time. Because
\begin{equation}
  \ln\left[-{d\theta\over dt}(t)\right]
  =\ln[k_{\rm des}\theta(0)]-k_{\rm des}t,
\end{equation}
we can obtain the rate constant which equals minus the slope of the
straight line. The same would hold if we would plot the logarithm of the
coverage as a function of time. Because of the equality this immediately
also yields the transition probability to be used in a simulation.

If the rate constant depends on time then solving the rate equation is
often much more difficult. We can always rewrite the rate equation as
\begin{equation}
  {1\over\theta}{d\theta\over dt}
  {d\ln\theta\over dt}
  =-k_{\rm des}.
\end{equation}
Integrating this equation yields
\begin{equation}
  \ln\theta(t)-\ln\theta(0)
  =-\int_0^t\!\!dt^\prime k_{\rm des}(t^\prime),
\end{equation}
or
\begin{equation}
  \theta(t)
  =\theta(0)\exp\left[-\int_0^t\!\!dt^\prime
  k_{\rm des}(t^\prime)\right].
\end{equation}
Whether of not we can get an analytical solution depends on whether we
can determine the integral. In Temperature-Programmed Desorption
experiments we have
\begin{equation}
  k_{\rm des}(t)
  =\nu\exp\left[-{E_{\rm act}\over k_{\rm B}(T_0+Bt)}\right]
\end{equation}
with $E_{\rm act}$ an activation energy, $\nu$ a preexponential factor,
$k_{\rm B}$ the Boltzmann-factor, $T_0$ the temperature at time $t=0$,
and $B$ the heating rate. The integral can be calculated analytically.
The result is
\begin{equation}
  \int_0^t\!\!dt^\prime\,
  \nu\exp\left[-{E_{\rm act}\over k_{\rm B}(T_0+Bt^\prime)}\right]
  =\Omega(t)-\Omega(0)
\end{equation}
with
\begin{equation}
  \Omega(t)={\nu\over B}(T_0+Bt){\rm E}_2
  \left[{E_{\rm act}\over k_{\rm B}(T_0+Bt)}\right],
\label{eq:omega}
\end{equation}
where ${\rm E}_2$ is an exponential integral.\cite{abr65} Although
this solution has been derived some time ago,\cite{jan95b} it has not
yet been used in the analysis of experimental spectra, but there are
several numerical techniques that work well for such simple
desorption.\cite{dej90} Note that we have not made any approximations
here and the transition probability $W_{\rm des}$ that we obtain will be
exact except for experimental errors.
\subsection{Unimolecular adsorption}
\label{subsec:uniads}

We start with the simplest case in which the adsorption rate is
proportional to the number of vacant sites, which is called Langmuir
adsorption. We will only indicate in this section in what way in the
common situation in which the adsorption is higher than expected based
on the number of vacant sites differs.\cite{tho97,san95,som93}  This
so-called precursor-mediated adsorption is really a composite process,
and has to be treated with the knowledge presented in various sections
of this chapter.

Again suppose we have atoms or molecules that adsorb onto one particular
type of site. We assume that we have a surface of area $A$ with $S$
adsorption sites. If $N_\alpha$ is the number of atoms/molecules in
configuration $\alpha$ then again
\begin{equation}
  {d\expec{N}\over dt}
  =\sum_{\alpha\beta}W_{\alpha\beta}P_\beta
   \left[N_\alpha-N_\beta\right].
\end{equation}
Diffusion can again be ignored. For the summation over $\alpha$ we have
to distinguish between two types of terms; the ones in which $\alpha$
can originate from $\beta$ by a adsorption, and the ones it cannot. The
latter terms have $W_{\alpha\beta}=0$ and so they to not contribute to
the sum. The former do contribute and we have $W_{\alpha\beta}=W_{\rm
  ads}$, with $W_{\rm ads}$ the transition probability for adsorption,
and $N_\alpha-N_\beta=1$. So all these non-zero terms contribute equally
to the sum for a given configuration $\beta$. Moreover, the number of
these terms is equally to the number of vacant sites in $\beta$ onto
which the molecules can adsorb, because each adsorption yields a
different $\alpha$. The number of vacant sites in configuration $\beta$
equals $S-N_\beta$, so
\begin{equation}
  {d\expec{N}\over dt}
  =W_{\rm ads}\sum_\beta P_\beta (S-N_\beta)
  =W_{\rm ads}(S-\expec{N}).
\end{equation}
Dividing by the number of sites $S$ gives the rate equation for the
coverage $\theta=\expec{N}/S$.
\begin{equation}
  {d\theta\over dt}=-W_{\rm des}(1-\theta).
\end{equation}
If we compare this to the macroscopic rate equation $d\theta/dt=k_{\rm
  ads}(1-\theta)$ with $k_{\rm ads}$ the macroscopic rate constant, we
see that $k_{\rm ads}=W_{\rm ads}$.

So far adsorption is almost the same as desorption. The only difference
is where we had $\theta$ for desorption we have $1-\theta$ for
adsorption on the right-hand-side of the rate equation. An importance
difference now arises however. Whereas the macroscopic rate constant for
desorption $k_{\rm des}$ is an basic quantity in kinetics of surface
reactions, $k_{\rm ads}$ is generally related to other properties. This
is because the adsorption process consists of atoms or molecules
impinging on the surface, and that is something that can be described
very well with kinetic gas theory.

Suppose that the pressure of the gas is $P$ and its temperature $T$,
then the number of molecules $F$ hitting a surface of unit area per unit
time is given by\cite{mcq76,bec85}
\begin{equation}
  F={P\over\sqrt{2\pi mk_{\rm B}T}}
\end{equation}
with $m$ the mass of the atom or molecule.  Not every atom or molecule
that hits a surface will stick to it. The sticking coefficient $\sigma$
is defined as the ratio of the number of molecules that stick to the
total number hitting the surface. It can also be looked upon as the
probability that an atom or molecule hitting the surface sticks. The
change in the number of molecules in an area $A$ due to adsorption can
then be written as the vacant area times the flux $F$ times the sticking
coefficient $\sigma$. The vacant area equals to area $A$ times the
fraction of sites in that area that is not occupied.  This all leads to
\begin{equation}
  {d\expec{N}\over dt}=A(1-\theta)F\sigma.
\end{equation}
If we compare this to the equations above we find
\begin{equation}
  W_{\rm ads}={AF\sigma\over S}
    ={PA_{\rm site}\sigma\over\sqrt{2\pi mk_{\rm B}T}},
\label{eq:adflux}
\end{equation}
where $A_{\rm site}$ is the area of a single site.

Adsorption described so far is proportional to the number of vacant
sites. Experiments measure the rate of adsorption and with the
expressions derived above one can calculate the microscopic rate
constant $W_{\rm ads}$. However, it is often found that the rate of
adsorption starts at a certain value for a bare surface and then hardly
changes when particles adsorb until the surface is almost completely
covered at which time it suddenly drops to zero. This behavior is
generally explained by describing the adsorption as a composite
process.\cite{tho97,san95,som93} A molecule impinging unto the surface
adsorbs with the probability $\sigma$ when the site it hits is vacant
just as before.  However, a molecule that hits a site that is already
occupied need not be scattered. It can adsorb indirectly. It first
adsorbs, with a certain probability, in a second adsorption layer. Then
it starts to diffuse over the surface in this second layer. It can
desorb at a later stage, or, and that's the important part, it can
encounter a vacant site and adsorb there permanently. This last part can
increase the adsorption rate substantially when there are already many
sites occupied. The precise dependence of the adsorption rate on the
coverage $\theta$ is determined by the rate of diffusion, by the rate of
adsorption onto the second layer, and by the rate of desorption from the
second layer. If there are factors that affect the structure of the
first adsorption layer, e.g. lateral interaction, then these too
influence the adsorption rate. If the adsorption is not direct, one
talks about a precursor mechanism. A precursor on top of an adsorbed
particle is an extrinsic precursor. An intrinsic precursor can be found
on top of a vacant site.\cite{cas81} The precursor mechanism will not
always be operative for a bare surface; i.e., there is not always an
intrinsic precursor. This means that we can use
equation~(\ref{eq:adflux}) if we take for $\sigma$ the sticking
coefficient for adsorption on a bare surface.
\subsection{Unimolecular reactions}

With the knowledge of simple desorption and adsorption given above it is
now easy to derive an expression for the rate constant $W_{\rm uni}$ for
a unimolecular reaction in term of a macroscopic rate constant. In fact
the derivation is exactly the same as for the desorption. Desorption
changes a site from A to $*$, whereas a unimolecular reaction changes it to
B. Replace $*$ by B in the expression for the desorption (and $W_{\rm
  des}$ by $W_{\rm uni}$ of course) and you have the correct expression.
As the expression for desorption do not contain a $*$, the procedure is
trivial and we find $W_{\rm uni}=k_{\rm uni}$ where $k_{\rm uni}$ is the
rate constant from the macroscopic rate equation.
\subsection{Diffusion}
\label{subsec:diff}

We treat diffusion as any other reaction, but experimentally one doesn't
look at changes in coverages but at displacements of atoms and
molecules. We will therefore also look here at how the position of a
particle changes.

We assume that we have only one particle on the surface, so that the
particle's movement is not hindered by any other particle. We also
assume that we have a square grid with axis parallel to the $x$- and the
$y$-axis and that the distance between grid points is given by $a$. We
will later look at other grids. If $x_\alpha$ is the $x$-coordinate of
the particle in configuration $\alpha$, then
\begin{equation}
  {d\expec{x}\over dt}
  =\sum_{\alpha\beta}W_{\alpha\beta}P_\beta[x_\alpha-x_\beta].
\end{equation}
The $x$-coordinate change because the particle hops from one to another
site. When it hops we have $x_\alpha-x_\beta=a,-a,\hbox{ and }0$ for a
hop along the $x$-axis towards larger $x$, a hop along the $x$-axis
towards smaller $x$, or a hop perpendicular to the $x$-axis,
respectively.  All these hops have a rate constant $W_{\rm hop}$ and are
equally likely. This means $d\expec{x}/dt=0$. The same holds for the
$y$-coordinate.

More useful is to look at the square of the coordinates. We then find
\begin{equation}
  {d\expec{x^2}\over dt}
  =\sum_{\alpha\beta}W_{\alpha\beta}P_\beta[x_\alpha^2-x_\beta^2].
\end{equation}
Now we have $x_\alpha^2-x_\beta^2=2ax_\beta+a^2,-2ax_\beta+a^2,\hbox{
  and }0$, respectively. Because the hops are still equally likely, we
have
\begin{equation}
  {d\expec{x^2}\over dt}=2W_{\rm hop}a^2.
\end{equation}
We find the same for the $y$-coordinate. The macroscopic equation for
diffusion is
\begin{equation}
  {d\expec{x^2+y^2}\over dt}=4D,
\end{equation}
with $D$ the diffusion coefficient. From this we see that we have
$W_{\rm hop}=D/a^2$.

On a hexagonal grid a particle can hop in six different directions for
which $x_\alpha-x_\beta=a,a/2,-a/2,-a,-a/2,\hbox{ and } a/2$ and
$y_\alpha-y_\beta=0,a\sqrt{3}/2,a\sqrt{3}/2,0,-a\sqrt{3}/2,\hbox{ and }
-a\sqrt{3}/2$. From this we get again $d\expec{x}/dt=0$. For the squared
displacement we find $x_\alpha^2-x_\beta^2=2ax_\beta+a^2,
ax_\beta+a^2/4, -ax_\beta+a^2/4, -2ax_\beta+a^2, -ax_\beta+a^2/4,
ax_\beta+a^2/4$. This yields again $d\expec{x^2}/dt=2W_{\rm hop}a^2$. We
find the same expression for the $y$-coordinate, so that also for a
hexagonal grid $W_{\rm hop}=D/a^2$. The same expression holds for a
trigonal grid. The derivation is identical to the ones for the square
and hexagonal grids.

\subsection{Bimolecular reactions}
\label{subsec:bimolrx}

For all of the processes we have looked at so far it was possible to
derive the macroscopic equations from the the Master Equation exactly.
This is not the case for bimolecular reactions. Bimolecular reactions
will give rise to an infinite hierarchy of macroscopic rate equations.
There are two bimolecular reactions we will consider: ${\rm A}+{\rm B}$
and ${\rm A}+{\rm A}$. The problem we have mentioned above is the same
for both reactions, but there is a small difference in the derivation of
a numerical factor in the macroscopic rate equation. We will start with
the ${\rm A}+{\rm B}$ reaction.

We look at the number of A's. The expressions for the number of B's can
be obtained by replacing A's by B's and B's by A's in the following
expressions. We have
\begin{equation}
  {d\expec{N^{\rm (A)}}\over dt}
  =\sum_{\alpha\beta}W_{\alpha\beta}P_\beta
   \left[N_\alpha^{\rm (A)}-N_\beta^{\rm (A)}\right],
\end{equation}
where $N_\alpha^{\rm (A)}$ stands for the number of A's. If $\alpha$ can
originate from $\beta$ by a ${\rm A}+{\rm B}$ reaction, then
$W_{\alpha\beta}=W_{\rm rx}$, otherwise $W_{\alpha\beta}=0$. If such a
reaction is possible, then $N_\alpha^{\rm (A)}-N_\beta^{\rm (A)}=-1$.
The problem now is with the number of configurations $\alpha$ that can
be obtained from $\beta$ by a reaction. This number is equal to the
number of AB pairs $N_\beta^{\rm (AB)}$. This leads then to
\begin{equation}
  {d\expec{N^{\rm (A)}}\over dt}
  =-W_{\rm rx}\sum_\beta P_\beta N_\beta^{\rm (AB)}
  =-W_{\rm rx}\expec{N^{\rm (AB)}}.
\end{equation}
We get the same right-hand-side for the change in the number of B's. We
see that on the right-hand-side we have obtained a quantity that we
didn't have before. This means that the rate equations are not closed.
We can now proceed in two ways. The first is to write down rate
equations for the new quantity $\expec{N^{\rm (AB)}}$ and hope that this
will lead to equations that are closed. If we do this, we find that this
will not happen. Instead we will get a right-hand-side that depends on
the number of certain combinations of three particles. We can write down
rate equations for these as well, and hope that this will lead finally
to a closed set of equations. But that too won't happen. Proceeding by
writing rate equations for the new quantities that we obtain will lead
to an infinite hierarchy of equations.

The second way to proceed is to introduce an approximation that will
make a finite set of these equations into a closed set. We can do this
at different levels. The crudest approximation, and the one that will
lead to the common macroscopic rate equations, is to approximate
$\expec{N^{\rm (AB)}}$ in terms of $\expec{N^{\rm (A)}}$ and
$\expec{N^{\rm (B)}}$. This actually turns out to involve two
approximations. The first one is that we assume that the number of
adsorbates are randomly distributed over the surface. In this case we
have $N_\beta^{\rm (AB)}=ZN_\beta^{\rm (A)}[N_\beta^{\rm (B)}/S-1]$, with
$Z$ the coordination number of the lattice: i.e., the number of nearest
neighbors of a site. ($Z=4$ for a square lattice, $Z=6$ of a hexagonal
lattice, and $Z=3$ for a trigonal lattice.) The quantity between square
brackets is the probability that a neighboring site of an A is occupied
by a B. This approximation leads to
\begin{equation}
  {d\expec{N^{\rm (A)}}\over dt}
  =-{Z\over S-1}W_{\rm rx}\sum_\beta P_\beta
   N_\beta^{\rm (A)}N_\beta^{\rm (B)}
  =-{Z\over S-1}W_{\rm rx}\expec{N^{\rm (A)}N^{\rm (B)}}.
\end{equation}
This is still not a closed expression. We have
\begin{equation}
  \expec{N^{\rm (A)}N^{\rm (B)}}
  =\expec{N^{\rm (A)}}\expec{N^{\rm (B)}}
  +\expec{[N^{\rm (A)}-\expec{N^{\rm (A)}}]
          [N^{\rm (B)}-\expec{N^{\rm (B)}}]}.
\end{equation}
The second term on the right stands for the correlation between
fluctuations in the number of A's and the number of B's. In general this
is not zero. Because the number of A's and B's decrease because of the
reaction simultaneously, this term is expected to be positive.
Fluctuations however decrease when the system size is increased. In the
thermodynamic limit $S\to\infty$ we can set it to zero. We finally get
\begin{equation}
  {d\expec{N^{\rm (A)}}\over dt}
  =-{Z\over S}W_{\rm rx}\expec{N^{\rm (A)}}\expec{N^{\rm (B)}}
\end{equation}
with $S-1$ replaced by $S$ because $S\gg 1$. Dividing by the number of
sites $S$ leads then to
\begin{equation}
  {d\theta_{\rm A}\over dt}
  =-ZW_{\rm rx}\theta_{\rm A}\theta_{\rm B}.
\end{equation}
This should be compared to the macroscopic rate equation
\begin{equation}
  {d\theta_{\rm A}\over dt}
  =-k_{\rm rx}\theta_{\rm A}\theta_{\rm B}.
\end{equation}
We see from this that we have $W_{\rm rx}=k_{\rm rx}/Z$, but only if the
two approximations are valid. This may not be the case when the
adsorbates form some kind of structure (e.g. islands or a
superstructure) or when the system is small (e.g. a small cluster of
metal atoms).

The derivation for the ${\rm A}+{\rm A}$ reaction is almost the same. We
have
\begin{equation}
  {d\expec{N^{\rm (A)}}\over dt}
  =\sum_{\alpha\beta}W_{\alpha\beta}P_\beta
   \left[N_\alpha^{\rm (A)}-N_\beta^{\rm (A)}\right].
\end{equation}
If $\alpha$ can originate from $\beta$ by a ${\rm A}+{\rm A}$ reaction,
then $W_{\alpha\beta}=W_{\rm rx}$, otherwise $W_{\alpha\beta}=0$. If
such a reaction is possible, then $N_\alpha^{\rm (A)}-N_\beta^{\rm
  (A)}=-2$, because now two A's react. The number of configurations
$\alpha$ that can be obtained from $\beta$ by a reaction is equal to the
number of AA pairs $N_\beta^{\rm (AA)}$. This leads then to
\begin{equation}
  {d\expec{N^{\rm (A)}}\over dt}
  =-2W_{\rm rx}\sum_\beta P_\beta N_\beta^{\rm (AA)}
  =-2W_{\rm rx}\expec{N^{\rm (AA)}}.
\end{equation}
If we do not want to get an infinite hierarchy of equations with rate
equations for quantities of more and more A's, we have to make an
approximation again. We approximate $\expec{N^{\rm (AA)}}$ in terms of
$\expec{N^{\rm (A)}}$. We first assume that the number of adsorbates are
randomly distributed over the surface. In this case we have
$N_\beta^{\rm (AA)}=(1/2)ZN_\beta^{\rm (A)}[N_\beta^{\rm (A)}/S]$. Note
the factor $1/2$ that avoids double counting of the number of AA pairs.
The quantity between square brackets is the probability that a
neighboring site of an A is occupied by a A. This approximation leads to
\begin{equation}
  {d\expec{N^{\rm (A)}}\over dt}
  =-{Z\over S-1}W_{\rm rx}\sum_\beta P_\beta
   (N_\beta^{\rm (A)})^2
  =-{Z\over S-1}W_{\rm rx}\expec{(N^{\rm (A)})^2}.
\end{equation}
The factor $2$ that we had previously has canceled against the factor
$1/2$ in the expression for the number of AA pairs. To proceed we note
that
\begin{equation}
  \expec{(N^{\rm (A)})^2}
  =\expec{N^{\rm (A)}}^2
  +\expec{(N^{\rm (A)}-\expec{N^{\rm (A)}})^2}.
\end{equation}
The second term on the right stands for the fluctuations in the number
of A's. This is clearly not zero, but positive. Setting it to zero is
again the thermodynamic limit. We finally get
\begin{equation}
  {d\expec{N^{\rm (A)}}\over dt}
  =-{Z\over S}W_{\rm rx}\expec{N^{\rm (A)}}^2.
\end{equation}
Dividing by the number of sites $S$ leads then to
\begin{equation}
  {d\theta_{\rm A}\over dt}
  =-ZW_{\rm rx}\theta_{\rm A}^2.
\end{equation}
This should be compared to the macroscopic rate equation
\begin{equation}
  {d\theta_{\rm A}\over dt}
  =-2k_{\rm rx}\theta_{\rm A}^2.
\end{equation}
Note that there is a factor 2 on the right-hand-side, which is used
because a reactions removes two A's. We see from this that we have
$W_{\rm rx}=2k_{\rm rx}/Z$.
\subsection{Bimolecular adsorption}
\label{subsec:bimolads}

We deal here with the quite common case of a molecule of the type ${\rm
  B}_2$ that adsorbs dissociatively on two neighboring sites. An example
of such adsorption is oxygen adsorption on many transition metal
surfaces. We will see this adsorption when we will discuss the
Ziff-Gulari-Barshad model in Chapter~\ref{ch:examples}. We will see here
that it is often convenient to look at limiting cases to derive an
expression of the rate constant of adsorption.

We look at the number of B's. We have again
\begin{equation}
  {d\expec{N^{\rm (B)}}\over dt}
  =\sum_{\alpha\beta}W_{\alpha\beta}P_\beta
   \left[N_\alpha^{\rm (B)}-N_\beta^{\rm (B)}\right],
\end{equation}
where $N_\alpha^{\rm (B)}$ stands for the number of B's. If $\alpha$ can
originate from $\beta$ by an adsorption reaction, then
$W_{\alpha\beta}=W_{\rm ads}$, otherwise $W_{\alpha\beta}=0$. If such a
reaction is possible, then $N_\alpha^{\rm (B)}-N_\beta^{\rm (B)}=2$. The
problem now is with the number of configurations $\alpha$ that can be
obtained from $\beta$ by a reaction. This number is equal to the number
of pairs of neighboring vacant sites $N_\beta^{(**)}$. This leads then
to
\begin{equation}
  {d\expec{N^{\rm (B)}}\over dt}
  =2W_{\rm ads}\sum_\beta P_\beta N_\beta^{(**)}
  =2W_{\rm ads}\expec{N^{(**)}}.
\end{equation}
The right-hand-side can in general only be approximated, but such an
approximation is not needed for the case of a bare surface. In that case
we have $N^{(**)}=ZS/2$, where $Z$ is the coordination number of the
lattice and $S$ the number of sites in the system. This leads to
\begin{equation}
  {d\expec{N^{\rm (B)}}\over dt}=ZSW_{\rm ads}.
  \label{eq:biads}
\end{equation}

The change in the number of adsorbates for a bare surface is also equal to 
\begin{equation}
  {d\expec{N^{\rm (B)}}\over dt}=2AF\sigma,
\end{equation}
where $A$ is the area of the surface, $F$ is the number of particles
hitting a unit area of the surface per unit time, and $\sigma$ is the
sticking probability. The factor 2 is due to the fact that a molecule
that adsorbs yields two adsorbates. The flux $F$ we've seen before and
is given by
\begin{equation}
  F={P\over\sqrt{2\pi mk_{\rm B}T}}
\end{equation}
with $P$ the pressure, $T$ the temperature and $m$ the mass of a
molecule. This means that
\begin{equation}
  {d\expec{N^{\rm (B)}}\over dt}={2AP\sigma\over\sqrt{2\pi mk_{\rm B}T}}.
\end{equation}
If we compare this with expression~\ref{eq:biads}, we get
\begin{equation}
  W_{\rm ads}={2A_{\rm site}P\sigma\over Z\sqrt{2\pi mk_{\rm B}T}}
\end{equation}
with $A_{\rm site}$ the area of a single site.
\chapter[Monte Carlo Simulations]{Monte Carlo Simulations}
\label{ch:MC}

For most systems of interest deriving analytical results from the Master
Equation is not possible. Approximations like Mean Field can of course
be used, but they may not be satisfactory. In such cases one can resort
to Monte Carlo simulations.

Monte Carlo methods have been known already for several decades for the
general Master Equations.\cite{bin86} Following Gillespie they have
become quite popular to simulate reactions in
solutions.\cite{gil76,gil77,hon90} A configuration $\alpha$ in that case
is defined as a set $\{N_1,N_2,\ldots\}$ where $N_i$ is the number of
molecules of type $i$ in the solution. There is no specification of
where the molecules are, as in our case for surface reactions. When
simulating reactions one talks about Dynamic Monte Carlo simulations, a
term that we will use as well. Many of the algorithms developed in that
area can be used for surface reactions as well. However, the efficiency
of the various algorithms (i.e., the computer time and memory) can be
vary different. There are also tricks to increase the efficiency of
simulations of reactions in solutions that do not work for surface
reactions and vice versa.\cite{jan99a,gil01,res01}

Kinetic Monte Carlo methods essentially form a subset of the algorithms
mentioned above. The different name is used because they have a
different origin. They were specifically developed for surface reactions
and are based on a dynamic interpretation of equilibrium Monte Carlo
simulations.\cite{fic91,kan94,men94} They will be treated in
section~\ref{sec:others}, whereas the Dynamic Monte Carlo methods are
discussed in section~\ref{sec:sme}. Section~\ref{sec:car} describes
CARLOS, a general purpose code to simulate surface reactions.
\section{Solving the Master Equation}
\label{sec:sme}
\subsection{The integral formulation of the Master Equation.}

To start with the derivation of the Monte Carlo algorithms for the
Master Equation it is convenient to cast the Master Equation in an
integral form. First we simplify the notation of the Master Equation. We
define a matrix
${\bf W}$ by
\begin{equation}
  {\bf W}_{\alpha\beta}\equiv W_{\alpha\beta},
\end{equation}
which has vanishing diagonal elements, because
$W_{\alpha\alpha}=0$ by definition, and a diagonal matrix ${\bf R}$ by
\begin{equation}
  {\bf R}_{\alpha\beta}
  \equiv\cases{0,&if $\alpha\ne\beta$,\cr
  \sum_\gamma W_{\gamma\beta},&if $\alpha=\beta$.\cr}
\end{equation}
If we put the probabilities of the configurations $P_\alpha$ in a vector
${\bf P}$, we can write the Master Equation as
\begin{equation}
  {d{\bf P}\over dt}=-({\bf R}-{\bf W}){\bf P}.
\end{equation}
This equation can be interpreted as a time-dependent
Schr\"odinger-equation in imaginary time with Hamiltonian ${\bf R}-{\bf
W}$. This interpretation can be very fruitful,\cite{alc94} and leads,
among others, to the integral formulation we present here.

We do not want to be distracted by technicalities at this point, so we
assume that ${\bf R}$ and ${\bf W}$ are time independent. We also
introduce a new matrix ${\bf Q}$, which is defined by
\begin{equation}
  {\bf Q}(t)\equiv\exp[-{\bf R}t].
\end{equation}
This matrix is time dependent by definition. With this definition we
can rewrite the Master Equation in the following integral form, as can
be seen by substitution.
\begin{equation}
  {\bf P}(t)={\bf Q}(t){\bf P}(0)
  +\int_0^t\!\!dt^\prime{\bf Q}(t-t^\prime){\bf W}{\bf P}(t^\prime).
\end{equation}
The equation is implicit in ${\bf P}$. By substitution of the
right-hand-side for ${\bf P}(t^\prime)$ again and again we get
\begin{eqnarray}
\label{eq:intf1}
  {\bf P}(t)=\bigg[{\bf Q}(t)
  &+&\int_0^t\!\!dt^\prime{\bf Q}(t-t^\prime){\bf W}{\bf Q}(t^\prime)\\
  &+&\int_0^t\!\!dt^\prime\int_0^{t^\prime}\!\!dt^{\prime\prime}
  {\bf Q}(t-t^\prime){\bf W}{\bf Q}(t^\prime-t^{\prime\prime})
  {\bf W}{\bf Q}(t^{\prime\prime})+\ldots\bigg]{\bf P}(0).\nonumber
\end{eqnarray}
This equation is valid also for other definitions of ${\bf R}$ and ${\bf
  W}$, but the definition we have chosen leads to a useful
interpretation.  Suppose at $t=0$ the system is in configuration
$\alpha$ with probability $P_\alpha(0)$. The probability that at time
$t$ the system is still in $\alpha$ (i.e., no reaction has taken place)
is given by ${\bf Q}_{\alpha\alpha}(t)P_\alpha(0)=\exp(-{\bf
  R}_{\alpha\alpha}t)P_\alpha(0)$. This shows that the first term in
equation.(\ref{eq:intf1}) represents the contribution to the
probabilities when no reaction takes place up to time $t$. The matrix
${\bf W}$ determines how the probabilities change when a reaction takes
place. The second term of equation.(\ref{eq:intf1}) represents the
contribution to the probabilities when no reaction takes place between
times $0$ and $t^\prime$, some reaction takes place at time $t^\prime$,
and then no reaction takes place between times $t^\prime$ en $t$. So the
second term stands for the contribution to the probabilities when a
single reaction takes place.  Subsequent terms represent contributions
when two, three, four, etc.\ reactions take place.
\subsection{The Variable Step Size Method.}

The idea of the Dynamic Monte Carlo method is not to compute
probabilities $P_\alpha(t)$ explicitly, but to start with some
particular configuration, representative for the initial state of the
experiment one wants to simulate, and then generate a sequence of other
configurations with the correct probability. The integral formulation
gives us directly a useful algorithm to do this.

Let's call the initial configuration $\alpha$, and let's set the initial
time to $t=0$. Then the probability that the system is still in $\alpha$
at a later time $t$ is given by
\begin{equation}
  {\bf Q}_{\alpha\alpha}(t)=\exp[-{\bf R}_{\alpha\alpha}t].
\end{equation}
The probability distribution that the first reaction takes place at time
$t$ is minus the derivative with respect to time of this expression:
i.e.,
\begin{equation}
  {\bf R}_{\alpha\alpha}\exp[-{\bf R}_{\alpha\alpha}t].
\label{eqprt}
\end{equation}
This can be seen by taking the integral of this expression from $0$ to
$t$, which yields the probability that a reaction has taken place in this
interval, which equals $1-{\bf Q}_{\alpha\alpha}(t)$. We generate a
time $t^\prime$ when the first reaction actually occurs according to
this probability distribution. This can be done by solving
\begin{equation}
  \exp[-{\bf R}_{\alpha\alpha}t^\prime]=r_1,
\end{equation}
where $r_1$ is a uniform deviate on the unit interval.\cite{fel70}

At time $t^\prime$ a reaction takes place. According to
equation~(\ref{eq:intf1}) the different reactions that transform
configuration $\alpha$ to another configuration $\beta$ have transition
probabilities $W_{\beta\alpha}$. This means that the probability that
the system will be in configuration $\beta$ at time $t^\prime+dt$ is
$W_{\beta\alpha}dt$, where $dt$ is some small time interval. We
therefore generate a new configuration $\alpha^\prime$ by picking it out
of all possible new configurations $\beta$ with a probability
proportional to $W_{\alpha^\prime\alpha}$. This gives us a new
configuration $\alpha^\prime$ at time $t^\prime$. At this point we're in
the same situation as when we started the simulation, and we can proceed
by repeating the previous steps. So we generate a new time
$t^{\prime\prime}$, using
\begin{equation}
  \exp[-{\bf R}_{\alpha^\prime\alpha^\prime}
  (t^{\prime\prime}-t^\prime)]=r_2,\label{eq:rxtime}
\end{equation}
for the time of the new reaction, and a new configuration
$\alpha^{\prime\prime}$ with a probability proportional to
$W_{\alpha^{\prime\prime}\alpha^\prime}$. In this manner we continue
until some preset condition is met that signals the end of the interval
we want to simulate.

We call this whole procedure the Variable Step Size Method (VSSM). It's
a simple yet very efficient method. The algorithm is as follows.

\bigskip
\noindent Variable Step Size Method: concept (VSSMc)
\begin{enumerate}
\setlength{\itemsep}{0pt}
\setlength{\parsep}{0pt}
\setlength{\parskip}{0pt}
\setlength{\partopsep}{0pt}
\setlength{\topsep}{0pt}
\item Initialize
  \begin{description}
\setlength{\itemsep}{0pt}
\setlength{\parsep}{0pt}
\setlength{\parskip}{0pt}
\setlength{\partopsep}{0pt}
\setlength{\topsep}{0pt}
  \item Generate an initial configuration $\alpha$.
  \item Set the time $t$ to some initial value.
  \item Choose conditions when to stop the simulation.
  \end{description}
\item \label{algst:vssmcrt} Reaction time
  \begin{description}
\setlength{\itemsep}{0pt}
\setlength{\parsep}{0pt}
\setlength{\parskip}{0pt}
\setlength{\partopsep}{0pt}
\setlength{\topsep}{0pt}
  \item Generate a time interval $\Delta t$ when no reaction takes place
  \begin{equation}
    \Delta t =-{1\over\sum_\beta W_{\beta\alpha}}\ln r,
  \end{equation}
  \item where $r$ is a random deviate on the unit interval.
  \item Change time to $t\to t+\Delta t$.
  \end{description}
\item \label{algst:vssmcrx} Reaction
  \begin{description}
\setlength{\itemsep}{0pt}
\setlength{\parsep}{0pt}
\setlength{\parskip}{0pt}
\setlength{\partopsep}{0pt}
\setlength{\topsep}{0pt}
  \item Change the configuration to $\alpha^\prime$ with probability
  $W_{\alpha^\prime\alpha}/\sum_\beta W_{\beta\alpha}$: i.e., do the
  reaction $\alpha\to\alpha^\prime$.
  \end{description}
\item Continuation
  \begin{description}
\setlength{\itemsep}{0pt}
\setlength{\parsep}{0pt}
\setlength{\parskip}{0pt}
\setlength{\partopsep}{0pt}
\setlength{\topsep}{0pt}
  \item If the stop conditions are fulfilled then stop. If not repeat at
    step~\ref{algst:vssmcrt}.
  \end{description}
\end{enumerate}

\smallskip
\noindent We see that the algorithm yields an ordered set of
configurations and reaction times that can be written as
\begin{equation}
  (\alpha_0,t_0)
  \mathop{\to}^{t_1}\alpha_1
  \mathop{\to}^{t_2}\alpha_2
  \mathop{\to}^{t_3}\alpha_3
  \mathop{\to}^{t_4}\ldots
\end{equation}
Here $\alpha_0$ is the initial configuration and $t_0$ is the time at
the beginning of the simulations. The changes $\alpha_{n-1}\to\alpha_n$
are caused by reactions taking place at time $t_n$. We will see that all
other algorithms that we will present also give such a result. They are
all equivalent because all give at time $t$ a configuration $\alpha$
with probability $P_\alpha(t)$ which is the solution of the Master
Equation with boundary condition
$P_\alpha(t_0)=\delta_{\alpha\alpha_0}$.
\subsection{Enabled and disabled reactions.}

Although all algorithms we will discuss in this section yield the same
result, they often do so at very different computational costs. We are
in particular interested in how computer time and memory scale with
system size. It is clear that in general the number of reactions in a
system is proportional to the size of the system (and also to the length
of the simulation in real time). The computational costs will therefore
scale at least linear with system size. We will focus not on costs for
the whole system, but instead on costs per reaction

Looking at the VSSMc algorithm above, we see that it scales in the worse
possible way with system size. In step~\ref{algst:vssmcrt}, for example,
we have to sum over all possible configurations. For a simple lattice
with $S$ sites and each grid point having $N$ possible labels we have a
total number of configurations equal to $N^S$. This means that VSSMc
scales exponentially with system size. Fortunately, it is easy to
improve this. Most of the terms in the summation are zero because there
is no reaction that changes $\alpha$ into $\beta$ and hence
$W_{\beta\alpha}=0$. So we should only use those changes that can
actually occur; i.e., we should keep track of the possible reactions.
Reactions that can actually occur at a certain location we call {\it
  enabled\/}. The total number of (enabled) reactions is proportional to
the system size, so we can reduce the scaling of computer time per
reaction at least to $O(S)$.\cite{knu73b} Actually, we can reduce the
costs even further because we need not determine all enabled reactions
every time at steps~\ref{algst:vssmcrt} and \ref{algst:vssmcrx}. A
reaction has only a local effect and does not affect reactions far away.
If a reaction takes place, this causes a local change in the
configuration. This change makes new reactions possible only locally,
whereas other reactions are not possible anymore. We say that such
reactions are {\it disabled\/}. The number of newly enabled and disabled
reactions only depends on what the configuration looks like at the
location where a reaction has just occurred, but it does not depend on
the system size (see figure~\ref{fig:endis}). So instead of determining
all enabled reactions again and again we do this only at the
initialization and then update a list of all enabled reactions. The
algorithms then becomes as follows.
\begin{figure}[ht]
\includegraphics[width=\hsize]{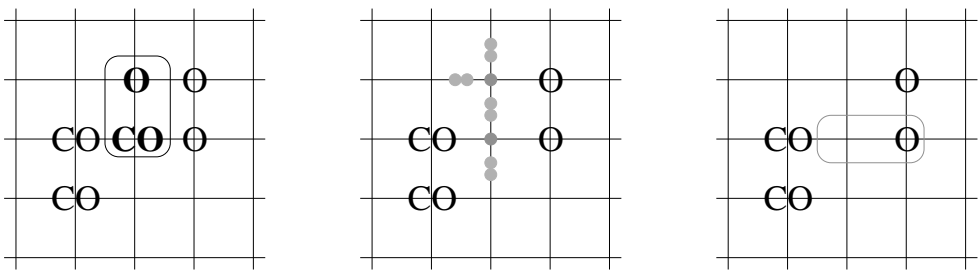}
\caption{The left shows part of a configuration for a model of CO
  oxidation. The fat CO and oxygen form ${\rm CO}_2$ which is removed
  from the surface. The middle shows newly enabled reaction; CO can
  adsorb at the sites marked by dark grey circles, oxygen can adsorb
  dissociatively on neighboring sites indicated by two light grey
  circles on the line connecting the sites. The right indicates a
  disabled reaction; the two encircled sites had a CO and an oxygen that
  could form a ${\rm CO}_2$.}
\label{fig:endis}
\end{figure}

\bigskip
\noindent Variable Step Size Method: improved version (VSSMi)
\begin{enumerate}
\setlength{\itemsep}{0pt}
\setlength{\parsep}{0pt}
\setlength{\parskip}{0pt}
\setlength{\partopsep}{0pt}
\setlength{\topsep}{0pt}
\item Initialize
  \begin{description}
\setlength{\itemsep}{0pt}
\setlength{\parsep}{0pt}
\setlength{\parskip}{0pt}
\setlength{\partopsep}{0pt}
\setlength{\topsep}{0pt}
  \item Generate an initial configuration $\alpha$.
  \item Make a list $L_{\rm rx}$ of all reactions.
  \item Calculate $k_\alpha\equiv\sum_\beta W_{\beta\alpha}$, with the sum
  being done only over the reactions in $L_{\rm rx}$.
  \item Set the time $t$ to some initial value.
  \item Choose conditions when to stop the simulation.
  \end{description}
\item \label{algst:vssmirt} Reaction time
  \begin{description}
\setlength{\itemsep}{0pt}
\setlength{\parsep}{0pt}
\setlength{\parskip}{0pt}
\setlength{\partopsep}{0pt}
\setlength{\topsep}{0pt}
  \item Generate a time interval $\Delta t$ when no reaction takes place
  \begin{equation}
    \Delta t =-{1\over k_\alpha}\ln r,
  \end{equation}
  \item where $r$ is a random deviate on the unit interval.
  \item Change time to $t\to t+\Delta t$.
  \end{description}
\item \label{algst:vssmirx} Reaction
  \begin{description}
\setlength{\itemsep}{0pt}
\setlength{\parsep}{0pt}
\setlength{\parskip}{0pt}
\setlength{\partopsep}{0pt}
\setlength{\topsep}{0pt}
  \item Pick the reaction $\alpha\to\alpha^\prime$ from $L_{\rm rx}$ with
  probability $W_{\alpha^\prime\alpha}/k_\alpha$: i.e., do the reaction
  $\alpha\to\alpha^\prime$.
  \end{description}
\item Update
  \begin{description}
\setlength{\itemsep}{0pt}
\setlength{\parsep}{0pt}
\setlength{\parskip}{0pt}
\setlength{\partopsep}{0pt}
\setlength{\topsep}{0pt}
  \item Remove the reaction $\alpha\to\alpha^\prime$ from $L_{\rm rx}$.
  \item Add new enabled reactions to $L_{\rm rx}$ and remove disabled
  reactions.
  \item Use these reactions to calculate $k_{\alpha^\prime}$ from
  $k_\alpha$.
  \end{description}
\item Continuation
  \begin{description}
\setlength{\itemsep}{0pt}
\setlength{\parsep}{0pt}
\setlength{\parskip}{0pt}
\setlength{\partopsep}{0pt}
\setlength{\topsep}{0pt}
  \item If the stop conditions are fulfilled then stop. If not repeat at
    step~\ref{algst:vssmirt}.
  \end{description}
\end{enumerate}

The reasoning leading to VSSMi suggests that the computer time per
reaction of this algorithm does not depend on system size. However, that
is still not true. There are two problems. First, picking the reaction
in step~\ref{algst:vssmirx} cannot be done in constant time just with
the list of all reactions. Second, adding new enabled reactions to the
list of reactions can be done easily in constant time, but removing
disabled reactions presents a problem. One can scan the list of all
reactions and in remove all disabled reactions from the list, but that
is an $O(S)$ operation. It may be possible to make links from the sites
where the last reaction has occurred to the places in the list of all
reactions where the possible disabled reactions reside, but that is very
complicated and has never been done.
\subsection{Weighted and uniform selection.}
\label{subsec:weiuni}

The selection in step~\ref{algst:vssmirx} is a weighted selection. To
make this selection one has to define cumulative rate constants
$C_{\alpha^\prime\alpha} \equiv \sum_{\beta\le\alpha^\prime}
W_{\beta\alpha^\prime}$.  The configurations that can be reached by a
reaction from $\alpha$ need to be ordered, and the summation is over all
configurations preceding $\alpha^\prime$ ($\beta<\alpha^\prime$) and
$\alpha^\prime$ itself. The reaction $\alpha\to\alpha^\prime$ can then
be picked by choosing $\alpha^\prime$ using $C_{\alpha^\prime-1\alpha}<
rk_\alpha \le C_{\alpha^\prime\alpha}$ where $r$ is a random deviate on
the unit interval and $\alpha^\prime-1$ is the configuration before
$\alpha^\prime$ in the ordering of the configurations. This {\it
  weighted\/} selection scales linearly with the number of reactions;
i.e., it scales as $O(S)$. The reason for this is that we have to scan
all the cumulative rate constants $C_{\alpha^\prime\alpha}$.

To pick a reaction in constant time we split the list of all reactions
in groups containing reactions of the same type (or more general with
the same rate constant). Two reactions are of the same type if they
differ only in their position and/or orientation. So CO adsorption, NO
dissociation (${\rm NO}\to{\rm N}+{\rm O}$ and ${\rm N}_2$ associative
desorption ($2{\rm N}\to{\rm N}_2$) are examples of reactions types. If
$L_{\rm rx}^{(i)}$ is the list of $N^{(i)}$ reactions with rate constant
$W^{(i)}$, then we proceed as follows. First, we pick a type of reaction
$j$ with probability $N^{(j)}W^{(j)}/\sum_iN^{(i)}W^{(i)}$, and then we
pick from $L_{\rm rx}^{(j)}$ a reaction at random. The first part scales
linearly with the number of lists $L_{\rm rx}^{(i)}$, because it is a
weighted selection. This number does not depend on the system size. The
second part is a {\it uniform\/} selection, and can be done in constant
time. So the second part also does not depend on the system size. If the
number of reaction types is small, and it often is, this method is very
efficient (see figure~\ref{fig:select}).
\begin{figure}[ht]
\includegraphics[width=\hsize]{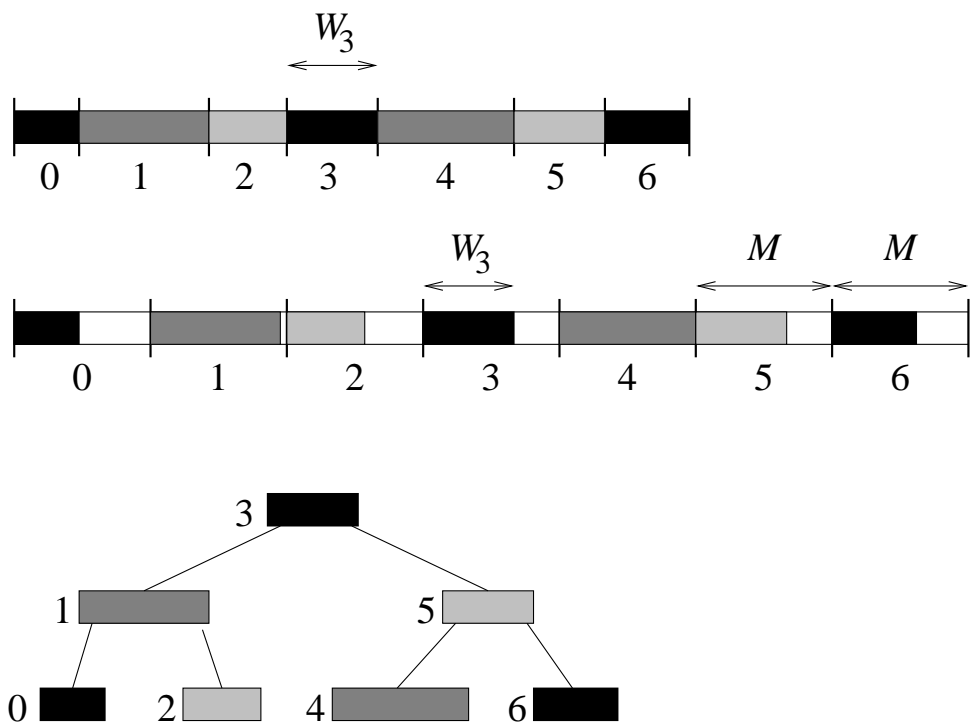}
\caption{Weighted (top), uniform (middle), and hierarchical (bottom)
  selection. For the weighted selection the cumulative length of the
  bars has to be added. In the uniform selection a bar is randomly
  selected and then accepted with probability $W_n/M$. In the
  hierarchical selection each node has the sum of the lengths of all
  bars in the subtree, and we need only to go down the tree to make the
  selection.}
\label{fig:select}
\end{figure}

It is possible to do the weighted selection of the reactions also in
$O(\log S)$ time by using a binary tree.\cite{knu73b} Each node of the
tree has a reaction and the cumulative rate constant of all reactions of
the node and both branches below the node. After $rk_\alpha$ has been
determined we look for the node with $C_{\rm left}<rk_\alpha\le C_{\rm
  left}+W_{\rm node}$ where $W_{\rm node}$ is the rate constant of the
reaction of the node and $C_{\rm left}$ is the cumulative rate constant
of the top node of the left branch. If there is no left branch then we
define $C_{\rm left}=0$. To find the node we do we do the following.
\begin{enumerate}
\setlength{\itemsep}{0pt}
\setlength{\parsep}{0pt}
\setlength{\parskip}{0pt}
\setlength{\partopsep}{0pt}
\setlength{\topsep}{0pt}
\item Start
  \begin{description}
\setlength{\itemsep}{0pt}
\setlength{\parsep}{0pt}
\setlength{\parskip}{0pt}
\setlength{\partopsep}{0pt}
\setlength{\topsep}{0pt}
  \item Set $X=rk_\alpha$.
  \item Take the top node of the tree.
  \end{description}
\item \label{step:found} Reaction found?
  \begin{description}
\setlength{\itemsep}{0pt}
\setlength{\parsep}{0pt}
\setlength{\parskip}{0pt}
\setlength{\partopsep}{0pt}
\setlength{\topsep}{0pt}
  \item if $C_{\rm left}<X\le C_{\rm left}+W_{\rm node}$
  \item then stop; take the reaction of the node.
  \item else go to the next step
  \end{description}
\item Continue in the left branch?
  \begin{description}
\setlength{\itemsep}{0pt}
\setlength{\parsep}{0pt}
\setlength{\parskip}{0pt}
\setlength{\partopsep}{0pt}
\setlength{\topsep}{0pt}
  \item if $X\le C_{\rm left}$
  \item then take the top node of the left branch and continue at
    step~\ref{step:found}
  \item else go to the next step
  \end{description}
\item Continue in the right branch.
  \begin{description}
\setlength{\itemsep}{0pt}
\setlength{\parsep}{0pt}
\setlength{\parskip}{0pt}
\setlength{\partopsep}{0pt}
\setlength{\topsep}{0pt}
  \item Set $X\to X-(C_{\rm left}+W_{\rm node})$.
  \item Take the top node of the right branch and continue at
    step~\ref{step:found}.
  \end{description}
\end{enumerate}

The number of nodes we have to inspect is equal to the depth of the
tree. If the tree is well-balanced this is $O(\log S)$. We can use this
method also for a weighted selection of the reaction type, if the number
of reaction types is large. In fact this occurs in DMC simulations of
reactions in solutions and the method we describe here is one method
that is used for these simulations.\cite{hon90}
\subsection{Handling disabled reactions.}

The problem of removing disabled reactions has a surprisingly simple
solution, although it is a bit more difficult to see that it is also a
correct solution. Instead of removing the disabled reactions, we simply
leave them in the list of all reactions, but when a reaction has to
occur we check if it is disabled. If it is we remove it. If it is an
enabled reaction, we treat it as usual. That this is correct can be
proven as follows. Suppose that $k_{\rm en}$ is the sum of the rate
constants of all enabled reactions, and we have one disabled reaction
with rate constant $k_{\rm dis}$. Also suppose without loss of
generality that the system is at time $t=0$. The probability
distribution for the first reaction to occur is $k_{\rm en}\exp(-k_{\rm
en}t)$ (see equation~(\ref{eqprt})). If we work with the list that
includes the disabled reaction than the probability distribution for the
first reaction occurring at time $t$ being also an enabled reaction is
$k_{\rm en}\exp(-(k_{\rm en}+k_{\rm dis})t)$, which is the probability
that no reaction occurs until time $t$ and then an enabled reaction
occurs. This is the first contribution to the probability distribution
for the enabled reaction if the disabled reaction is not removed. The
probability distribution that the first reaction is not but the second
reaction is enabled and occurs at time $t$ is given by
\begin{eqnarray}
  &&k_{\rm en}
  \int_0^t\!\!dt^\prime\,
  e^{-k_{\rm en}(t-t^\prime)}
  k_{\rm dis}e^{-(k_{\rm en}+k_{\rm dis})t^\prime}\nonumber\\
  &&=k_{\rm en}e^{-k_{\rm en}t}k_{\rm dis}
  \int_0^t\!\!dt^\prime\,
  e^{-k_{\rm dis}t^\prime}\nonumber\\
  &&=k_{\rm en}e^{-k_{\rm en}t}
  \left[1-e^{-k_{\rm dis}t}\right].
\end{eqnarray}
Adding this to $k_{\rm en}\exp(-(k_{\rm en}+k_{\rm dis})t)$ gives us
$k_{\rm en}\exp(-k_{\rm en}t)$, which is what we should have.

This shows that adding a single disabled reaction does not change the
probability distribution for the time that the first enabled reaction
occurs. In the same way we can show that adding a second disabled
reaction gives the same probability distribution as having a single
disabled reaction, and that adding a third disabled reaction is the same
as having two disabled reactions, etc. So by induction we see that
disabled reactions do not change the probability distribution for the
occurrence of the enabled reactions. Also step~\ref{algst:vssmirx} of
VSSMi is no problem. The enabled reactions are chosen with a probability
proportional to their rate constant whether or not disabled reactions
are present in he list.

The VSSM algorithm now gets the following form.

\bigskip
\noindent Variable Step Size Method with an approximate list of reactions
(VSSMa)
\begin{enumerate}
\setlength{\itemsep}{0pt}
\setlength{\parsep}{0pt}
\setlength{\parskip}{0pt}
\setlength{\partopsep}{0pt}
\setlength{\topsep}{0pt}
\item Initialize
  \begin{description}
\setlength{\itemsep}{0pt}
\setlength{\parsep}{0pt}
\setlength{\parskip}{0pt}
\setlength{\partopsep}{0pt}
\setlength{\topsep}{0pt}
  \item Generate an initial configuration $\alpha$.
  \item Make lists $L_{\rm rx}^{(i)}$ containing all reactions of type $i$.
  \item Calculate $k^{(i)}\equiv N^{(i)}W^{(i)}$, with $N^{(i)}$ the
  number of reactions of type $i$ and $W^{(i)}$ the rate constant of these
  reactions.
  \item Set the time $t$ to some initial value.
  \item Choose conditions when to stop the simulation.
  \end{description}
\item \label{algst:vssmart} Reaction time
  \begin{description}
\setlength{\itemsep}{0pt}
\setlength{\parsep}{0pt}
\setlength{\parskip}{0pt}
\setlength{\partopsep}{0pt}
\setlength{\topsep}{0pt}
  \item Generate a time interval $\Delta t$ when no reaction takes place
  \begin{equation}
    \Delta t =-{1\over\sum_i k^{(i)}}\ln r,
  \end{equation}
  \item where $r$ is a random deviate on the unit interval.
  \item Change time to $t\to t+\Delta t$.
  \end{description}
\item \label{algst:vssmarx} Reaction
  \begin{description}
\setlength{\itemsep}{0pt}
\setlength{\parsep}{0pt}
\setlength{\parskip}{0pt}
\setlength{\partopsep}{0pt}
\setlength{\topsep}{0pt}
  \item Pick a type of reaction $j$ with probability $k^{(j)}/\sum_i
  k^{(i)}$, and then pick the reaction $\beta\to\alpha^\prime$ from
  $L_{\rm rx}^{(j)}$ at random. If the reaction is enabled go to
  step~\ref{algst:vssmaer}. If it is disabled go to
  step~\ref{algst:vssmadr}. (Note that during the simulation $k^{(j)}$ may
  have obtained contributions from disabled reactions starting from
  configurations different from $\alpha$.)
  \end{description}
\item \label{algst:vssmaer} Enabled reaction
  \begin{description}
\setlength{\itemsep}{0pt}
\setlength{\parsep}{0pt}
\setlength{\parskip}{0pt}
\setlength{\partopsep}{0pt}
\setlength{\topsep}{0pt}
  \item Change the configuration to $\alpha^\prime$.
  \end{description}
\item Enabled update
  \begin{description}
\setlength{\itemsep}{0pt}
\setlength{\parsep}{0pt}
\setlength{\parskip}{0pt}
\setlength{\partopsep}{0pt}
\setlength{\topsep}{0pt}
  \item Remove the reaction $\beta\to\alpha^\prime$ from $L_{\rm
  rx}^{(j)}$. Change $k^{(j)}\to k^{(j)}-W^{(j)}$. (Note that because the
  reaction is enabled $\beta=\alpha$.)
  \item Add new enabled reactions to the lists $L_{\rm rx}^{(i)}$.
  \item Use these reactions to calculate the new values for $k^{(i)}$ from
  the old ones.
  \item Skip to step~\ref{algst:vssmacn}.
  \end{description}
\item \label{algst:vssmadr} Disabled reaction
  \begin{description}
\setlength{\itemsep}{0pt}
\setlength{\parsep}{0pt}
\setlength{\parskip}{0pt}
\setlength{\partopsep}{0pt}
\setlength{\topsep}{0pt}
  \item Do not change the configuration; $\alpha^\prime$ is the same
  configuration as $\alpha$.
  \end{description}
\item Disabled update
  \begin{description}
\setlength{\itemsep}{0pt}
\setlength{\parsep}{0pt}
\setlength{\parskip}{0pt}
\setlength{\partopsep}{0pt}
\setlength{\topsep}{0pt}
  \item Remove the disabled reaction from $L_{\rm rx}^{(j)}$.
  \item Change $k^{(j)}\to k^{(j)}-W^{(j)}$.
  \end{description}
\item \label{algst:vssmacn} Continuation
  \begin{description}
\setlength{\itemsep}{0pt}
\setlength{\parsep}{0pt}
\setlength{\parskip}{0pt}
\setlength{\partopsep}{0pt}
\setlength{\topsep}{0pt}
  \item If the stop conditions are fulfilled then stop. If not repeat at
    step~\ref{algst:vssmart}.
  \end{description}
\end{enumerate}

The computer time per reaction of algorithm VSSMa scales as $O(1)$. This
is achieved by working with an approximate list of all reactions; a list
that can contain disabled reactions. We will see that the same trick can
be used with other algorithms as well. Note that picking the reaction
type at step~\ref{algst:vssmarx} can be done hierarchically as explained
at the end of section~\ref{subsec:weiuni}.
\subsection{Reducing memory requirements.}

There are other ways we might want to change the VSSM algorithm, but
then because of memory considerations. The description of the algorithms
uses a list of all reactions. This list is quite large and scales with
system size. We can do away with this list at the cost of increasing
computer time, although the algorithm will still scale as $O(1)$.

Instead of keeping track of all individual reactions we only keep track
of how many reactions there are of each different type; i.e., no lists
$L_{\rm rx}^{(i)}$ but only the numbers $N^{(i)}$. Because there are no
lists we have to count how many reactions become disabled after a
reaction has occurred. This is similar to adding enabled reactions and
can be done in constant time. (Note that this is only because we are
using no lists. Removing disabled reactions from lists is what costs
time.) This means that the number $N^{(i)}$ will be exact. The only
problem is, after the type of reaction is determined, how to determine
which particular reaction will take place. This can be done by randomly
searching on the surface. The number of places one has to look does not
depend on the system size, but on the probability that the reaction can
occur on a randomly selected site. This random search for the location
of the reaction is another form of uniform selection. Another
application of such a uniform selection will be given in
section~\ref{ssec:prac}. If the type of reaction can take place on many
places, then a particular reaction should rapidly be found.

To make the formulation of the new algorithm not too difficult we use a
more restrictive definition of a reaction type, in the sense that two
reactions are of the same type if one can be obtained from to other
by a translation. Previously we also talked about the same reaction type
if the orientation was different. We don't do this here, because we want
to have an unambiguous meaning if we say that a reaction occurs at a
particular site (see step~\ref{algst:vssmsrl} of VSSMs) even if more
sites are involved. For example, if we have a reaction type of an A
reacting with a B where the B is at a site to the right of the A, then
by the site of this reaction we mean the site where the A is.  The new
algorithm then becomes as follows.

\bigskip
\noindent Variable Step Size Method with random search for the location
of reactions (VSSMs).
\begin{enumerate}
\setlength{\itemsep}{0pt}
\setlength{\parsep}{0pt}
\setlength{\parskip}{0pt}
\setlength{\partopsep}{0pt}
\setlength{\topsep}{0pt}
\item Initialize
  \begin{description}
\setlength{\itemsep}{0pt}
\setlength{\parsep}{0pt}
\setlength{\parskip}{0pt}
\setlength{\partopsep}{0pt}
\setlength{\topsep}{0pt}
  \item Generate an initial configuration.
  \item Count how many reactions $N^{(i)}$ of type $i$ there are.
  \item Calculate $k^{(i)}\equiv N^{(i)}W^{(i)}$, with $W^{(i)}$ the
  rate constant of the reactions of type $i$.
  \item Set the time $t$ to some initial value.
  \item Choose conditions when to stop the simulation.
  \end{description}
\item \label{algst:vssmsrt} Reaction time
  \begin{description}
\setlength{\itemsep}{0pt}
\setlength{\parsep}{0pt}
\setlength{\parskip}{0pt}
\setlength{\partopsep}{0pt}
\setlength{\topsep}{0pt}
  \item Generate a time interval $\Delta t$ when no reaction takes place
  \begin{equation}
    \Delta t =-{1\over\sum_i k^{(i)}}\ln r,
  \end{equation}
  \item where $r$ is a random deviate on the unit interval.
  \item Change time to $t\to t+\Delta t$.
  \end{description}
\item Reaction type
  \begin{description}
\setlength{\itemsep}{0pt}
\setlength{\parsep}{0pt}
\setlength{\parskip}{0pt}
\setlength{\partopsep}{0pt}
\setlength{\topsep}{0pt}
  \item Pick a type of reaction $j$ with probability $k^{(j)}/\sum_i
  k^{(i)}$.
  \end{description}
\item \label{algst:vssmsrl} Reaction location \begin{description} \item
\setlength{\itemsep}{0pt}
\setlength{\parsep}{0pt}
\setlength{\parskip}{0pt}
\setlength{\partopsep}{0pt}
\setlength{\topsep}{0pt}
  Pick randomly a site for the reaction, until a site is found where the
  reaction can actually occur.
  \end{description}
\item Update
  \begin{description}
\setlength{\itemsep}{0pt}
\setlength{\parsep}{0pt}
\setlength{\parskip}{0pt}
\setlength{\partopsep}{0pt}
\setlength{\topsep}{0pt}
  \item Change the configuration.
  \item Determine the new enabled reactions, and change the $N^{(i)}$'s
  accordingly.
  \item Determine the disabled reactions, and change the $N^{(i)}$'s
  accordingly.
  \end{description}
\item Continuation
  \begin{description}
\setlength{\itemsep}{0pt}
\setlength{\parsep}{0pt}
\setlength{\parskip}{0pt}
\setlength{\partopsep}{0pt}
\setlength{\topsep}{0pt}
  \item If the stop conditions are fulfilled then stop. If not repeat at
  step~\ref{algst:vssmsrt}.
  \end{description}
\end{enumerate}
\subsection{Oversampling and the Random Selection Method.}
\label{subsec:oversampl}

The determination of a reaction and its time can be split in three
parts; the time of the reaction, the type of the reaction, and the site
of the reaction. The last two parts were combined in the previous
versions of VSSM. The determination of the reaction type has to be done
before the determination of the location of the reactions in VSSMs
(otherwise one doesn't know when to stop searching in
step~\ref{algst:vssmsrl}), but the time of the reaction can be
determined independently from which reactions occurs where. It is also
possible to determine all three parts independently. This has the
advantage that even less bookkeeping is necessary; adding and removing
reactions to update lists or numbers of types of reaction is not
necessary. The drawback is however the same as in VSSMs, only worse; as
in step~\ref{algst:vssmsrl} of VSSMs reactions will be attempted at
certain locations where the reactions cannot take place. If this does
not occur to often, then this drawback may be small.

The trick to do away with the bookkeeping is to use a technique called
{\it oversampling\/}.  Suppose we have just one type of reaction and that we
have $N$ of them. (A reaction type is defined here in the same way as
for VSSMs.) The time to the next occurrence of a reaction is then given
the probability distribution $NW\exp(-NWt)$ where $W$ is the rate
constant of the reaction. If we assume, however, that we have $M$ of these
reactions with $M>N$ then we can also generate the time of the next
reaction from the distribution $MW\exp(-MWt)$, but then accept the
reaction with probability $N/M$. To prove this we need to add the
contributions that the reaction has to be generated one, two, three
etc. times before one is found that is accepted.
\begin{eqnarray}
  &&{N\over M}MW\exp(MWt)\nonumber\\
  &&+\int_0^t\!\!dt^\prime\,
     {N\over M}MWe^{-MW(t-t^\prime)}
     \left[1-{N\over M}\right]MWe^{-MWt^\prime}\nonumber\\
  &&+\int_0^t\!\!dt^\prime\int_0^{t^\prime}\!\!dt^{\prime\prime}\,
     {N\over M}MWe^{-MW(t-t^\prime)}\nonumber\\
  &&\qquad\times\left[1-{N\over M}\right]MWe^{-MW(t^\prime-t^{\prime\prime})}
     \left[1-{N\over M}\right]MWe^{-MWt^{\prime\prime}}
     +\ldots\nonumber\\
  &&=NWe^{-MWt}
     \left[1+\left[1-{N\over M}\right]MWt
     +{1\over 2}\left[1-{N\over M}\right]^2M^2W^2t^2+\ldots\right]\nonumber\\
  &&=NWe^{-MWt}e^{[1-N/M]MWt}
    =NWe^{-NWt}.
\end{eqnarray}

The following algorithm is only useful if we do not need to determine
$N$ explicitly. This can be accomplished if we assume that all reaction
types can take place everywhere on the surface. In terms of lists this
means that each list $L_{\rm rx}^{(i)}$ has the same $S$ reactions
during an entire simulation. Because the lists do not change and they
have a simple definition, we do not need to determine them explicitly.
Also the times of the reactions are always taken from the same
probability distribution, and the probabilities to choose a reaction
type do not change. The algorithm looks as follows.

\bigskip
\noindent Random Selection Method (RSM).
\begin{enumerate}
\setlength{\itemsep}{0pt}
\setlength{\parsep}{0pt}
\setlength{\parskip}{0pt}
\setlength{\partopsep}{0pt}
\setlength{\topsep}{0pt}
\item Initialize
  \begin{description}
\setlength{\itemsep}{0pt}
\setlength{\parsep}{0pt}
\setlength{\parskip}{0pt}
\setlength{\partopsep}{0pt}
\setlength{\topsep}{0pt}
  \item Generate an initial configuration.
  \item Set the time $t$ to some initial value.
  \item Define $k\equiv SW_{\rm max}$ where $W_{\rm max}$ is the maximum
  of the rate constants $W^{(i)}$'s of type $i$.
  \item Choose conditions when to stop the simulation.
  \end{description}
\item \label{algst:rsmrti} Reaction time
  \begin{description}
\setlength{\itemsep}{0pt}
\setlength{\parsep}{0pt}
\setlength{\parskip}{0pt}
\setlength{\partopsep}{0pt}
\setlength{\topsep}{0pt}
  \item Generate a time interval $\Delta t$ when no reaction takes place
  \begin{equation}
    \Delta t =-{1\over k}\ln r,
  \end{equation}
  \item where $r$ is a random deviate on the unit interval.
  \end{description}
\item \label{algst:rsmrty} Reaction type
  \begin{description}
\setlength{\itemsep}{0pt}
\setlength{\parsep}{0pt}
\setlength{\parskip}{0pt}
\setlength{\partopsep}{0pt}
\setlength{\topsep}{0pt}
  \item Pick a type of reaction randomly.
  \end{description}
\item \label{algst:rsmrl} Reaction location
  \begin{description}
\setlength{\itemsep}{0pt}
\setlength{\parsep}{0pt}
\setlength{\parskip}{0pt}
\setlength{\partopsep}{0pt}
\setlength{\topsep}{0pt}
  \item Pick a site randomly.
  \end{description}
\item \label{algst:rsmup} Update
  \begin{description}
\setlength{\itemsep}{0pt}
\setlength{\parsep}{0pt}
\setlength{\parskip}{0pt}
\setlength{\partopsep}{0pt}
\setlength{\topsep}{0pt}
  \item Change time to $t\to t+\Delta t$.
  \item If the reaction is possible at the site from
  step~\ref{algst:rsmrl}, then accept the reaction with probability
  $W^{(i)}/W_{\rm max}$ where $i$ is the type of reaction from
  step~\ref{algst:rsmrty}
  \item If the reaction is possible and accepted, change the
  configuration.
  \end{description}
\item Continuation
  \begin{description}
\setlength{\itemsep}{0pt}
\setlength{\parsep}{0pt}
\setlength{\parskip}{0pt}
\setlength{\partopsep}{0pt}
\setlength{\topsep}{0pt}
  \item If the stop conditions are fulfilled then stop. If not repeat at
  step~\ref{algst:rsmrti}.
  \end{description}
\end{enumerate}

This algorithm is called the Random Selection Method (RSM). Note that
the reaction time, the type of the reaction, and the location of the
reaction can be done in any order. Only time and the configuration of
the system needs to be updated. The method is therefore very efficient,
provided that in step~\ref{algst:rsmup} reaction are accepted often.
\subsection{The First Reaction Method.}

Instead of splitting the time, the type, and the location of a reaction,
it is also possible to combine them. This is done in the First Reaction
Method.

\bigskip
\noindent The First Reaction Method (FRM)
\begin{enumerate}
\setlength{\itemsep}{0pt}
\setlength{\parsep}{0pt}
\setlength{\parskip}{0pt}
\setlength{\partopsep}{0pt}
\setlength{\topsep}{0pt}
\item Initialize
  \begin{description}
\setlength{\itemsep}{0pt}
\setlength{\parsep}{0pt}
\setlength{\parskip}{0pt}
\setlength{\partopsep}{0pt}
\setlength{\topsep}{0pt}
  \item Generate an initial configuration $\alpha$.
  \item Set the time $t$ to some initial value.
  \item Make a list $L_{\rm rx}$ containing all reactions.
  \item Generate for each reaction $\alpha\to\beta$ in $L_{\rm rx}$ a
  time of occurrence
  \begin{equation}
    t_{\beta\alpha}=t-{1\over W_{\beta\alpha}}\ln r
  \end{equation}
  with $W_{\beta\alpha}$ the rate constant for the reaction and $r$ a
  random deviate on the unit interval.
  \item Choose conditions when to stop the simulation.
  \end{description}
\item \label{algst:frmrx} Reaction
  \begin{description}
\setlength{\itemsep}{0pt}
\setlength{\parsep}{0pt}
\setlength{\parskip}{0pt}
\setlength{\partopsep}{0pt}
\setlength{\topsep}{0pt}
  \item Take the reaction $\alpha\to\alpha^\prime$ with
  $t_{\alpha^\prime\alpha}\le t_{\beta\alpha}$ for all $\beta$.
  \item If the reaction is enabled go to step~\ref{algst:frmeu}. If not
    go to step~\ref{algst:frmdu}.
  \end{description}
\item \label{algst:frmeu} Enabled update
  \begin{description}
\setlength{\itemsep}{0pt}
\setlength{\parsep}{0pt}
\setlength{\parskip}{0pt}
\setlength{\partopsep}{0pt}
\setlength{\topsep}{0pt}
  \item Change the configuration to $\alpha^\prime$.
  \item Change time to $t\to t_{\alpha^\prime\alpha}$
  \item Remove the reaction $\alpha\to\alpha^\prime$ from $L_{\rm rx}$.
  \item Add new enabled reactions to $L_{\rm rx}$ and generate for each
  reaction $\alpha^\prime\to\beta$ a time of occurrence
  \begin{equation}
    t_{\beta\alpha^\prime}=t-{1\over W_{\beta\alpha^\prime}}\ln r.
  \end{equation}
  \item Skip to step~\ref{algst:frmcn}.
  \end{description}
\item \label{algst:frmdu} Disabled update
  \begin{description}
\setlength{\itemsep}{0pt}
\setlength{\parsep}{0pt}
\setlength{\parskip}{0pt}
\setlength{\partopsep}{0pt}
\setlength{\topsep}{0pt}
  \item Do not change the configuration: $\alpha^\prime$ is the same
  configuration as $\alpha$.
  \item Remove the disabled reaction from $L_{\rm rx}$.
  \end{description}
\item \label{algst:frmcn} Continuation
  \begin{description}
\setlength{\itemsep}{0pt}
\setlength{\parsep}{0pt}
\setlength{\parskip}{0pt}
\setlength{\partopsep}{0pt}
\setlength{\topsep}{0pt}
  \item If the stop conditions are fulfilled then stop. If not set
  $\alpha$ to $\alpha^\prime$ and repeat at step~\ref{algst:frmrx}.
  \end{description}
\end{enumerate}

This algorithm is called Discrete Event simulation in computer
science.\cite{mit82} In FRM the determination of the type and the site
of a reaction is replaced by comparing times of occurrences for
individual reactions. That this is correct can be seen as follows.
Suppose we have two reactions with rate constants $W_1$ and $W_2$. The
probability that no reaction occurs in the interval $[0,t]$ is then
$\exp[-(W_1+W_2)t]$, whereas the probability that neither reaction 1 nor
reaction 2 occurs in that interval equals $\exp(-W_1t)\exp(-W_2t)$,
which is obviously the same as the previous expression. This proves that
FRM generates correct reaction times. It's a bit more work to show that
the reactions are chosen with the correct probability. The probability
distribution for the reactions times of the reactions are
$W_i\exp(-W_it)$ with $i=1,2$. The probability that reaction 1 occurs
before reaction 2 in the FRM algorithm is given by
\begin{eqnarray}
  &&\int_0^\infty\!\! dt\,W_1e^{-W_1t}
    \int_t^\infty\!\! dt^\prime W_2e^{-W_1t^\prime}\\
  &&\quad=W_1\int_0^\infty\!\! dt\,e^{-W_1t}e^{-W_2t}
         ={W_1\over W_1+W_2}.\nonumber
\end{eqnarray}
For reaction 2 we find $W_2/(W_1+W_2)$, which shows that FRM also picks
the reactions with the correct probability. So we see that one can
either generate one time for all reactions and then choose one reaction,
or generate times for all reactions and then take the first that will
occur. We will use this later on in another
way.

The disadvantage of FRM is the determination of the reaction with the
smallest time of occurrence. Scanning a list of all reaction for each
new reaction scales as $O(S)$. More efficient is to make the list of all
reactions an ordered one, and keep it ordered during a simulation.
Getting the next reaction scales then as $O(1)$, but inserting new
reactions in $L_{\rm rx}$ scales as $O(\log S)$.\cite{knu73b} This is not
as good as constant time, but it is not particularly bad either. Still
VSSM is often more efficient than FRM, but VSSM cannot always be used as
we will show later, whereas FRM can always be used.

Note that disabled reactions are not removed from the list of all
reactions. Note also that we only have to generate reaction times for
the new enabled reactions. Times for reactions already in $L_{\rm rx}$
need not be generated again. Suppose that at time $t=t_1$ a time has
been generated for a reaction with rate constant $W$. The probability
distribution for that time is $W\exp[-W(t-t_1)]$. Now assume that at
time $t=t_2>t_1$ the reaction has not occurred. We might generate a new
time using the new probability distribution $W\exp[-W(t-t_2)]$. However,
the ratio of the values of these probability distributions for times
$t>t_2$ is $W\exp[-W(t-t_2)]/W\exp[-W(t-t_1)]=\exp[W(t_2-t_1)]$ is a
constant. Hence relative probabilities for the times $t>t_2$ that the
reaction can occur are the same for both probability distributions, and
no new time need to be generated.
\subsection{Practical considerations}
\label{ssec:prac}

There are different aspects to consider by people who just want to use
the algorithms above to simulate a particular reactions system, and by
people who want to implement them. For the implementation the efficiency
of the methods described above depends very much on details of the
algorithm that we have not discussed. However, some general guidelines
can and will be given here. The interested reader is referred to
references~\cite{luk98} and \cite{seg99} for a more extensive analysis.

An important point is that memory and computation time depend mainly on
the data structures that are used. Except for the time steps there is
relatively little to really calculate. This involves the generation of a
random number. Random numbers are also needed to pick reactions or
reaction types and sites. More critical are the data structures that
contain the reactions and/or reaction types. These lists are priority
queues,\cite{knu73b} and in particular for FRM these may become quite
large. A problem are the disabled reactions. Removing them depends
linearly on the size of the lists and is generally inefficient, and
should not be done after each reaction. It is better to remove them only
when they should occur, and it is found that they have become disabled.
Alternatively, one can do garbage collection when the size of the list
becomes too large.\cite{knu73a} The determination of the next reaction
that should occur depends only logarithmically on the size of the list
in FRM. In VSSM and RSM this can even be done in constant time.

There are a few other aspects that are important and that we haven't
mentioned yet. A central step in all algorithms is the determination of
what are the new reactions that have become possible just after a
reaction has occurred. There are dependencies between the reactions that
may be used to speed up the simulation. A small example may make this
clearer. Suppose we have just adsorption of A or B onto vacant sites,
and formation of AB from an A next to a B leaving two vacant sites. The
formation of an AB will allow new A and B adsorptions, but no new AB
formation. So it is not necessary to check if any AB formations have
become enabled.

Testing if a reaction is disabled is not trivial. It won't do to see if
the occupation of the relevant sites allows the reaction to occur. It
may be that the occupation of the sites has changed a few times but then
converted back to a situation so that the reaction can occur again. What
has happened then is that when the reaction became enabled for the
second time it was added to the list of reactions for the second time
too. If the first instance of the reaction on the list is not recognized
as disabled, then the reaction will take place at the first time of
occurrence. This means that effectively the reaction has a double rate
constant. This is similar to oversampling
(section~\ref{subsec:oversampl}) and accepting each reaction with
probability 1. (This problem does not occur, of course, in VSSMs and in
RSM.)

Recognizing that a reaction is disabled can be done by keeping track of
when a reaction became enabled and when the occupation of a site last
changed. If a site involved in the reaction changed after the reaction
became enabled, then the reaction should be regarded as being disabled.
Using the times of these changes may however lead to problems because of
rounding errors in the representation of real (floating point) numbers.
Instead one can use integers that count reactions when they become
enabled. Each reaction is assigned then its count number, and each site
is assigned the number of the reaction that last changed it. It a site
involved in a reaction has a number larger than the number of the
reaction, then the reaction is disabled.

From the point of view of using the algorithms to simulate a system an
important aspect seems to be the scaling with system size. The important
difference between FRM on the one, and VSSM (i.e., VSSMa and VSSMs) and
RSM on the other hand is the dependence on the system size. Computer
time per reaction in VSSM and RSM does not depend on the size of the
system. This is because in these methods picking a reaction is done
using uniform selection, which does not depend on the size of the list
of reactions. In RSM there is not even such a list. In FRM the computer
time per reaction depends logarithmically on the system size. Here we
have to determine which of all reactions will occur first. So for large
systems VSSM and RSM are generally to be preferred. The data structure
of FRM is so time consuming that FRM should only be used if really
necessary.

There are however a number of cases that occur quite frequently in which
VSSM and RSM are not efficient. This is when there are many reaction
types and when the rate constants depend on time. Time-dependent rate
constants will be discussed in section~\ref{ssec:tdtp}. Many reaction
types arise, for example, when there are lateral interactions. In this
case VSSM becomes inefficient because it will take a lot of time to
determine the reaction type. If with the lateral interactions many
adsorbates affect the rate constant of a reaction, then the number of
reaction types easily becomes larger than the number of sites (see
section~\ref{sec:latint}). RSM can be used for lateral interactions,
provided that the effect of them is small. With RSM one need only
include in the reaction description those sites for which the occupancy
changes. If one also includes the sites with the adsorbates affecting
the rate constants then the probability that one picks a reaction type
that can occur at the randomly chosen site is too low. The adsorbates
affecting the rate constants should, of course, be included when one
calculates the rate constant for the determination of the acceptance of
a reaction. If the effect of lateral interactions is large then this
acceptance will often be low, and RSM will not be very efficient. This
is very often the case. In general, one should realize that simulations
of systems with lateral interactions are always costly.

If VSSM and RSM can be used, then the choice between them depends on
how many sites in the system the reactions can occur. RSM is efficient
for reactions that occur on many sites. The probability that a reaction
is possible on the randomly chosen location is then high. If this is not
the case then VSSM should be used.

The choice between FRM, VSSM, and RSM need not be made for all reactions
in a system together, but can be made on a per reaction type basis,
because it is easy to combine the different methods. Suppose that
reaction type 1 is best treated by VSSM, but reaction type 2 best by
RSM. We then determine the first reaction of type 1 using VSSM, and the
first of type 2 by RSM. The first reaction to actually occur is then
simply the first reaction of these two. The proof that this is correct
is identical to the proof of the correctness of FRM. Combining
algorithms in this way can be particularly advantageous for models with
many reaction types.

To summarize, VSSM is generally the best method to use unless the number
of reaction types is very large. In that case use FRM. If you have a
reaction that occurs almost everywhere, RSM should be
considered. Simply doing the simulation with different methods and
comparing is of course best. One should also look for alternative ways
to model the system (see chapter~\ref{ch:modeling}).
\subsection{Time-dependent transition probabilities.}
\label{ssec:tdtp}

If the transition probabilities $W_{\alpha\beta}$ are themselves time
dependent, then the integral formulation above needs to be adapted. This
situation arises, for example, when dealing with Temperature-Programmed
Desorption or Reactions (TPD/TPR),\cite{jan95b,jan95c,nie97} and when
dealing with voltammetry.\cite{kop98} The definition of the
matrices ${\bf W}$ and ${\bf R}$ remains the same, but instead of a
matrix ${\bf Q}(t)$ we get
\begin{equation}
  {\bf Q}(t^\prime,t)\equiv\exp\left[
  -\int_t^{t^\prime}\!\!dt^{\prime\prime}\,{\bf R}(t^{\prime\prime})
  \right].
\label{eq:tdepnot}
\end{equation}
With this new ${\bf Q}$ matrix the integral formulation of the Master
Equation becomes
\begin{eqnarray}
  {\bf P}(t)=\bigg[{\bf Q}(t,0)
  &+&\int_0^t\!\!dt^\prime
  {\bf Q}(t,t^\prime){\bf W}(t^\prime){\bf Q}(t^\prime,0)\\
  &+&\int_0^t\!\!dt^\prime\int_0^{t^\prime}\!\!dt^{\prime\prime}
  {\bf Q}(t,t^\prime){\bf W}(t^\prime){\bf Q}(t^\prime,t^{\prime\prime})
  {\bf W}(t^{\prime\prime}){\bf Q}(t^{\prime\prime},0)
  +\ldots\bigg]{\bf P}(0).\nonumber
  \label{eq:intf2}
\end{eqnarray}
The interpretation of this equation is the same as that of
equation~(\ref{eq:intf1}). This means that it is also possible to use
VSSM to solve the Master Equation. The relevant equation to determine
the times of the reactions becomes
\begin{equation}
  {\bf Q}(t_n,t_{n-1})=r,\label{eq:nonrx}
\end{equation}
where $t_{n-1}$ is the time of the last reaction that has occurred, and
the equation should be solved for $t_n$, which is the time of the next
reaction. If just after $t_{n-1}$ the system is in configuration
$\alpha_{n-1}$, then the next reaction leading to configuration
$\alpha_n$ should be picked out off all possible reaction with
probability proportional to $W_{\alpha_n\alpha_{n-1}}(t_n)$.

The drawback of VSSM for time-dependent transition probabilities is that
the equation for the times of the reactions is often very difficult to
be solved efficiently. Equation~(\ref{eq:nonrx}) can in general not be
solved analytically, but a numerical solution with a $Q$ of the form
shown in figure~\ref{fig:qtime} is also not easy. The problem is that
${\bf R}$ in equation~(\ref{eq:tdepnot}) can contain many terms or terms
that have a very different time dependence due to reactions with
different activation energy. A possible solution is to use VSSM for each
reaction type separately; i.e., we solve equation~(\ref{eq:nonrx}) for
each reaction type separately. The next reaction is then of the type
with the smallest value for $t_n$, and the first reaction is chosen from
those of that type as in VSSM.\cite{seg99} This works provided the
number of reaction types is small.
\begin{figure}[ht]
\includegraphics[width=\hsize]{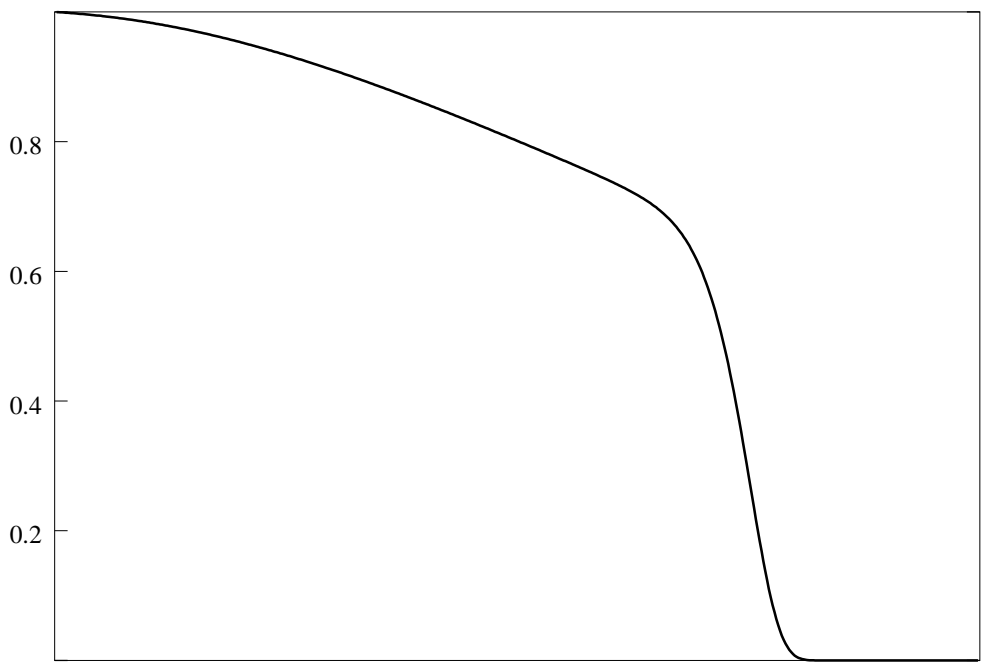}
\caption{Sketch of the probability that no reaction has taken place as a
  function of time typically for a Temperature-Programmed Desorption
  spectrum.  The slow initial decrease is due to a reaction with a small
  activation energy and preexponential factor. The large decrease at the
  end is due to a reaction with a high activation energy and
  preexponential factor.}
\label{fig:qtime}
\end{figure}

Instead of computing a time for the next reaction using the sum of the
transition probabilities of all possible reactions, we can also compute
a time for each reaction. So if we're currently at time $t$ and in
configuration $\alpha$, then we compute for each reaction
$\alpha\to\beta$ a time $t_{\beta\alpha}$ using
\begin{equation}
  \exp\left[-\int_t^{t_{\beta\alpha}}\!\!dt^\prime\,W_{\beta\alpha}(t^\prime)
  \right]=r,\label{eq:rxtime2}
\end{equation}
where $r$ is again a uniform deviate on the unit interval. The first
reaction to occur is then the one with the smallest $t_{\beta\alpha}$.
It can be shown that this time has the same probability distribution as
that of VSSM just as for time-independent rate constants. This method is
FRM for time-dependent reaction rate constants.\cite{jan95b}

The equations defining the times for the reactions,
equation~(\ref{eq:rxtime2}), are often much easier to solve than
equation~(\ref{eq:nonrx}). It may seem that this is offset by the fact
that the number of equations~(\ref{eq:rxtime2}) that have to be solved is
very large, but that is not really the case. Once one has computed the
time of a certain reaction, it is never necessary to compute the time of
that reaction again, but one can use the time that one has computed at
the moment during the simulation when the reaction has become
possible just as for the case of time-independent rate constants.

Equation~(\ref{eq:rxtime2}) can have an interesting property, which is that
it may have no solution. The expression
\begin{equation}
  P_{\rm not}(t)\equiv\exp\left[-\int_{t_{\rm now}}^t\!\!dt^{\prime}\,
  W_{\beta\alpha}(t^{\prime})\right]
\end{equation}
is the probability that the reaction $\alpha\to\beta$ has not occurred
at time $t$ if the current time is $t_{\rm now}$. As $W_{\beta\alpha}$
is a non-negative function of time, this probability decreases with
time. It is bound from below by zero, but it need not go to zero for
$t\to\infty$. If it does not, then there is no solution when $r$ is
smaller than $\lim_{t\to\infty}P_{\rm not}(t)$. This means that there is
a finite probability that the reaction will never occur. This is the
case with some reaction in cyclic voltammetric
experiments.\cite{kop98} There is always a solution if the integral
goes to infinity. This is the case when $W_{\beta\alpha}$ goes slower to
zero than $1/t$, or does not go to zero at all.
\section{A comparison with other methods.}
\label{sec:others}

There are a few other approaches that we want to mention here. The fixed
time step method discretized time. The algorithmic methods are older
methods that are still being used. Kinetic Monte Carlo is very similar
to VSSM and is quite popular. Cellular Automata are mentioned also but
only briefly, as they are really outside the scope of this introduction
to MC methods.
\subsection{The fixed time step method}

If we discretize time then the Master Equation can be written as
\begin{equation}
  P_\alpha(t+\Delta t)
  =P_\alpha(t)
  +\sum_\beta\left[W_{\alpha\beta}\Delta t\,P_\beta(t)
  -W_{\beta\alpha}\Delta t\,P_\alpha(t)\right].
\end{equation}
This means that if at time $t$ we are in configuration $\alpha$, then at
time $t+\Delta t$ we are still in configuration $\alpha$ with
probability $1-\sum_\beta W_{\beta\alpha}\Delta t$ and in configuration
$\beta$ different from $\alpha$ with probability $W_{\beta\alpha}\Delta
t$. This leads to the following algorithm. For each reaction
$\alpha\to\beta$ we generate a random number between 0 and 1. If $r\le
W_{\beta\alpha}\Delta t$ then we change the configuration to $\beta$ and
time to $t+\Delta t$. If $r>W_{\beta\alpha}\Delta t$ then we only change
time to $t+\Delta t$.

The algorithm above can be very efficient if it works. For example, if
the rate constants are time dependent, then the fixed time step method
assumes that it is constant during the interval $[t,t+\Delta t]$. This
avoids the evaluation of integrals like (\ref{eq:rxtime2}). However, it
is obviously an approximation, which might necessitate small time steps.
The time step also has to be small because the probabilities
$W_{\beta\alpha}\Delta t$ and $1-\sum_\beta W_{\beta\alpha}\Delta t$
must be between 0 and 1. Determining the maximum $\Delta t$ that gives
correct results might be cumbersome.

A more subtle problem is that one should avoid the possibility that two
reactions cannot both occur but for both holds $r\le
W_{\beta\alpha}\Delta t$. For example, an adsorbate can desorb or react
with a neighbor. The adsorbate cannot do both, but if for both reactions
we find $r\le W_{\beta\alpha}\Delta t$, then we have a problem. In
practice this means that $\Delta t$ should be chosen so small that at
most one reaction occurs during each time step.
\subsection{Algorithmic approach}

Almost all older Dynamic Monte Carlo methods are based on an algorithm
that defines in what way the configuration changes. (A nice review with
many references to work with these methods is reference~\cite{lom91}.)
The generic form of that algorithm consists of two steps. The first step
is to pick a site. The second step is to try all reactions at that site.
(This may involve picking additional neighboring sites.) If a reaction
is possible at that site, then it is executed with some probability that
is characteristic for that reaction. These two steps are repeated many
times. The sites are generally picked at random. In a variant of this
algorithm just one reaction is tried until on average all sites have
been visited once, and then the next reaction is tried, etc. This
variant is particular popular in situations with fast diffusion; the
``real'' reactions are tried first on average once on all sites, and
then diffusion is used to equilibrate the system before the next cycle
of ``real'' reactions.

These algorithmic Dynamic Monte Carlo methods have provided very
valuable insight in the way the configuration of the adsorbates on a
catalyst evolves, but they have some drawbacks. First of all there is no
real time. Instead time is specified in so-called Monte Carlo steps
(MCS). One MCS is usually defined as the cycle in which every site has
on average been visited once for each reaction. The second drawback is
how to choose the probabilities for reactions to occur. It is clear that
faster reactions should have a higher probability, but it is not clear
how to quantify this. This drawback is related to the first. Without a
link between these probabilities and microscopic reaction rate constants
it is not possible {\it a priori\/} to tell how many real seconds one
MCS corresponds to. The idea is that Monte Carlo time in MCS and real
time are proportional. We have used the similarity with RSM and shown
that this is indeed the case provided temporal fluctuations are
disregarded and one has a steady state. In the case of, for example,
oscillations the two time scales are not proportional.\cite{luk98} In
practice people have used the algorithmic approach to look for
qualitative changes in the behavior of the system when the reaction
probabilities are varied, or they have fitted the probabilities to
reproduce experimental results.

The third drawback is that it is difficult with this algorithmic
definition to compare with other kinetic theories. Of course, it is
possible to compare results, but an analysis of discrepancies in the
results is not possible as a common ground (e.g., the Master Equation in
our approach) is missing. The generic form of the algorithm described
above resembles the algorithm of RSM. Indeed one may look upon RSM as a
method in which the drawbacks of the algorithmic approach have been
removed.
\subsection{Kinetic Monte Carlo}

The problem of real time in the algorithmic formulation of Dynamic Monte
Carlo has also been solved by Fichthorn and Weinberg.\cite{fic91} Their
method is called Kinetic Monte Carlo (KMC) and has become quite popular.
They replaced the reaction probabilities by rate constants, and assumed
that the probability distribution $P_{\rm rx}(t)$ of the time that a
reaction occurs is a Poisson process; i.e., it is given by
\begin{equation}
  P_{\rm rx}(t)=k\exp[-k(t-t_{\rm now})],
\end{equation}
where $t_{\rm now}$ is the current time, and $k$ is the rate constant.
Using the properties of this distribution they derived a method that is
really identical to our VSSM, expect in two aspects. One aspect is that
the Master Equation is absent, which makes it again difficult to make a
comparison with other kinetic theories. Instead the method was derived
by asking under which conditions an equilibrium MC simulation can be
interpreted as a temporal evolution of a system. The other aspect is
that time is incremented deterministically using the expectation value
of the probability distribution of the first reaction to occur; i.e.,
\begin{equation}
  \Delta t={1\over\sum_iN_ik_i},\label{eq:rxtime3}
\end{equation}
where $k_i$ is the rate constant of reaction type $i$ (this is the same
as our transition probabilities $W$ in equation~(\ref{eq:MEdef})), and
$N_i$ is the number of reaction of type $i$. This avoids having to solve
equation~(\ref{eq:rxtime}), and has been used subsequently by many
others.  However, as solving that equation only involves generating a
random number and a logarithm, which is a negligible contribution to the
computer time, this is not really an advantage.
Equation~(\ref{eq:rxtime3}) does neglect temporal fluctuations, which
may be incorrect for systems of low dimensionality.\cite{pri97}

Although the derivation of Fichthorn and Weinberg only holds for Poisson
processes, their method has also been used to simulate TPD
spectra.\cite{men94} In that work it was assumed that, when $\Delta t$
computed with equation~(\ref{eq:rxtime3}) is small, the rate constants are
well approximated over the interval $\Delta t$ by their values at the
start of that interval. This seems plausible, but, as the rate constants
increase with time in TPD, equation~(\ref{eq:rxtime3}) systematically
overestimates $\Delta t$, and the peaks in the simulated spectra are
shifted to higher temperatures. In general, if the rate constants are
time dependent then it may not even be possible to define the
expectation value. We have already mentioned the case of cyclic
voltammetry where there is a finite probability that a reaction will not
occur at all. The expectation value is then certainly not defined. Even
if a reaction will occur sooner or later the distribution $P_{\rm
  rx}(t)$ has to go faster to zero for $t\to\infty$ than $1/t^2$ for the
expectation value to be defined. Solving equations~(\ref{eq:rxtime}) or
(\ref{eq:rxtime2}) does not lead to such problems.
\subsection{Cellular Automata}

There is an extensive literature on Cellular Automata. A discussion of
this is outside the scope of this chapter, and we will restrict
ourselves to some general remarks. We will also restrict ourselves to
Cellular Automata in which each cell corresponds to one site. The
interested reader is referred to
references\cite{mai91,mai93a,dan96,boo96,wei97a,wol02} for an overview
of the application of Cellular Automata to surface reactions.

The main characteristic of Cellular Automata is that each cell, which
corresponds to a grid point in our model of the surface, is updated
simultaneously. This allows for an efficient implementation on massive
parallel computers. It also facilitates the simulation of pattern
formation, which is much harder to simulate with some asynchronous
updating scheme as in Dynamic Monte Carlo.\cite{dro95a} The question is
how realistic a simultaneous update is, as a reaction seems to be a
stochastic process. One has tried to incorporate this randomness by
using so-called probabilistic Cellular Automata, in which updates are
done with some probability. These Cellular Automata differ little from
Dynamic Monte Carlo. In fact, probabilistic Cellular Automata can be
made that are equivalent to the RSM algorithm.\cite{seg99}
\section{The CARLOS program}
\label{sec:car}

CARLOS is a general purpose program for surface reactions.\cite{carlos}
It was developed by Johan Lukkien at the Eindhoven University of
Technology in the Netherlands. General purpose means that it does not
have any reaction hard-coded, but one can specify almost any type of
reaction on input. For time-independent rate constants one can choose
between VSSM, RSM, and FRM. With time-independent rate constants the
different methods can be combined in one simulation. For time-dependent
rate constants only FRM can be used. Each rate constant $W$ can be
specified as a constant or as
\begin{equation}
  W=\nu e^{-E{\rm act}/k_{\rm B}T},
\end{equation}
and the preexponential factor $\nu$, the activation energy $E_{\rm
  act}$, and the temperature $T$ are to be given on input. These
quantities can be linear functions of time. In this way one can simulate
TPD/TPR experiments ($T$ a function of time) and linear sweep
voltammetry experiments ($E_{\rm act}$ a function of time). One can also
change any quantity in discrete steps. This allows the use of VSSM and
RSM for rate constants with arbitrary time dependence, although strictly
speaking only approximately.

CARLOS uses pattern recognition for finding which reactions are
possible. Each reaction has to be specified as a list of sites involved
in the reaction, the occupation of the sites before, and the
occupations of them after the reaction has taken place. For example, in
\begin{equation}
  (0,0),(1,0):\hbox{ A B}\to\hbox{$*$ $*$}
\end{equation}
an integer pair $(n_1,n_2)$ indicates a unit cell, A and B are
reactions, and $*$ the product. This can be interpreted as an molecule A
reacting with a molecule B at a neighboring site to form AB which
immediately desorbs leaving two vacant sites. This interpretation,
however, is based on the meaning of the labels A, B, and $*$. CARLOS does
not know this meaning so the interpretation is ultimately up to the
user. CARLOS only searches for a pattern on the left of the arrow and
changes it to the pattern on the right. The labels can also be used to
indicate different substrate atoms, types of sites (sites with different
coordination numbers, defect sites, step sites, etc.), or surface
reconstruction.

The indices specifying the unit cells are relative. The specification
above stands not just for a reaction at $(0,0)$ and $(1,0)$, but at
$(n_1,n_2)$ and $(n_1+1,n_2)$ for any integers $n_1$ and $n_2$. Having
more than one site per unit cell is also possible. For example in
\begin{equation}
  (0,0/0),(0,0/1):\hbox{ A B}\to\hbox{$*$ $*$}
\end{equation}
the reaction takes place on two sites in the same unit cell. The
integers after the slash indicate which sites are involved. CARLOS does
not make any assumptions on distances or angles. So
\begin{equation}
  (0,0),(1,0):\hbox{ A B}\to\hbox{$*$ $*$}
\end{equation}
may refer to a square, a hexagonal, or any other type of
lattice. However, to simplify the specification of many symmetry related
reactions, one can specify this symmetry, and then CARLOS does assume
certain conventions in the specification of the sites. For example, the
lines
\halign{\quad\hfil#&\quad#\hfil\cr
Active sites: & top on hex\cr
$\vdots$ & $\vdots$ \cr
Symmetry: & hex 120\cr
$\ldots$ & $(0,0),(1,0):\hbox{ A B}\to\hbox{$*$ $*$}$\cr}
\noindent on input make CARLOS assume that the lattice is hexagonal, and
that the primitive translation vectors have the same length and make an
angle of $60^\circ$. Apart from the reaction explicitly stated above,
CARLOS will also generate
\begin{equation}
  (0,0),(-1,1):\hbox{ A B}\to\hbox{$*$ $*$}
\end{equation}
and
\begin{equation}
  (0,0),(0,-1):\hbox{ A B}\to\hbox{$*$ $*$},
\end{equation}
which are obtained by rotations of $120^\circ$.

As output CARLOS can generate a list of coverages and rates as a
function of time. This list contains all possible information on the
kinetics. The configuration of a system as a function of time can be
shown on screen and/or sent to a file either in graphic format or in a
format that CARLOS can use to start a new simulation.

CARLOS always produces output concerning the performance of itself. This
includes how long it took to do a simulation, but also information on
the efficiency with which each type of reaction was simulated. This
information can then be used to change the method, or to change the way
a reaction system is modeled. The efficiency can also be influenced
through CARLOS's garbage collection that removes disabled reactions. One
can specify how much memory CARLOS should use. This makes it possible to
run CARLOS with limited computer resources, but it can also be used to
increase the frequency with which CARLOS does garbage collection. CARLOS
also has an option that tries to optimize its memory usage. CARLOS does
not have any hard coded limits on the number of reactants, the number of
reaction types, or the system size. This depends only on computer
limits.

A special class of reactions are formed by the so-called immediate
reactions. They take place as soon as they become enabled; time is not
advanced. These reactions were originally introduced to simulate models
with infinitely fast reactions like the ${\rm CO}_2$ formation in the
ZGB-model. They proved to be also very useful, however, for making efficient
models of other reaction systems. CARLOS distinguishes even immediate
reactions with different priorities; reactions with higher priorities
take place before immediate reactions with lower priorities.

One common application of these immediate reactions is repulsive
interactions that are so strong that they are always avoided by a
system. Suppose that a molecule A strongly repels other A's on
neighboring sites. Instead of implementing this strong repulsion as a
situation with high energy, it may be more efficient to block all
neighboring sites of each A. This can be accomplished by an immediately
reaction.
\begin{equation}
  (0,0),(1,0):\hbox{ A $*$}\to\hbox{A blocked}
\label{eq:blk}
\end{equation}
where $*$ is a vacant site and ``blocked'' is also a vacant site, but one
that cannot be occupied by another adsorbate. We assume for simplicity
that there is just one site per unit cell, and that the other
neighboring sites will be blocked by similar immediate reactions. Things
become a bit complicated when the A is removed by, for example, a
desorption. Using a normal reaction of the form
\begin{equation}
  (0,0),(1,0):\hbox{ A blocked}\to\hbox{$*$ $*$}
\end{equation}
will not work. One reason is that this will change only one neighboring
site from ``blocked'' to $*$. Another is that the neighboring site might
have another A neighbor, so that it should remain blocked.

What does work are a normal reaction for the desorption and two
immediate reactions to update the ``blocked'' labels. The desorption is
specified by
\begin{equation}
  (0,0):\hbox{ A}\to\hbox{removed}
\end{equation}
where ``removed'' specifies a vacant site where an A has been
removed. This label is used to trigger the first new immediate reaction
\begin{equation}
  (0,0),(1,0):\hbox{ removed blocked}\to\hbox{removed $*$}.
\label{eq:deblk1}
\end{equation}
After this reaction has changed all ``blocked'' labels of the
neighboring sites a second new immediate reaction
\begin{equation}
  (0,0):\hbox{ removed}\to\hbox{$*$}
\label{eq:deblk2}
\end{equation}
removes it. This procedure may seem to have the same error that there
might be another A neighbor so that the site should retain the
``blocked'' label. However, this can be corrected if now the original
immediate reaction~(\ref{eq:blk}) takes place again. This procedure is
correct provided that the immediate reactions are done in a certain
order. Reaction~(\ref{eq:blk}) should have the lowest priority as it
should occur last. Reaction~(\ref{eq:deblk2}) should have a higher
priority, and reaction~(\ref{eq:deblk1}) should have the highest
priority. Note that there is an obvious overhead; vacant sites with two
or more neighboring A's will change their label from ``blocked'' to $*$
and then immediately back to ``blocked'' again. This overhead is
often negligible.

Lateral interactions can also be treated specially by CARLOS. Suppose
that we have a one-dimensional system with molecules A that only desorb.
To include the influence of neighboring A's one can specify desorption
from all possible configurations of an A and its neighboring sites. For
a one-dimensional system this involves just four configuration: $*$A$*$,
AA$*$, $*$AA, and AAA. If we have, however, interactions with $Z$
neighboring sites that can either be vacant or occupied by $N$ different
adsorbates, then we have $(N+1)^Z$ different configurations. We see that
we easily get very many reactions. So this way is the most general, but
it is not efficient.

If the lateral interactions can be assumed to be pairwise additive, then
CARLOS has another way to handle them. Each neighboring adsorbate is
assigned an energy value that tells CARLOS how much its presence changes
the activation energy of the reaction. So for the example above
\begin{equation}
  (-1,0),(0,0),(1,0):\hbox{ \{ $*$ A : -10.0\} A \{ $*$ A : -10.0\}}
  \to\hbox{\# $*$ \#}
\end{equation}
would mean that a neighboring A would decrease the activation energy by
-10.0 units of energy. The curly braces group all labels for the site
that are allowed for the reaction. Here the neighbor must be $*$ or A.
The label \# is shorthand meaning that the occupation of the site does
not change.
\section{Dynamic Monte Carlo simulations of rate equations}

Macroscopic rate equations may have their shortcomings with respect to
surface reactions, but they are perfectly acceptable in other situation;
e.g., for reactions in the gas phase or in solutions where the reactants
are well mixed. The rate equations specify how the concentrations of the
reactants and products change as a function of these concentrations. In
\begin{equation}
  {dc_{\rm A}\over dt}
  =\sum_{{\rm X}\ne{\rm A}}k_1^{({\rm X})}c_{\rm X}
  -k_1^{({\rm A})}c_{\rm A}
  +\sum_{{\rm X},{\rm Y}\ne{\rm A}}k_2^{({\rm X}{\rm Y})}c_{\rm X}c_{\rm Y}
  -\sum_{{\rm X}\ne{\rm A}}k_2^{({\rm A}{\rm X})}c_{\rm A}c_{\rm X}
\end{equation}
the terms on the right-hand-side correspond to changes in the
concentration of A due to reactions of the form ${\rm X}\to{\rm
  A}+\ldots$, ${\rm A}\to\ldots$, ${\rm X}+{\rm Y}\to{\rm A}+\ldots$,
and ${\rm X}\to{\rm A}+\ldots$, respectively. The $k$'s are rate
constants.

The standard procedure to solve these equations is to use numerical
methods for sets of ordinary differential equations.\cite{pre89,sto93}
This is, however, generally not trivial. Because the rate constants
often differ enormously, the rate equations form a stiff set. Such sets
are known to be difficult to solve. An alternative method is to use DMC.
This approach has been pioneered by Gillespie and is now used
widely.\cite{gil76,gil77,hon90}

The idea is to replace the rate equations by a Master Equation, and then
to solve that Master Equation with one of the methods that we have
discussed in this chapter. The first step is to use discrete numbers of
reactant and product molecules instead of concentrations. This can
simply be done by specifying the number of molecules in a fixed volume
$V$. If the concentration of molecule X is $c_{\rm X}$, then the number
of these molecules is $N_{\rm X}=Vc_{\rm X}$. A configuration $\alpha$
is now a specification of the numbers of all molecule types in the
system. For example, if we have the reactions ${\rm A}\to{\rm B}$ and
${\rm B}\to{\rm C}$, then there are three types of molecules (A, B, and
C) and the configuration consists of the number $N_{\rm A}$, $N_{\rm
  B}$, and $N_{\rm C}$.

The next step is to pose a Master Equation from which the rate equations
can be derived. So we do not derive a Master Equation here, but use it
as a mathematical model from which the properties of the rate equations
can be derived. The Master Equation has also a rather obvious chemical
interpretation. It has, of course, the usual form
\begin{equation}
  {dP_\alpha\over dt}=\sum_\beta
  \left[W_{\alpha\beta}P_\beta-W_{\beta\alpha}P_\alpha\right],
\end{equation}
with the configurations corresponding to the sets $\{N_{\rm A},N_{\rm
  B},N_{\rm C},\ldots\}$. We now need to determine the transition
probabilities. (Here we clearly distinguish between the transitions
probabilities of the Master Equation and the rate constants of the rate
equations.) Suppose we have a reaction ${\rm A}\to{\rm B}$ that changes
configuration $\alpha=\{N_{\rm A}^{(\alpha)},N_{\rm
  B}^{(\alpha)},\ldots\}$ to configuration $\beta=\{N_{\rm
  A}^{(\beta)},N_{\rm B}^{(\beta)},\ldots\}$. We take
$W_{\beta\alpha}=w_{{\rm A}\to{\rm B}}N_{\rm A}^{(\alpha)}$ with $N_{\rm
  A}^{(\beta)}=N_{\rm A}^{(\alpha)}-1$, $N_{\rm B}^{(\beta)}=N_{\rm
  B}^{(\alpha)}+1$, and $N_{\rm X}^{(\beta)}=N_{\rm X}^{(\alpha)}$ for
other molecules. The quantity $w_{{\rm A}\to{\rm B}}$ does not depend on
configurations and is characteristic for the reaction. From
\begin{eqnarray}
  {d\expec{N_{\rm A}}\over dt}
  &=&\sum_{\beta\alpha}W_{\beta\alpha}P_\alpha
     \left[N_{\rm A}^{(\beta)}-N_{\rm A}^{(\alpha)}\right]\\
  &=&w_{{\rm A}\to{\rm B}}\sum_\alpha N_{\rm A}^{(\alpha)}P_\alpha[-1]
  =-w_{{\rm A}\to{\rm B}}\expec{N_{\rm A}}
  \nonumber
\end{eqnarray}
we see that we get the rate equation
\begin{equation}
  {dc_{\rm A}\over dt}=-k_1c_{\rm A}
\end{equation}
if we identify $c_{\rm A}$ with $\expec{N_{\rm A}}/V$ and take $w_{{\rm
A}\to{\rm B}}=k_1$.

Similarly when we have the bimolecular reaction ${\rm A}+{\rm B}\to{\rm
  C}$ and we take $W_{\beta\alpha}=w_{{\rm A}+{\rm B}\to{\rm C}}N_{\rm
  A}^{(\alpha)}N_{\rm B}^{(\alpha)}$ with $N_{\rm A}^{(\beta)}=N_{\rm
  A}^{(\alpha)}-1$, $N_{\rm B}^{(\beta)}=N_{\rm B}^{(\alpha)}-1$,
$N_{\rm C}^{(\beta)}=N_{\rm C}^{(\alpha)}+1$, and $N_{\rm
  X}^{(\beta)}=N_{\rm X}^{(\alpha)}$ with ${\rm X}\ne{\rm A},{\rm
  B},{\rm C}$, we get
\begin{equation}
  {d\expec{N_{\rm A}}\over dt}
  =w_{{\rm A}+{\rm B}\to{\rm C}}\sum_\alpha
   N_{\rm A}^{(\alpha)}N_{\rm B}^{(\alpha)}P_\alpha[-1]
  =w_{{\rm A}+{\rm B}\to{\rm C}}
   \expec{N_{\rm A}N_{\rm B}}.
\end{equation}
If we compare this with the derivations in section~\ref{subsec:bimolrx}
we see that we do not have the problem of the number of A-B pairs. This
is because the transition probability is chosen to avoid this
problem. We do have the same problem that we need to make the
approximation $\expec{N_{\rm A}N_{\rm B}}=\expec{N_{\rm A}}\expec{N_{\rm
B}}$ just as in section~\ref{subsec:bimolrx}. This approximation holds
in the thermodynamic limit. Gillespie has argued that, if the
approximation does not hold, the Master Equation is a more realistic
description than the rate equations.\cite{gil76,gil77} The fluctuations
that cause  the approximation to break down are real, and they are
neglected in the rate equations, but properly taken into account in the
Master Equation. With the approximation we get
\begin{equation}
  {dc_{\rm A}\over dt}=-k_2c_{\rm A}c_{\rm B}
\end{equation}
with $k_2=w_{{\rm A}+{\rm B}\to{\rm C}}V$.

Other reactions can be handled similarly and we see that the Master
Equation gives the solution of the rate equations in the form
$\expec{N_{\rm X}}/V$. Because the Master Equation is solved via a Monte
Carlo procedure, the numerical problems are avoided. On the other hand
the Monte Carlo simulations that solve the Master Equation have to be
repeated a number of times, because a single simulation gives $N_{\rm
  X}(t)$ whereas we want to know $\expec{N_{\rm X}(t)}$.

The VSSM, RSM, and FRM algorithms can be used here just as we did for
surface reactions. References to the place of a reaction should be
ignored, and transitional probabilities in the algorithms should be
replaced by expressions like $W_{\beta\alpha}=w_{{\rm A}\to{\rm
    B}}N_{\rm A}^{(\alpha)}$ and $W_{\beta\alpha}=w_{{\rm A}+{\rm
    B}\to{\rm C}}N_{\rm A}^{(\alpha)}N_{\rm B}^{(\alpha)}$. Monte Carlo
simulations to solve rate equations have mainly be used to solve systems
with many different reactants and reaction types. This means that there
is another aspect of the Monte Carlo algorithms that determines the
efficiency than there is for surface reactions. A good book on the
subject is the one by Honerkamp.\cite{hon90} Recent work in this area
has focussed on an idea to replace the molecule by effective particles
that represent more than one molecule.\cite{gil01,res01} This means that
one reaction of these effective particles correspond to many reactions
between the real molecules, which can speed up the simulation
substantially. A similar procedure has been suggested for surface
reactions.\cite{jan99a}
\chapter[Modeling Reaction Systems]{Modeling Reaction Systems}
\label{ch:modeling}

In this chapter we will look at how to model reactions. This may seem
rather trivial. One ``just'' has to specify which sites are involved in
a reaction, and the occupation of these sites before and after the
reaction. It turns out that this is often indeed all one needs to do to
simulate a system, but many times there are various ways to model a
system, and then the question is which one gives the most efficient
simulation. This is typically the case when one has a system with
different types of site, reactions with very different rate constants,
diffusion, and/or lateral interactions. We will discuss here a large
number of different reactions. We will start with simple ones that are
straightforward to model, and then go to more complicated cases. In the
description of the models in this chapter we will have in mind the way
models are used by the CARLOS code (see section~\ref{sec:car}). There
will probably be few differences with other general-purpose codes, but
programs that simulate only a limited set of reactions systems will very
likely have some modeling approach hard-coded. If you work with such a
code, it will not be possible to change the model. Nevertheless we still
think that this chapter is useful for people having to work with such
codes, if only to become aware of the different modeling possibilities.
\section[Unimolecular adsorption, $\ldots$]{Unimolecular adsorption,
  desorption, and conversion}

Modeling unimolecular adsorption, desorption, and conversion is
essentially the same. In each case the reaction can be written as ${\rm
  A}\to{\rm B}$. For adsorption A is a vacant site and B the site
occupied by an adsorbate, for desorption A is the site occupied by an
adsorbate and B a vacant site, and for conversion A is the unconverted
adsorbate and B the converted one. Note that we only look at the site
and its occupation. We ignore the fact that prior to adsorption the
adsorbate is in the gas phase or dissolved in a solute. For the
simulation this is irrelevant. In the following we use ${\rm A}\to{\rm
  B}$ as a generic form for all three cases.

Modeling this reaction is very simple. In the notation of the CARLOS
program (see section~\ref{sec:car}) we have
\begin{equation}
  (0,0):\hbox{ A}\to\hbox{B}.
\end{equation}
This means that each site which has a label A, this label can change to
B. Apart from the reaction we also have to specify its rate constant.
This can either be done explicitly, or by specifying an activation
energy plus a preexponential factor. The latter procedure has to be used
for temperature-dependent processes.

Figure~\ref{fig:unirx1} show a snapshot of a simulation of unimolecular
desorption. A small square grid is used, and about one third of the
sites are occupied. Note that the adsorbates are randomly distributed
over the grid. There is no mechanism that can lead to any kind of
ordering. Figure~\ref{fig:unirx2} shows how the coverage and the
desorption rate change in time. The coverage is given in monolayers (ML)
which is the fraction of all sites that is occupied. For the isothermal
desorption the coverage and the desorption rate are simple exponential
decreasing functions of time. Because the system is not very big, there
is a clear difference in the figures for the coverage and for the
desorption rate. Whereas the coverage shows quite smooth curves, there
is substantial noise in the desorption rate. This is because the
desorption rate $R$ and the coverage $\theta$ are related via
$R=-d\theta/dt$. So the desorption rate shows the fluctuations in the
slope of the coverage. These fluctuations are hard to see in the plot of
the coverage itself. The results of the simulations do not change when
diffusion is included. This would only increase the computer time.
\begin{figure}[ht]
\includegraphics[width=0.5\hsize]{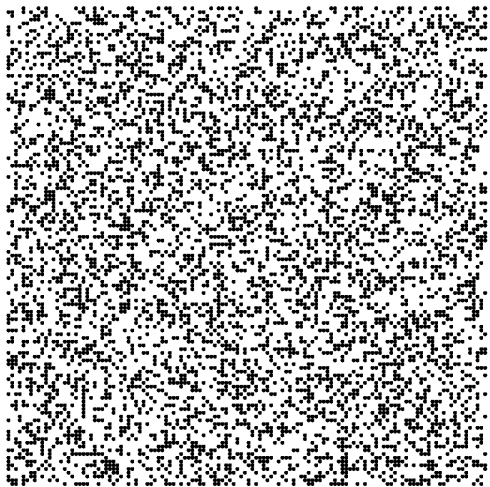}
\caption{Snapshot of unimolecular desorption on a square grid of size
  $128\times 128$. The initial configuration was a grid with each grid
  point occupied, and the snapshot shows a situation in which about two
  third of the sites has been vacated. Adsorbates are in black, vacant
  sites in white.}
\label{fig:unirx1}
\end{figure}
\begin{figure}[ht]
\includegraphics[width=\hsize]{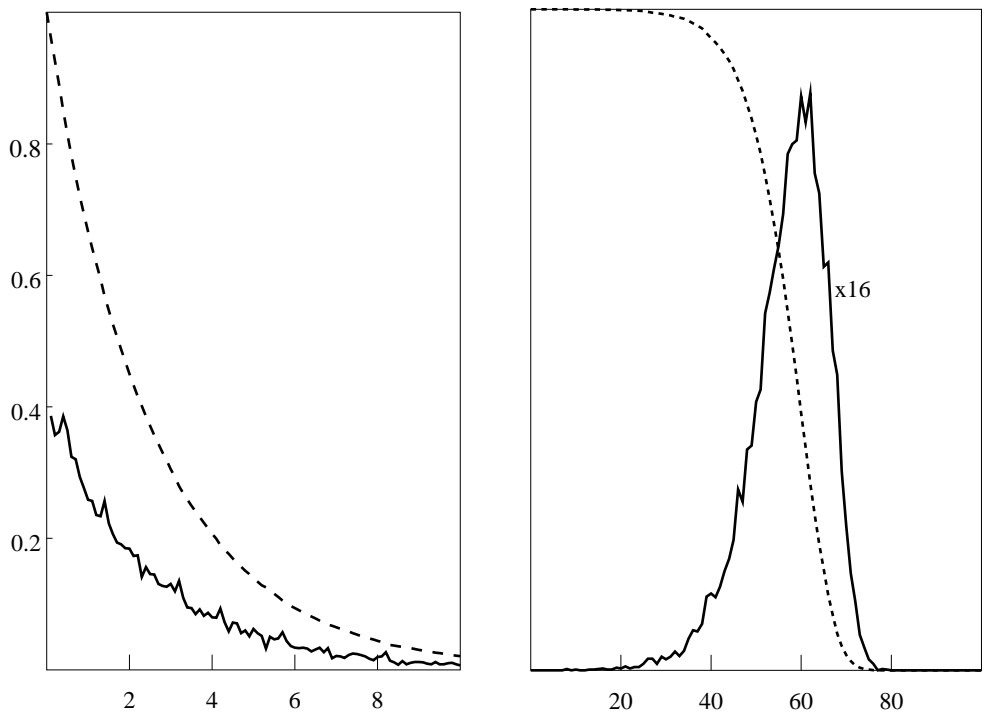}
\caption{Change of the coverage (in ML; dashed curves) and the
  desorption rate (in reactions per second per site; solid curves) as a
  function of time (in seconds) for the same system as the previous
  figure. Isothermal (left) and Temperature-Programmed (right)
  Desorption are shown. The grid size is $128\times 128$. On the left the
  rate constant is $0.4\,\hbox{sec}^{-1}$. On the right the activation
  energy is $15000\,$K, preexponential factor is
  $10^{13}\,\hbox{sec}^{-1}$, heating rate is $2\,$K/sec, and the
  initial temperature is $350\,$K.}
\label{fig:unirx2}
\end{figure}
\section{Bimolecular reactions}

Bimolecular reactions are not really more difficult to model than
unimolecular reactions, but there are a few differences. The first one
is that there are different orientations that the reactants can take
in with respect to each other. One should also be aware that there is a
difference if in ${\rm A}+{\rm B}$ we have ${\rm B}\ne{\rm A}$ or ${\rm
  B}={\rm A}$. We have already seen this difference in
section~\ref{subsec:bimolrx}.

We start with the case with ${\rm A}+{\rm A}$ and for simplicity we
assume that we have a square grid and that the A's react to form a
molecule that immediately desorbs so that we have $2{\rm A}\to 2*$. This
is called associative desorption. In the notation for CARLOS
(section~\ref{sec:car}) we get
\begin{eqnarray}
   (0,0),(1,0)&:&\hbox{ A A}\to\hbox{$*$ $*$},\nonumber\\
   (0,0),(0,1)&:&\hbox{ A A}\to\hbox{$*$ $*$}.
\end{eqnarray}
We see that we have to specify two ways in which the A's can react
corresponding to the different relative orientations of a pair of AA
neighbors. Again we also have to specify the rate constant. This can
again either be done explicitly, or by specifying an activation energy
plus a preexponential factor.

Figure~\ref{fig:birx1} show a snapshot of a simulation of associative
desorption. We see that there are isolated A's. If
we have no diffusion then these will remain on the surface indefinitely.
Figure~\ref{fig:birx2} shows how the coverage and the desorption rate
change in time. Again the coverage shows quite smooth curves, and there
is noise in the desorption rate. We see that the coverage does not go to
zero because not all A's react.
\begin{figure}[ht]
\includegraphics[width=0.5\hsize]{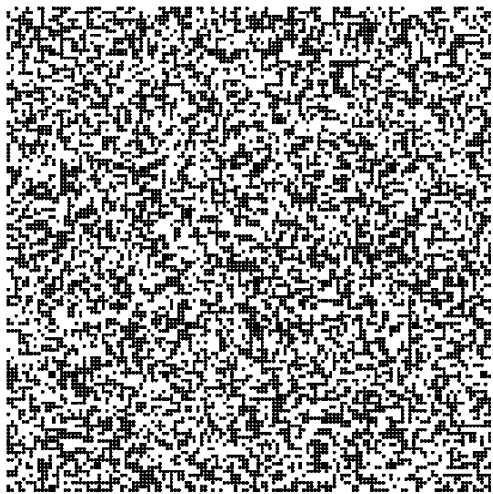}
\caption{Snapshot of associative desorption on a square grid of size
  $128\times 128$. The initial configuration was a grid with each grid
  point occupied, and the snapshot shows a situation in which about half
  of the sites has been vacated. We get horizontal and vertical rows of
  adsorbates, because the adsorbates (in black) desorb in pairs.}
\label{fig:birx1}
\end{figure}
\begin{figure}[ht]
\includegraphics[width=\hsize]{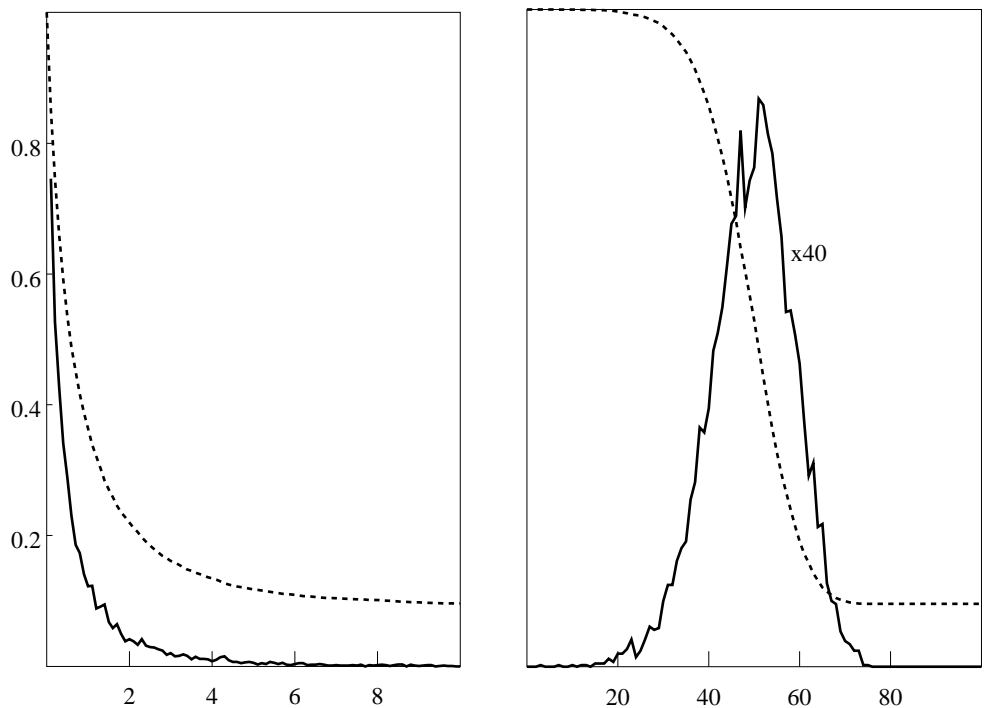}
\caption{Change of the coverage (in ML; dashed curves) and the
  desorption rate (in reaction per second per site; solid curves) as a
  function of time (in seconds) for the same system as the previous
  figure. Isothermal (left) and Temperature-Programmed (right)
  Desorption are shown. The grid size is $128\times 128$. On the left the
  rate constant is $0.4\,\hbox{sec}^{-1}$. On the right the activation
  energy is $15000\,$K, preexponential factor is
  $10^{13}\,\hbox{sec}^{-1}$, heating rate is $2\,$K/sec, and the
  initial temperature is $350\,$K.}
\label{fig:birx2}
\end{figure}

If we include diffusion all the A's will eventually react.
Figure~\ref{fig:birx3} shows that the indeed the coverage goes to zero
for large time $t$. We model the diffusion by
\begin{eqnarray}
   (0,0),(1,0)&:&\hbox{ A $*$}\to\hbox{$*$ A},\nonumber\\
   (0,0),(0,1)&:&\hbox{ A $*$}\to\hbox{$*$ A},\nonumber\\
  (0,0),(-1,0)&:&\hbox{ A $*$}\to\hbox{$*$ A},\nonumber\\
  (0,0),(0,-1)&:&\hbox{ A $*$}\to\hbox{$*$ A}.
\label{eq:moddiff}
\end{eqnarray}
There are four reaction because an A can hop to one of four neighboring
sites if vacant. If the diffusion is so fast that the particles are
randomly mixed then we have
\begin{equation}
  {d\theta\over dt}=-ZW_{\rm des}\theta^2.
\end{equation}
This was shown in section~\ref{subsec:bimolrx}. For the isothermal case
this yields
\begin{equation}
  \theta(t)={\theta(0)\over 1+ZW_{\rm des}\theta(0)t},
\end{equation}
and for the Temperature-Programmed Desorption case
\begin{equation}
  \theta(t)={\theta(0)\over 1+Z[\Omega(t)-\Omega(0)]\theta(0)}
\end{equation}
with $\Omega$ given by equation~(\ref{eq:omega}).  These rate of
desorption $-d\theta/dt$ derived from these analytical solutions are
also shown in Fig.~\ref{fig:birx3}.
\begin{figure}[ht]
\includegraphics[width=\hsize]{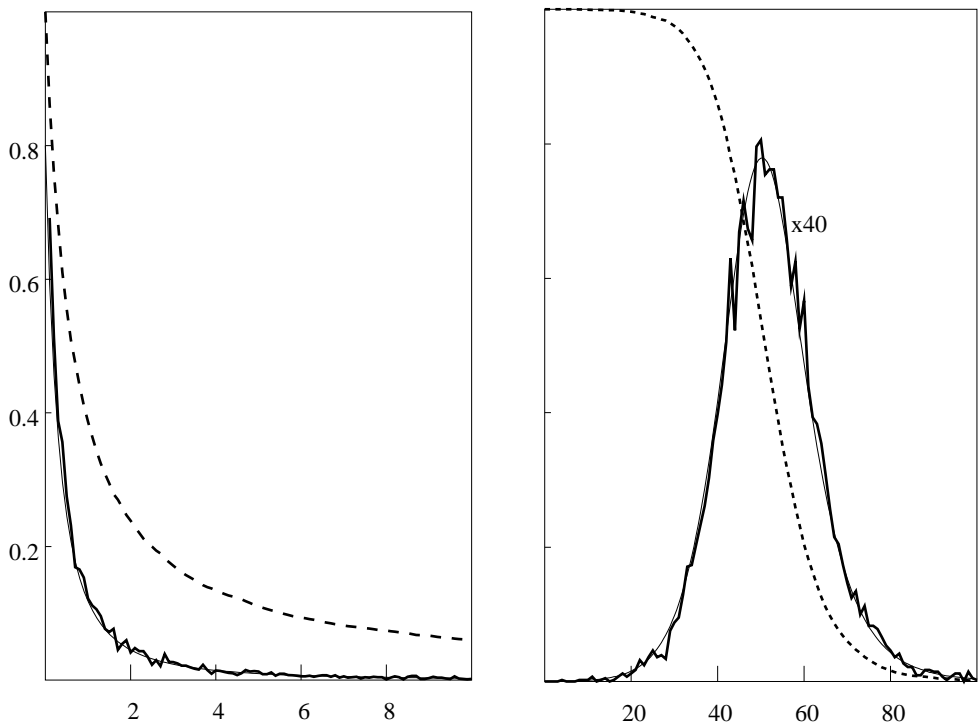}
\caption{Change of the coverage (in ML; dashed curves) and the
  desorption rate (in reaction per second per site; solid curves) as a
  function of time (in seconds) for associative desorption on a square
  grid of size $128\times 128$ with diffusion with a rate constant that
  is 100 times the one of desorption. The initial configuration was a
  grid with each grid point occupied. Isothermal (left) and
  Temperature-Programmed (right) Desorption are shown. On the left the
  rate constant is $0.4\,\hbox{sec}^{-1}$. On the right the activation
  energy is $15000\,$K, preexponential factor is
  $10^{13}\,\hbox{sec}^{-1}$, heating rate is $2\,$K/sec, and the
  initial temperature is $350\,$K. The thin solid lines show the
  analytical solution.}
\label{fig:birx3}
\end{figure}

Next we deal with the case with ${\rm B}\ne{\rm A}$ and for simplicity
we again assume that we have a square grid and that A and B react to
form a molecule that immediately desorbs so that we have ${\rm A}+{\rm
  B}\to 2*$. This is called also associative desorption. In the notation
for CARLOS (section~\ref{sec:car}) we get
\begin{eqnarray}
   (0,0),(1,0)&:&\hbox{ A B}\to\hbox{$*$ $*$},\nonumber\\
   (0,0),(0,1)&:&\hbox{ A B}\to\hbox{$*$ $*$},\nonumber\\
  (0,0),(-1,0)&:&\hbox{ A B}\to\hbox{$*$ $*$},\nonumber\\
  (0,0),(0,-1)&:&\hbox{ A B}\to\hbox{$*$ $*$}.
\end{eqnarray}
We see that we have to specify four ways in which an A can react with a
B corresponding to the different relative orientations of a pair of AB
neighbors. (In CARLOS we have only to specify one reaction and the
information that the others have to be generated by subsequent rotations
over $90^\circ$.) These are two more than for $2{\rm A}\to 2*$, because
AB and BA are of course the same when A and B are the same. We also
include diffusion of A and B, which is modeled as in (\ref{eq:moddiff}).

Figure~\ref{fig:birx4} show a snapshot of a simulation. The initial
configuration corresponds to a random mixture of equal numbers of A's
and B's. It can be noted that there are areas that have almost no B's
and others that have almost no A's. The reason is that locally the
number of A's and the number of B's are not the same. After some time
then the particles in the minority have react and the only particles of
the other type are left.\cite{pri97} The size of the areas with only A's
or only B's depends on the ratio between the rate constant of the
reaction and the hopping rate constant, and it increase as $\sqrt{t}$
with time.\cite{red96} As Fig.~\ref{fig:birx5} shows the coverages
decrease as $1\sqrt{t}$ for large $t$.\cite{pri97,ovc78,tou83}
Initially, however, the particles are still randomly mixed, and the
coverage decreases according to the macroscopic rate equations which
yields $\theta(0)/(1+ZW_{\rm rx}\theta(0)t)$.
\begin{figure}[ht]
\includegraphics[width=0.5\hsize]{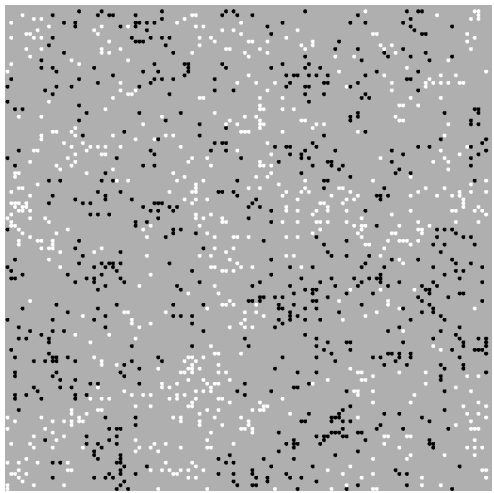}
\caption{Snapshot of ${\rm A}+{\rm B}\to 2*$ on a square grid of size
  $128\times 128$. A's are black, B's are white, and vacant sites are
  grey. The initial configuration was a fully occupied grid with equal
  amount of A's and B's. The rate constant for the reaction is $W_{\rm
    rx}=10\,\hbox{sec}^{-1}$ and for hopping $W_{\rm
    hop}=1\,\hbox{sec}^{-1}$.}
\label{fig:birx4}
\end{figure}
\begin{figure}[ht]
\includegraphics[width=\hsize]{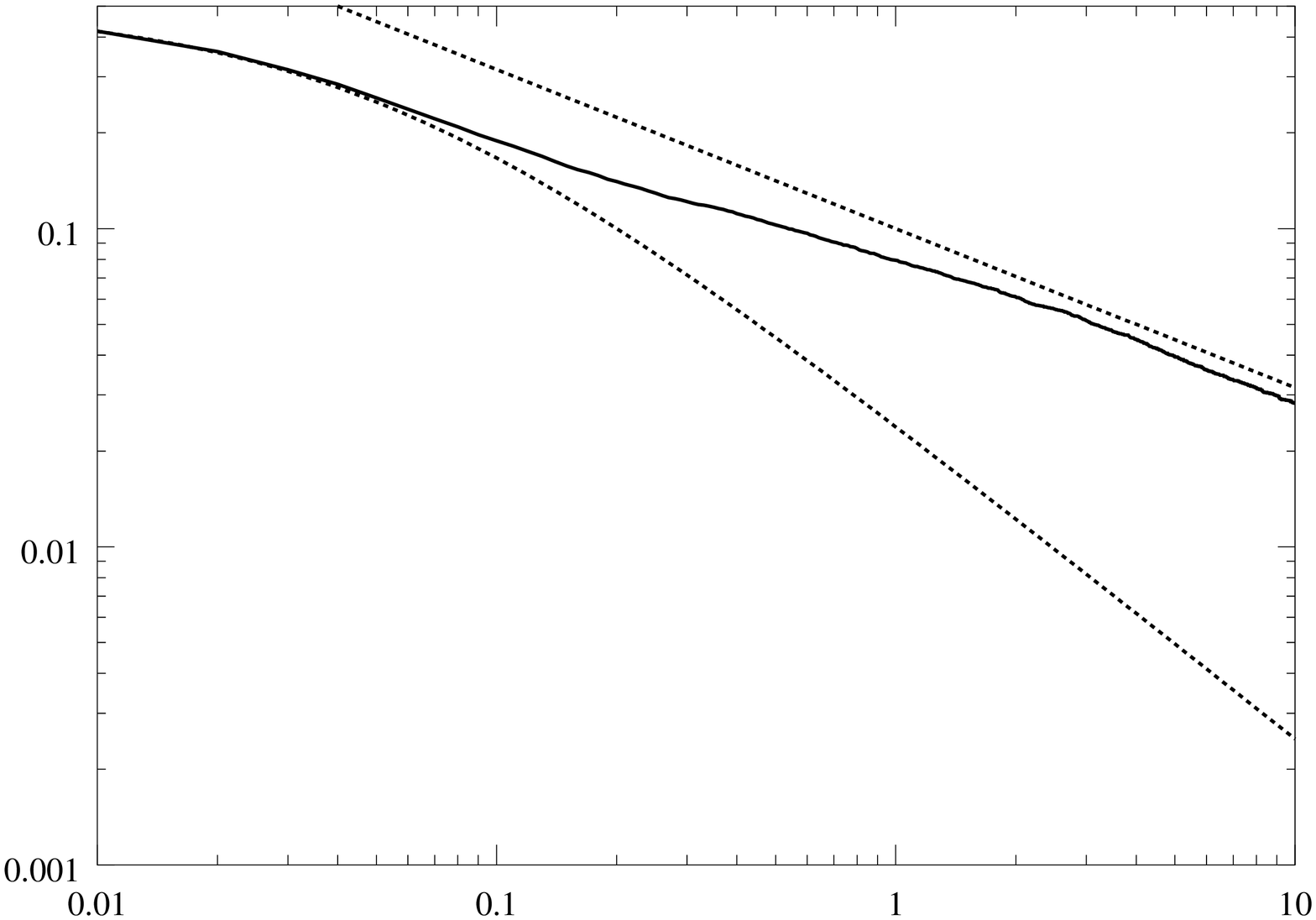}
\caption{Change of the coverage (solid line) as a function of time (in
  seconds) for the same system as the previous figure. The lower dashed
  line depicts the change in coverage obtained from the macroscopic rate
  equations; i.e., when the adsorbates would be randomly distributed at
  all times. The upper dashed line is proportional to $t^{-1/2}$, and
  shows the dependence of the coverage at long times.}
\label{fig:birx5}
\end{figure}
\section{Multiple sites}

The previous sections show how to model all uni- and bimolecular
reactions on a surface with just one site in a unit cell. If there is
more than one site per unit cell, then there are different possibilities
to model them depending on the substrate and the reactions.

Suppose that we are dealing with a (100) surface of an fcc metal and
that we have reaction involving a top (1-fold) and a hollow (4-fold)
site. Such system has two sites per unit cell. However, a first approach
ignores this difference. The main advantage is that we can work with a
smaller unit cell with just one site. This can be seen as follows. If
${\bf a}_1$ and ${\bf a}_2$ are the primitive vectors which have the
same length and orthogonal and which span the unit cell of the (100)
surface, then the sites in the unit cell can be given positions ${\bf
  0}$ and $({\bf a}_1+{\bf a}_2)/2$. If we can ignore the difference
between the two sites, then these sites form a simple lattice with
translation vectors $({\bf a}_1+{\bf a}_2)/2$ and $({\bf a}_1-{\bf
  a}_2)/2$. The advantage of working with such a simple lattice is that
in general we have fewer reactions to specify because we do not
distinguish between the two sites, and it will be easier (and
computationally cheaper) to do the calculations of position on the
surface where a reaction will take place. Whether or not this is correct
depends on the reactions.

Even if we need to distinguish between the two sites, then it is still
possible to work with the lattice with just one site per unit cell. We
need, however, to add a label to the sites to distinguish them. For example,
we are dealing with CO oxidation with CO adsorbing on top sites and
atomic oxygen on 4-fold sites. If the reaction to form ${\rm CO}_2$
occurs when a CO and oxygen are at neighboring sites, then we can model
this with
\begin{eqnarray}
   (0,0),(1,0)&:&\hbox{ CO O}\to\hbox{t h},\nonumber\\
   (0,0),(0,1)&:&\hbox{ CO O}\to\hbox{t h},\nonumber\\
  (0,0),(-1,0)&:&\hbox{ CO O}\to\hbox{t h},\nonumber\\
  (0,0),(0,-1)&:&\hbox{ CO O}\to\hbox{t h}.
\end{eqnarray}
Here position $(1,0)$ is at direction $({\bf a}_1+{\bf a}_2)/2$ from
$(0,0)$, whereas $(0,1)$ is at direction $({\bf a}_1-{\bf a}_2)/2$ from
$(0,0)$. The labels ``t'' and ``h'' indicate a top or hollow site,
respectively, being vacant. It is important to make sure that a
``t'' site is always associated with CO, and a ``h'' site always with
oxygen. This means that CO adsorption should be modeled by
\begin{equation}
   (0,0):\hbox{ t}\to\hbox{CO},
\end{equation}
and dissociative oxygen adsorption by
\begin{eqnarray}
   (0,0),(1,1)&:&\hbox{ h h}\to\hbox{O O},\nonumber\\
  (0,0),(1,-1)&:&\hbox{ h h}\to\hbox{O O}.
\end{eqnarray}
The relative positions $(1,1)$ and $(1,-1)$ here correspond to
translations ${\bf a}_1$ and ${\bf a}_2$, respectively. It is also
important to make sure that the initial configuration corresponds to a
checkerboard pattern of ``t'' or CO and ``h'' or O. This kind of bookkeeping
has the obvious drawback of being error-prone, and takes computer time.

Instead of using labels to distinguish sites, it is also possible to
work with a unit cell with two sites. If ${\bf a}_1$ and ${\bf a}_2$ are
the translations for the same CO oxidation as above we now get for the
CO adsorption
\begin{equation}
   (0,0/0):\hbox{ $*$}\to\hbox{CO},
\end{equation}
with the third ``O'' in $(0,0/0)$ referring to the top site in the unit cell.
The dissociative adsorption of oxygen becomes
\begin{eqnarray}
  (0,0/1),(1,0/1)&:&\hbox{ $*$ $*$}\to\hbox{O O},\nonumber\\
  (0,0/1),(0,1/1)&:&\hbox{ $*$ $*$}\to\hbox{O O}.
\end{eqnarray}
The ``1'' in $(0,0/1)$ indicates the hollow site. Note that we can now use
the same label for any vacant site. The oxidation reaction becomes
\begin{eqnarray}
    (0,0/0),(0,0/1)&:&\hbox{ CO O}\to\hbox{$*$ $*$},\nonumber\\
   (0,0/0),(-1,0/1)&:&\hbox{ CO O}\to\hbox{$*$ $*$},\nonumber\\
   (0,0/0),(0,-1/1)&:&\hbox{ CO O}\to\hbox{$*$ $*$},\nonumber\\
  (0,0/0),(-1,-1/1)&:&\hbox{ CO O}\to\hbox{$*$ $*$}.
\end{eqnarray}

Whether it is better to use labels to distinguish the sites or to work
with a unit cell with two sites depends on the reactions and the
substrate. Also coding may play a role. Calculations of the positions of
where reactions can take place can often be done more efficient when the
grid sizes are powers of two. If we use labels, then this might not be
possible. For example, suppose we have a (111) surface of an fcc metal
and we are dealing with top and the two hollow sites. These form again a
simple lattice, but if we use this simple lattice then we have to work
with system sizes that are multiples of three.
\section{Systems without translational symmetry}

The simulations always assume that the sites form a lattice. The
previous section on multiple sites has shown that by adding labels
specifying extra properties to a site we can modify the lattice. We can
use this even to model a system that has no translational symmetry while
still using a lattice.

Suppose we want to model a stepped surface. We can model such a surface
with a large unit cell and multiple sites. This may not be a good idea,
however. We will be dealing with many sites per unit cell, and we will
have to specify for each its reactions. This will generally lead to a
long list of reactions, even if the different sites on the terraces have
the same properties, and only the sites at the steps behave differently.
In such a case it is better to use a label to distinguish to sites at
the step. For example, if we are dealing with simple desorption of an
adsorbate A then we can model desorption from a terrace site as
\begin{equation}
   (0,0):\hbox{ A}\to\hbox{$*$}.
\end{equation}
Desorption from a site at the step becomes
\begin{equation}
   (0,0):\hbox{ As}\to\hbox{$*$s}.
\end{equation}
We add an ``s'' to distinguish the adsorbate on a step site and a vacant
step site. Note that there is a difference with the example of CO
oxidation on a (100) surface of the section on multiple sites. There the
label ``CO'' already implied one type of site and ``O'' another. Here the
same adsorbate can be found on both types of sites. We also need to
specify different rate constants for the desorption for both types of
sites, because otherwise distinguishing them would not be meaningful.
The precise position of the steps has to be specified in the initial
configuration.

For a unimolecular reaction we need to specify just two reactions. If we
have a bimolecular reaction we need to specify reactions for all
possible combinations of occupations of step and terrace sites. So if we
have a reaction ${\rm A}+{\bf B}\to 2*$, then we will have $\hbox{ A
  B}\to\hbox{$*$ $*$}$, $\hbox{ As B}\to\hbox{$*$s $*$}$, $\hbox{ A
  Bs}\to\hbox{$*$ $*$s}$, and $\hbox{ As Bs}\to\hbox{$*$s $*$s}$.
Diffusion can also be regarded as a bimolecular reaction, and we need to
specify $\hbox{ A $*$}\to\hbox{$*$ A}$, $\hbox{ As $*$}\to\hbox{$*$s
  A}$, $\hbox{ A $*$s}\to\hbox{$*$ As}$, and $\hbox{ As
  $*$s}\to\hbox{$*$s As}$ for the diffusion of A. Needless to say that
all these possibilities will have in general different rate constants.

The procedure above will be unavoidable if we have point defects that
are not regularly distributed over the surface. The specification of the
initial configuration determines where the defects are. It is of course
possible to have more than one type of defect.

On a surface with defects most sites usually are normal sites and only a
minority is a defect site. The procedure above can also be used when
that is not the case. This means that we can use it to model a surface
of a bimetallic catalyst. Only the interpretation of the label changes;
it will not indicate a normal or step site, but a site on one or the
other metal. If there are reactions that are not affected by the type of
metal on which it occurs, another model may be more efficient. Suppose
that we have one site per unit cell disregarding the difference between
the metals and all adsorbates occupy top sites. We then specify an
imaginary second site. The first site gets the adsorbate and the second
has to label specifying the metal. The second site doesn't actually
exist. There is only one site, but we split the information on this site
in two. With adsorbate A and metal M1 and M2 we have
\begin{equation}
   (0,0/0),(0,0/1):\hbox{ A M1}\to\hbox{$*$ M1}
\end{equation}
and
\begin{equation}
   (0,0/0),(0,0/1):\hbox{ A M2}\to\hbox{$*$ M2}
\end{equation}
for desorption. In the previous models we combined the information on
the adsorbate and the site in one label. Here we keep that information
apart. This is advantageous if there is a reaction that does not depend
on the metal. So if diffusion of the adsorbate is equally fast on both
metals, we can model this with
\begin{equation}
   (0,0/0),(1,0/0):\hbox{ A $*$}\to\hbox{$*$ A}
\end{equation}
and similar expressions for hops to other neighboring sites. We see that
there is no information on the metal. This would not be possible if the
information on the adsorbate and the information on the metal would have
been combined.

So far the occupation of the sites have been allowed to change through
reactions, but the properties of the sites themselves have been fixed.
This need not always be the case. The surface composition of a
bimetallic catalyst may change, or we might be dealing with a
reconstructing surface. For a reconstructing surface we can introduce a
label that specifies to which phase of the substrate a site belongs. A
change of surface composition of a bimetallic catalyst or a
reconstruction can be modeled with reactions that specify the changes in
the substrate. For example, there have been many studies of CO oxidation
on reconstructing platinum surfaces. In one of the simplest models of
this process there are reaction for the growth of the
phases.\cite{kuz99a,kor99b,kuz02} One phase, called the $\alpha$ phase,
growths if there are no adsorbates. This can be modeled by
\begin{equation}
  (0,0/1),(0,0/0),(1,0/1),(1,0/0):
  \hbox{ $\alpha$ $*$ $\beta$ $*$}\to\hbox{$\alpha$ $*$ $\alpha$ $*$}.
\end{equation}
The $\beta$ phase growths in an area with sites occupied by CO.
\begin{equation}
  (0,0/1),(0,0/0),(1,0/1),(1,0/0):
  \hbox{ $\beta$ CO $\alpha$ CO}\to\hbox{$\beta$ CO $\beta$ CO}.
\end{equation}
It should be realized that there are restrictions in what one can do
with a changing substrate in DMC simulations. One does need to be able to
put everything on a grid. Reconstructions that lead, for example, to
surface structure with a different density can only be modeled if this
change is ignored. Also changes of the local point-group symmetry may
not be possible to model.
\section{Infinitely fast reactions}

Sometimes a reaction is so much faster than the other reactions in a
system that one regard may as an infinitely fast reaction. This has the
advantage that one doesn't need to compute the time when that reaction
takes place. One simply checks if the sites involved are occupied by the
correct reactants and then replaces them by the products without
changing the time. More difficult are situations where more than one
infinitely fast reaction are possible but not all of them can actually
take place. Sometimes one can choose a reaction at random, sometimes one
needs to give different infinite reactions different priorities. There
are also situations when it is better to incorporate a fast reaction in
another reaction, and there are situations where one doesn't actually
have infinitely fast reactions, but they can be useful in the modeling.
CARLOS (see section~\ref{sec:car}) calls these infinitely fast
reactions immediate reactions.

Suppose we have a system with three reactions; simple adsorption of A's,
simple adsorption of B's, and when an A occupies a site next to a B they
react and the product desorbs. We will look at the case where the
reaction ${\rm A}+{\rm B}$ and the desorption is infinitely fast; i.e.,
we have with infinitely large rate constant ${\rm A}+{\rm B}\to 2*$ as
far as the occupation of the sites is concerned. If the rate constant
for adsorption of A (B) is larger than the one for adsorption of B (A),
then after some time the surface will become completely covered by A's
(B's). We will therefore only look at the situation in which A and B
have the same rate constant for adsorption. In this system it can occur
that there is a vacant site with two or more neighboring sites that are
occupied by B's. If an A adsorbs onto the vacant site, then it can react
with either of the neighboring B's. It seems obvious to pick the B with
which the A will react at random. Similarly, if a B adsorbs at a site
where it gets two or more A neighbors, then it will react with one of
them which should be picked at random.

Actually we have implicitly assumed in the previous paragraph that all
neighbors are equivalent. This is the case for example on a square grid
that represents a (100) surface of an fcc metal, or a hexagonal grid
that represents a (111) surface of an fcc metal. This is not the case
for a rectangular grid that represents a (110) surface of an fcc
metal. There is a different rate constant for a reaction with a
neighboring in one direction than for the direction perpendicular to
it. It is possible in this situation to have different priorities for
the infinitely fast reactions. Suppose an A has a B neighbor in
direction 1 and a B neighbor in direction 2. It will react with both of
them infinitely fast, but it prefers to react with the one in direction
1. In such a situation one should give the reaction with the neighbor in
direction 1 a higher priority. So first check if there are AB pairs
oriented in direction 1. If this is the case then have them react. If there
is more than one of such a pair then if necessary to choose at random among
the different pairs. After there are no more of such pairs, then check
for AB pairs oriented in direction 2, and react them.

One can also remove the infinitely fast reaction
altogether. Straightforward modeling as above yields
\begin{equation}
   (0,0):\hbox{ $*$}\to\hbox{A}
\end{equation}
for the adsorption, and
\begin{equation}
   (0,0),(1,0):\hbox{ A B}\to\hbox{$*$ $*$}
\end{equation}
plus symmetry-related expressions for the infinitely fast reaction. We
can now combine these two reactions. For example an adsorption on a site
with vacant neighboring sites can be modeled as
\begin{equation}
  \matrix{& (0,1),\cr
          (-1,0), & (0,0), & (1,0), \cr
          & (0,-1) \cr}:
  \matrix{& *\cr
          * & * & * \cr
          & * \cr}\to
  \matrix{& *\cr
          * & \hbox{A} & * \cr
          & * \cr}.
\end{equation}
For simplicity we have assumed that we have a square grid. The rate
constant for this reaction equals the rate constant for adsorption. More
interesting is the adsorption on a site with one B neighbor. We then
have
\begin{equation}
  \matrix{& (0,1),\cr
          (-1,0), & (0,0), & (1,0), \cr
          & (0,-1) \cr}:
  \matrix{& *\cr
          * & * & \hbox{B} \cr
          & * \cr}\to
  \matrix{& *\cr
          * & * & * \cr
          & * \cr}.
\end{equation}
The adsorption takes place at $(0,0)$, but there is an immediate
reaction with the B so that the sites at $(0,0)$ and $(1,0)$ become
vacant again. On a square grid there are three other symmetry-related
reactions. The rate constant for each of them is again the rate constant
for adsorption. If there are two B neighbors we get
\begin{equation}
  \matrix{& (0,1),\cr
          (-1,0), & (0,0), & (1,0), \cr
          & (0,-1) \cr}:
  \matrix{& \hbox{B}\cr
          * & * & \hbox{B} \cr
          & * \cr}\to
  \matrix{& \hbox{B}\cr
          * & * & * \cr
          & * \cr}
\end{equation}
and symmetry-related reactions or
\begin{equation}
  \matrix{& (0,1),\cr
          (-1,0), & (0,0), & (1,0), \cr
          & (0,-1) \cr}:
  \matrix{& *\cr
          \hbox{B} & * & \hbox{B} \cr
          & * \cr}\to
  \matrix{& *\cr
          \hbox{B} & * & * \cr
          & * \cr}
\end{equation}
and symmetry-related reactions. In both reactions there are two
possibilities for A to react. This affects the rate constant. Suppose we
have at time $t$ the situation
\begin{equation}
   \matrix{&*\cr \hbox{B} & * & \hbox{B}\cr &*\cr}
\end{equation}
and an A adsorbs in the middle. This adsorption takes on average a time
$W_{\rm ads}^{-1}$, where $W_{\rm ads}$ is the rate constant for
adsorption. So after adsorption we are on average at time $t+W_{\rm
  ads}^{-1}$ and have the situation
\begin{equation}
   \matrix{&*\cr \hbox{B} & \hbox{A} & \hbox{B}\cr &*\cr}.
\end{equation}
Next the A reacts with one of the B's to give
\begin{equation}
   \matrix{&*\cr * & * & \hbox{B}\cr &*\cr},
   \hbox { or }
   \matrix{&*\cr \hbox{B} & * & *\cr &*\cr}
\end{equation}
with equal probability, and we are still at time $t+W_{\rm ads}^{-1}$.
If use the second method of modeling the reactions, then we go directly
from the initial to one of the final situations. If the rate constant
for each of the reactions is $W_{\rm 2B}$, then on average the reaction
takes place at time $t+(2W_{\rm 2B})^{-1}$. We get a factor $1/2$,
because in the initial situations there are two reactions possible. The
total rate constant for that situation is the sum of the rate constants
of all possible reactions; i.e., $2W_{\rm 2B}$. To get the same result
as before we therefore must have $W_{\rm 2B}=W_{\rm ads}/2$. In general,
if we have $N$ B neighbors in the initial situation then the direct
reactions should have a rate constant that is equal to the rate constant
for adsorption divided by $N$.

Whether removing the infinitely fast reactions is efficient will depend
on the system, but it is more common to introduce such reactions rather
than to remove them. Introducing infinitely fast reactions need not be
restricted to reactions that actual occur. In fact, the reactions we
introduce are generally ones that exist only in the model and not in
reality. Suppose that we have adsorption of a somewhat bulky adsorbate
A. The adsorbate occupies not only a particular site, but also makes it
impossible for other adsorbates to occupy neighboring sites. On a square
grid we might try to model this as follows.
\begin{equation}
  \matrix{& (0,1),\cr
          (-1,0), & (0,0), & (1,0), \cr
          & (0,-1) \cr}:
  \matrix{& *\cr
          * & * & * \cr
          & * \cr}\to
  \matrix{& \hbox{At}\cr
          \hbox{Al} & \hbox{A} & \hbox{Ar} \cr
          & \hbox{Ab} \cr}
\end{equation}
where the labels on the right stand for the adsorbate, and blocked sites
above, left, right, and below, respectively. (Having all five
sites getting a label A is not a good idea, because after more
adsorbates have adsorbed you won't be able to tell which A stands for an
adsorbate and which stands for a blocked site.) The drawback of this
model is that effectively more than these four neighboring sites are
blocked. For example, it is not possible for another A to adsorb on site
$(1,1)$ because such adsorption above needs $*$ on $(1,0)$ and $(0,1)$
whereas there are Ar and Aa, respectively.

This problem can be solved with infinitely fast reactions. The
adsorption we model simply with
\begin{equation}
   (0,0):\hbox{ $*$}\to\hbox{A}.
\end{equation}
This reaction has a finite rate constant. The blocking of the
neighboring sites is modeled with
\begin{equation}
  (0,0),(1,0):\hbox{ A $*$}\to\hbox{A blocked}.
  \label{eq:rxblock}
\end{equation}
This reaction, and the symmetry-related ones, are infinitely fast. In
this way another A can adsorb on $(1,1)$. This seems quite
straightforward, but things become a bit more tricky when the adsorbate
can also diffuse and desorb, We look at desorption, Just
\begin{equation}
   (0,0):\hbox{ A}\to\hbox{$*$}
\end{equation}
does not work, because this leaves ``blocked'' labels for sites that are
not blocked any longer,
\begin{equation}
   (0,0),(1,0):\hbox{ A blocked}\to\hbox{$*$ $*$}
\end{equation}
doesn't work either, because this reaction can occur only once for each
adsorbate, This means only one label ``blocked'' is changed back to $*$,
but the adsorbate may have been blocking more than one site, What does
work is first
\begin{equation}
   (0,0):\hbox{ A}\to\hbox{vacated},
\end{equation}
The label ``vacated'' indicates that an adsorbate has just desorbed from
the site, It is used to remove the ``blocked'' labels by
\begin{equation}
   (0,0),(1,0):\hbox{ vacated blocked}\to\hbox{vacated $*$},
\end{equation}
This should be an infinitely fast reaction, Note that the label
``vacated'' stays so that it can remove all ``blocked'' labels, To get
rid of the ``vacated'' label we finally have another infinitely fast
reaction
\begin{equation}
   (0,0):\hbox{ vacated}\to\hbox{$*$},
\end{equation}
It is clear that this reaction should only occur after all ``blocked''
labels have been removed, so it should have a lower priority than the
previous reaction, The last thing that we now need to do is to give the
original blocking reaction~(\ref{eq:rxblock}) an even lower priority, If
we would not do that we could get the following infinite loop
\begin{equation}
  \matrix{* &* &* &* &\to &* &\hbox{A} &* &*\cr
    & & & &\to &\hbox{blocked} &\hbox{A} &* &*\cr
    & & & &\to &\hbox{blocked} &\hbox{A} &\hbox{blocked} &*\cr
    & & & &\to &\hbox{blocked} &\hbox{A} &\hbox{blocked} &\hbox{A}\cr
    & & & &\to &\hbox{blocked} &\hbox{vacated} &\hbox{blocked} &\hbox{A}\cr
    & & & &\to &\hbox{blocked} &\hbox{vacated} &* &\hbox{A}\cr
    & & & &\to &\hbox{blocked} &\hbox{vacated} &\hbox{blocked} &\hbox{A}\cr
    & & & &\to &\hbox{blocked} &\hbox{vacated} &* &\hbox{A}\cr
    & & & &\to &\ldots\cr}
\end{equation}
(We have looked only at the sites along one line for convenience,) With
the blocking reaction having the lowest priority sites only become blocked
again after the ``vacated'' label has been removed,
\section{Diffusion}

Diffusion is often mentioned when shortcomings of DMC are
discussed, This is not quite appropriate, It is even less appropriate to
say that DMC has problems when there are reactions with very different
rate constants, This is indeed a problem when a fix step size is used,
because the step size should be small enough so that the fastest
reactions is simulated correctly, For the slower reactions a small step
size is, however, inefficient, There is no problem with variable step
sizes, Diffusion is indeed a problem, because it is often much faster
than the other processes, DMC will happily simulate this fast diffusion,
but most computer time is spent on diffusion and only a very small
fraction on the other reaction, which are often the ones one really
interesting in, This is not really a shortcoming of DMC, however, but
reflects an intrinsic property of the system one is studying, Any method
that simulates all processes that take place will have to spent most
time on the diffusion simply because most events are adsorbates moving
from site to site, A similar situation arises when one has a very fast
adsorption-desorption equilibrium with most events adsorption and
desorption processes, Nevertheless, there are some ways to reduce the
fraction of computer time that one has to spend on diffusion,

To model diffusion one needs to have a reaction like
\begin{equation}
   (0,0),(1,0):\hbox{ A $*$}\to\hbox{$*$ A},
\end{equation}
The problem is in the rate constant and in a possible use of other
reactions as well, In the simplest case one simply gives the reaction
above a rate constant so that one obtains the correct diffusion constant
(see section~\ref{subsec:diff}), If the diffusion is not too fast (i,e,,
a substantial part of the simulation is spent on other reactions), then
this is all one needs to do, If the diffusion is too fast, then there are a
few other options, Very often diffusion is so fast that it equilibrates
the adlayer before another process/reaction has taken place, In such a
situation the precise rate of diffusion is not important, As long as
this equilibration occurs, then the simulation yields correct
results, This often means that it is possible to make diffusion much
slower than in reality, One has to think here about a reduction of
orders of magnitude, The precise value also depends on the system
size, The displacement of a particle through diffusion increase with the
square of time, If the system is small, then it will rapidly have moved
through the whole system, If the system is large, then this will take
longer, This means that for small systems the diffusion rate can be
reduced more than for large systems,

In some algorithmic approaches of DMC fast diffusion has been simulated
as follows, After a particle has adsorbed it starts to diffuse over the
surface, This diffusion consists of random hops from one site to
another, These hops continue until the particle encounters another
particle with which it can react, It then stops hopping and reacts with
the other particle,\cite{zhd97b,zhd98b} This method was used to model CO
oxidation, CO is the rapidly diffusing particle, and atomic oxygen is
fixed at an adsorption site, and is the particle with which CO reacts,
Diffusion was regarded as being faster than any of the other processes,
but, as the CO react with the very first oxygen that it encounters, it
seems that the formation of ${\rm CO}_2$ is really even faster, If we
work with infinitely fast reaction then we have
\begin{equation}
   (0,0),(1,0):\hbox{ CO $*$}\to\hbox{$*$ CO}
\end{equation}
for the diffusion, This reaction has a low priority, The formation of
${\rm CO}_2$ is represented by
\begin{equation}
   (0,0),(1,0):\hbox{ CO O}\to\hbox{$*$ $*$}
\end{equation}
with a high priority,

If we really want to have a diffusion faster than the oxidation, then
things become more complicated, For the diffusion we now have the
infinitely fast reaction
\begin{equation}
   (0,0),(1,0):\hbox{ CO $*$}\to\hbox{CO CO},
\end{equation}
The label ``CO'' does not mean that there is a CO at that site, but that
CO will visit the site during its diffusion over the surface, This
reaction has a high priority, say 2, The reaction is modeled via
\begin{equation}
   (0,0),(1,0):\hbox{ CO O}\to\hbox{r r},
\end{equation}
The label ``r'' stands for sites on which adsorbates have been removed,
This reaction is also infinitely fast, but has a lower priority, say 1,
This low priority insures that all sites that the CO can visit get a
label ``CO'' before the reaction takes place, After one reaction has
taken place we want all vacant site to get the usual label ``$*$''
again, We do this with an infinitely fast reaction
\begin{equation}
   (0,0),(1,0):\hbox{ r CO}\to\hbox{r r}
\end{equation}
with a priority 4, and an infinitely fast reaction
\begin{equation}
   (0,0):\hbox{ r}\to\hbox{$*$}
\end{equation}
with priority 3, These reaction have the following effect, After one
oxygen atom has reacted with CO, all ``CO'' labels are converted into
``r'' labels, and after that has occurred the ``r'' labels are converted back
into ``$*$'',

It should be clear that the second procedure is much more time consuming
because each CO adsorption is accompanied by a very large number of
reactions, How large the difference between the two procedure is not
clear, Figure~\ref{fig:fastCO} shows snapshots of an adlayer after about
a third of the oxygen atoms has been reacted away but that was initially
completely covered with oxygen except for one site , Note that the hole
in the oxygen layer has a smoother edge when the reaction is faster than
the diffusion,
\begin{figure}[ht]
\includegraphics[width=\hsize]{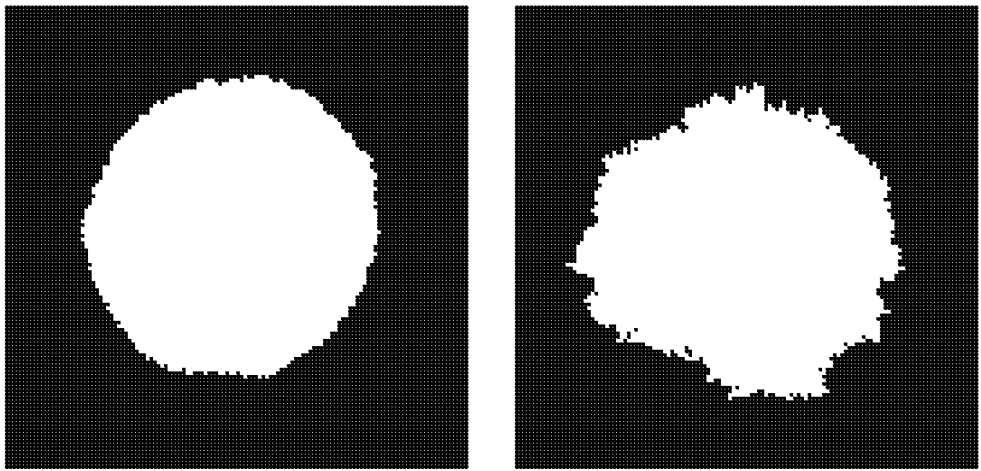}
\caption{Snapshots of holes in an oxygen layer that have been formed by
  reaction with CO, In both cases the initial situation was a surface
  completely covered with oxygen except for one site, Diffusion of CO
  and reaction with oxygen are infinitely fast, but on the left the
  reaction is infinitely faster than the diffusion and on the right it
  is the other way around,}
\label{fig:fastCO}
\end{figure}
\section{Lateral interactions}
\label{sec:latint}

Lateral interactions are interactions between adsorbates, It is
well-known that these interactions lead to structured adlayers at low
temperatures, Recently, people have realized that the kinetics of
surface reactions can be substantially affected by these interactions
even at high temperature, Little is known about the form of these
interactions and even less about the strength of them, In this section
we present a general but not very efficient method, the method that is
implemented in CARLOS to model pairwise additive interactions, and a few
tricks to model more complicated interactions with reasonable
efficiency, In all cases we assume that the lateral interactions are
short range,

A general method to model lateral interaction consists of specifying the
reaction and the occupation of the sites that may have adsorbates that
will affect the reaction, For example, if we have a simple desorption of
an adsorbate A on a square grid with a rate constant that depends on the
occupation of the four neighboring sites, then we have the reactions
\begin{eqnarray}
  \matrix{& (0,1),\cr
          (-1,0), & (0,0), & (1,0), \cr
          & (0,-1) \cr}&:&
  \matrix{& *\cr
          * & \hbox{A} & * \cr
          & * \cr}\to
  \matrix{& *\cr
          * & * & * \cr
          & * \cr},\nonumber\\
  \matrix{& (0,1),\cr
          (-1,0), & (0,0), & (1,0), \cr
          & (0,-1) \cr}&:&
  \matrix{& *\cr
          * & \hbox{A} & \hbox{A} \cr
          & * \cr}\to
  \matrix{& *\cr
          * & * & \hbox{A} \cr
          & * \cr},\nonumber\\
  \matrix{& (0,1),\cr
          (-1,0), & (0,0), & (1,0), \cr
          & (0,-1) \cr}&:&
  \matrix{& \hbox{A}\cr
          * & \hbox{A} & \hbox{A} \cr
          & * \cr}\to
  \matrix{& \hbox{A} \cr
          * & * & \hbox{A} \cr
          & * \cr},\\
  \matrix{& (0,1),\cr
          (-1,0), & (0,0), & (1,0), \cr
          & (0,-1) \cr}&:&
  \matrix{& *\cr
          \hbox{A} & \hbox{A} & \hbox{A} \cr
          & * \cr}\to
  \matrix{& *\cr
          \hbox{A} & * & \hbox{A} \cr
          & * \cr},\nonumber\\
  \matrix{& (0,1),\cr
          (-1,0), & (0,0), & (1,0), \cr
          & (0,-1) \cr}&:&
  \matrix{& \hbox{A} \cr
          \hbox{A} & \hbox{A} & \hbox{A} \cr
          & * \cr}\to
  \matrix{& \hbox{A} \cr
          \hbox{A} & * & \hbox{A} \cr
          & * \cr},\nonumber\\
  \matrix{& (0,1),\cr
          (-1,0), & (0,0), & (1,0), \cr
          & (0,-1) \cr}&:&
  \matrix{& \hbox{A} \cr
          \hbox{A} & \hbox{A} & \hbox{A} \cr
          & \hbox{A} \cr}\to
  \matrix{& \hbox{A} \cr
          \hbox{A} & * & \hbox{A} \cr
          & \hbox{A} \cr},\nonumber
\end{eqnarray}
Of all symmetry-related reactions only one is shown, Each of these
reactions can be given a different rate constant (or activation energy
and preexponential factor), By specifying all possible occupations of
the neighboring sites explicitly any form of the lateral interactions
can be modeled, The disadvantage should also be clear, The list of
reactions can be quite long, With $Z$ neighboring sites that may be
occupied by an adsorbate that affects the reaction and $A$ possible
occupations of each of these sites (including no adsorbate) there are
$A^Z$ reactions to specify,

CARLOS (see section~\ref{sec:car}) has the possibility to deal with
lateral interactions that change the activation energy of a reaction in
a pairwise additive manner, Let's take the previous example of a
desorbing adsorbate A again, In CARLOS desorption can be specified as
\begin{eqnarray}
\label{eq:rxlatint}
  &&\matrix{& (0,1),\cr
          (-1,0), & (0,0), & (1,0), \cr
          & (0,-1) \cr}:\\
  &&\matrix{& \hbox{\{ $*$ A : $\varphi$ \}}\cr
          \hbox{\{ $*$ A : $\varphi$ \}}
          & \hbox{A} 
          & \hbox{\{ $*$ A : $\varphi$ \}} \cr
          & \hbox{\{ $*$ A : $\varphi$ \}} \cr}\to
  \matrix{& \#\cr
          \# & * & \# \cr
          & \# \cr},\nonumber
\end{eqnarray}
The notation with the curly braces means that an adsorbate A at the site
changes the activation energy of the reaction by an amount $\varphi$, A
vacant site does not change the activation energy, The hashes (\#) on
the right mean that the occupation of the site does not change, This way
to model lateral interactions does not have the flexibility of the more
general approach, but it is definitely much easier,

Figure~\ref{fig:latints} shows the typical effect of lateral
interactions, Interactions between nearest neighbors are very often
repulsive, On a square grid this leads the checkerboard structure at low
temperatures, (Increasing the temperature leads to via an order-disorder
phase transition to a random structure,) When diffusion is slow
relaxation is slow and the checkerboard does not cover the whole system,
but domains are formed, If one wants to get a good idea of these domains
one can change the model in the following way, Instead of the
reaction~(\ref{eq:rxlatint}) one uses
\begin{eqnarray}
  &&\matrix{& (0,1),\cr
          (-1,0), & (0,0), & (1,0), \cr
          & (0,-1) \cr}:\\
  &&\matrix{& \hbox{\{ $*2$ A2 : $\varphi$ \}}\cr
          \hbox{\{ $*2$ A2 : $\varphi$ \}}
          & \hbox{A1} 
          & \hbox{\{ $*2$ A2 : $\varphi$ \}} \cr
          & \hbox{\{ $*2$ A2 : $\varphi$ \}} \cr}\to
  \matrix{& \#\cr
          \# & *1 & \# \cr
          & \# \cr},\nonumber
\end{eqnarray}
and
\begin{eqnarray}
  &&\matrix{& (0,1),\cr
          (-1,0), & (0,0), & (1,0), \cr
          & (0,-1) \cr}:\\
  &&\matrix{& \hbox{\{ $*1$ A1 : $\varphi$ \}}\cr
          \hbox{\{ $*1$ A1 : $\varphi$ \}}
          & \hbox{A2} 
          & \hbox{\{ $*1$ A1 : $\varphi$ \}} \cr
          & \hbox{\{ $*1$ A1 : $\varphi$ \}} \cr}\to
  \matrix{& \#\cr
          \# & *2 & \# \cr
          & \# \cr},\nonumber
\end{eqnarray}
In the initial condition if a site has a label ending in 1 (2) then a
neighboring site should have a label ending in 2 (1), The reason for
having two reactions representing effectively the same reaction is that
one can now give neighboring sites on output a different coloring, If one
gives ``A1'' to same color as ``$*$2'', and ``A2'' the same color as
``$*$1'', then all sites of one domain will have the same color as can
be seen in Fig,~\ref{fig:latints}, Completely covered or vacant areas
will appear as a checkerboard now,
\begin{figure}[ht]
\includegraphics[width=\hsize]{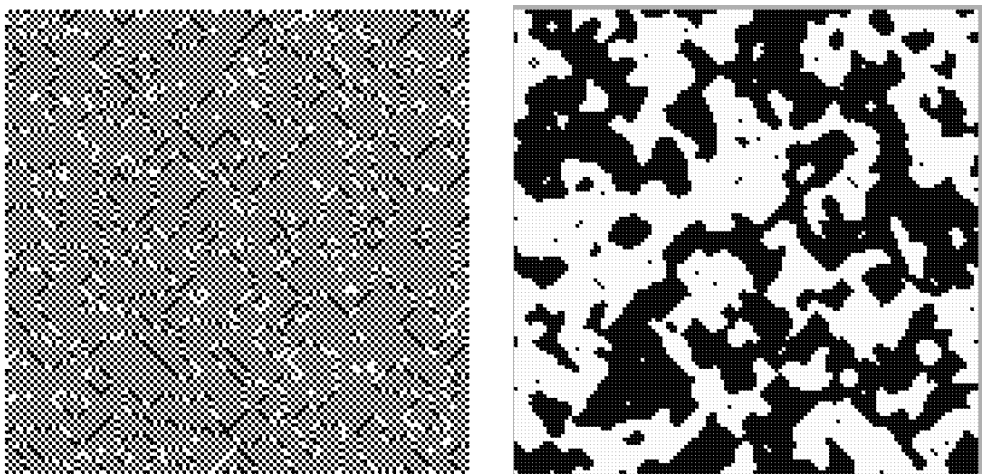}
\caption{Snapshots of simulations of simple desorption with lateral
  interactions and no diffusion, Nearest neighbors repel each other, but
  next-nearest neighbors attract each other, On the left the adsorbates
  (in black) are shown for situation with half of the sites are
  occupied, On the right a similar situation is shown, but there the
  coloring of is reversed on alternate sites, So if a site is black when
  occupied and white when vacant, then a neighboring site is black when
  vacant and white when occupied,}
\label{fig:latints}
\end{figure}

For lateral interactions that are not pairwise additive we need not
always use the general method, One should always look for some simple
prescription describing the energetics, For example, suppose that the
activation energy for simple desorption depends on the number of
nearest-neighbor sites being occupied, Then this might be modeled by the
reaction
\begin{equation}
   (0,0/0),(0,0/1):\hbox{ A $n$}\to\hbox{$*$ $n$},
\end{equation}
Here we have two sites per unit cell, The first is an actual site, but
the second has a label that specifies how many of the neighbors of the
actual site are occupied; i,e,, the label $n$ is number of adsorbates on
neighboring sites, In CARLOS one can use infinitely fast reactions to
update the labels specifying these numbers, but one can of course also
implement this explicitly,

Another example would be simple desorption on a hexagonal grid with an
activation energy that is affected by adsorbates that are nearest
neighbors and that form equilateral triangles, For each real site we can
defined two additional sites, If the translations of the unit cell are
given by $a(1,0)$ and $a(1/2,1/2\sqrt{3})$, where $a$ is the unit cell
parameter, then we can use site $(0,0/0)$ as real site, $(0,0/1)$ for
the occupation of sites $(0,0/0)$, $(1,0/0)$, and $(0,1/0)$, and
$(0,0/2)$ for the occupation of sites $(1,0/0)$, $(0,1/0)$, and
$(1,1/0)$, The desorption can then be modeled in the notation of CARLOS
as
\begin{eqnarray}
   &&(0,0/0),
     (0,0/1),(-1,0/2),(-1,0/1),(-1,-1/2),(0,-1/1),(0,-1/1):\nonumber\\
   &&\hbox{ A}
     \hbox{ \{ no yes : $\varphi$\}}
     \hbox{ \{ no yes : $\varphi$\}}
     \hbox{ \{ no yes : $\varphi$\}}\nonumber\\
   &&\hbox{ \{ no yes : $\varphi$\}}
     \hbox{ \{ no yes : $\varphi$\}}
     \hbox{ \{ no yes : $\varphi$\}}\nonumber\\
   &&\to\hbox{$*$ no no no no no no},
\end{eqnarray}
The label ``yes'' means that the three sites are all occupied, and the
label ``no'' means that at least one is vacant, The activation energy
depends linearly on the number of triangles formed by the adsorbates,
and each triangle changes the adsorption energy by $\varphi$,
\chapter[Examples]{Examples}
\label{ch:examples}

This chapter is somewhat similar to chapter~\ref{ch:modeling} on how to
model reaction systems, However, the emphasize here is more on the
information one can get from DMC simulations; i,e,, the reason why one
wants to do DMC simulations instead of using conventional macroscopic
rate equations,
\section{The Ziff-Gulari-Barshad model}
\label{sec:zgb}

Although the work by Ziff, Gulari, and Barshad does not represent the
first application of Monte Carlo to model surface reactions it is
probably the most influential work of its type,\cite{zif86b} There are
several reasons for that, It deals with CO oxidation which was and still
is a very important process in catalysis and surface science, It is a
very simple model, which makes it generic, Its simplicity also made it
possible to analyze in detail the relation between the microscopic
reactions and the macroscopic properties, It showed the shortcomings of
the macroscopic rate equations and what the origin of these shortcomings
were, It showed that these DMC simulations could yield so-called kinetic
or non-equilibrium phase transitions, In fact, apart from the
first-order phase transition also known from macroscopic rate equations,
it showed that there was a continuous phase transition as well,
\begin{figure}[t]
\includegraphics[width=\hsize]{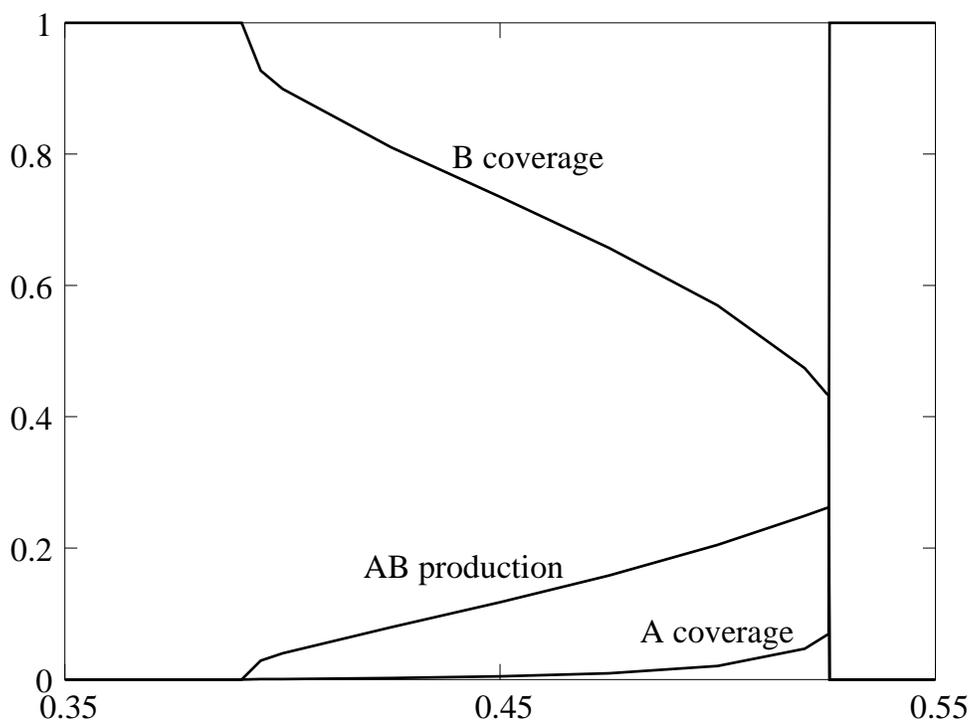}
\caption{Phase diagram of the Ziff-Gulari-Barshad model, The coverages
  and the AB formation per unit time per site are shown as a function of
  the $y$ parameter,}
\label{fig:zgb}
\end{figure}

We present here the model, which we will call the ZGB-model, in its
generic form, There are two adsorbates; A and B, If one wants to use the
ZGB-model for CO oxidation, then A stands for CO and B for atomic
oxygen, There are three reactions in the model, Adsorbate A can adsorb
at single vacant sites, Adsorbate B can adsorb, but as it forms diatomic
molecules in the gas phase, two neighboring vacant sites are needed, An
A will react with a B if they are nearest neighbors, This reactions is
infinitely fast, so this takes place immediately after an adsorption,
We can write the reaction as
\begin{eqnarray}
  {\rm A(gas)}+*&\to&{\rm A(ads)}\nonumber\\
  {\rm B}_2{\rm (gas)}+2*&\to&2{\rm B(ads)}\\
  {\rm A(ads)}+{\rm B(ads)}&\to&{\rm AB(gas)}+2*\nonumber
\end{eqnarray}
where $*$ is a vacant site, ``ads'' stands for an adsorbed species, and
``gas'' for a species in the gas phase, The reactions involving two
sites can only take place on neighboring sites, Focusing on the sites
only we have
\begin{eqnarray}
  *&\to&{\rm A}\nonumber\\
  2*&\to&2{\rm B}\\
  {\rm A}+{\rm B}&\to&2*\nonumber
\end{eqnarray}
The adsorbates do not diffuse in the original model, and the grid is a
square one, There are many extensions to this model dealing, amongst
others, with desorption of CO and oxygen,\cite{kau89} diffusion of the
adsorbates,\cite{kau89} an Eley-Rideal mechanism for the oxidation step,
physisorption of the reactants, lateral interactions between the
adsorbates,\cite{kau89} blocking of the sites due to poisoning with lead
or alloying,\cite{hov92,mai95} reconstruction of the surface (see
section~\ref{sec:copt}),\cite{gel98,kuz99a,kor99b,kuz02,gel99,kor98a,kuz98,kor99a}
and an inert adsorbate that causes oscillations,\cite{jan97}
\begin{figure}[t]
\includegraphics[width=0.22\hsize,angle=-90]{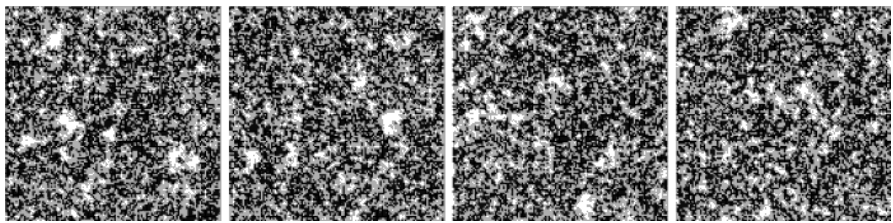}
\caption{Snapshots of the adlayer in the Ziff-Gulari-Barshad model for
  $y=0.5255$ a value just below the first-order transition point, The
  black islands are formed by B's, The A's form the white islands,}
\label{fig:zgb2}
\end{figure}

The rate constant for adsorption of A can be derived as in
section~\ref{subsec:uniads}, and for adsorption of B as in
section~\ref{subsec:bimolads}, The results are
\begin{eqnarray}
  W_{\rm A,ads}&=&{yPA_{\rm site}\sigma_{\rm A}
      \over\sqrt{2\pi mk_{\rm B}T}}\\
  W_{\rm B,ads}&=&{2(1-y)PA_{\rm site}\sigma_{\rm B}
      \over 4\sqrt{2\pi mk_{\rm B}T}},
\end{eqnarray}
The $\sigma$'s are sticking coefficients, The quantity $y$ is the
fraction of the molecules in the gas phase that are A's, If we assume
that the sticking coefficients are equal to each other, then we can
simplify the rate constants to $y$ and $(1-y)/2$, respectively, by
replacing time $t$ by $\tau=tPA_{\rm site}\sigma_{\rm A}/\sqrt{2\pi
  mk_{\rm B}T}$, We see that then the model depends only on one
parameter; i,e,, only on $y$,

Simulations show that there are three states for the system, One
possibility is that the surface is completely covered by A's, There are
no reactions that can take the system out of this state, Such a state is
called on absorbing state in non-equilibrium statistical physics, In
catalysis one talks in such a case of A poisoning, because it leads to
the undesirable situation that the reactivity is zero, There is another
absorbing state, but then with B poisoning, If the parameter $y$ is
below a critical value $y_1$, then the system will always evolve into
the B poisoning state, If the parameter $y$ is above another critical
value $y_2$, then the system will always evolve into the A poisoning
state, For $y<y_1$ there are so many ${\rm B}_2$ molecules in the gas
phase that B adsorption will outcompete the A's for the vacant sites
that are formed by the reaction between the A's and B's, To same thing
happens for $y>y_2$ except in this situations the A adsorption wins, At
$y_1$ and at $y_2$ there is a kinetic or non-equilibrium phase
transition,

For $y_1<y<y_2$ there is a third state with A's and B's on the surface
and a non-zero reactivity, Figure~\ref{fig:zgb} shows how the reactivity
and the coverage depend on $y$, Note that all quantities change
discontinuously at $y_2$, The phase transition at that value of $y$ is
therefore called a first-order transition, At $y_1$ the quantities
change continuously, so we have a continuous (or second-order) phase
transition, Macroscopic rate equations also predict the first-order phase
transition, but not the continuous one, However, the first-order phase
transition is predicted by the macroscopic rate equations to be at
$y_1=2/3$, whereas the best estimates from DMC simulations are
$y_2=0.52560\pm 0.00001$,\cite{zif92,bro93} The continuous phase
transition is estimated to be at $y_1=0.39065\pm 0.00010$,\cite{jen90}

Figure~\ref{fig:zgb2} shows the reason for the discrepancy between the
DMC results and those of macroscopic rate equations, The adlayer is
definitely not a random mixture of adsorbates, The reason is that the
fast reaction between the A's and B's, This reaction causes segregation
of the adsorbates, Isolated A's will not last long on the surface,
because a B may adsorb on one of the vacant neighboring sites that will
immediately react with the A which will remove the A from the surface,
For a similar reason isolated B's will be rare, Only islands of the same
kind of adsorbate can last, because the particles in the center of an
island have no neighboring sites onto which other particle can adsorb
with which they will react, These islands are formed randomly, Islands
of B are larger, They need be because A's need only one vacant site for
adsorption and can relatively easily break them up, Islands of A can be
smaller, B's need two neighboring vacant sites for adsorption and have
more difficulty to remove A's,
\section{TPD/TPR of $2{\rm AB}\to{\rm A}_2+{\rm B}_2$}

The model in this section illustrates how atomic details influence the
kinetics, The AB molecule is adsorbed on the surface and the first
reaction is the dissociation into an A and a B, This is only possible,
however, if a neighboring site of the AB molecule is vacant, In the
macroscopic rate equation
\begin{equation}
  {d\theta_{\rm AB}\over dt}=-k\theta_{\rm AB}\theta_*,
\end{equation}
we have dissociation as long as are molecule to dissociate and
$\theta_*>0$, This means that if initially less than half the sites are
occupied by AB's all molecules can dissociate, In a DMC simulation
things are not so straightforward, because the vacant sites might not be
accessible, For example, the model here is applicable to NO reduction on
rhodium, The nitrogen and oxygen atoms that are formed when NO
dissociates do hardly diffuse at all, and severely restrict the motion
of NO to find a vacant site,\cite{har99,jan01}
\begin{figure}[t]
\includegraphics[width=\hsize]{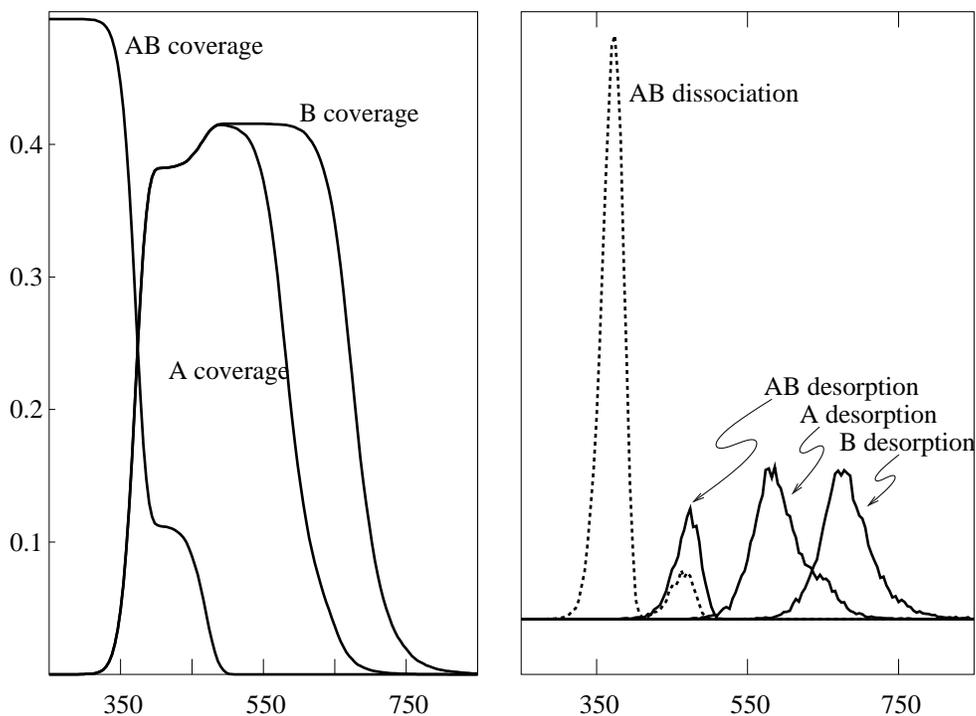}
\caption{Changes in the coverages (left) and rates of the reactions
  (right) for the $2{\rm AB}\to{\rm A}_2+{\rm B}_2$ process as a
  function of temperature (in Kelvin), The results are obtained from a
  simulation with a $256\times 256$ grid of a TPD/TPR experiment with a
  heating rate of $10\,{\rm K/sec}$ and an initial AB coverage of
  $0.4946\,$ML,}
\label{fig:abred}
\end{figure}

The model here has four reactions
\begin{eqnarray}
  {\rm AB(ads)}+*&\to&{\rm A(ads)}+{\rm B(ads)},\nonumber\\
  {\rm AB(ads)}&\to&{\rm AB(gas)}+*,\\
  2{\rm A(ads)}&\to&{\rm A}_2{\rm (ads)}+2*,\nonumber\\
  2{\rm B(ads)}&\to&{\rm B}_2{\rm (ads)}+2*,\nonumber
\end{eqnarray}
where $*$ is a vacant site, ``ads'' stands for an adsorbed species, and
``gas'' for a species in the gas phase, The reactions involving two
sites can only take place on neighboring sites, Focusing only on the
sites we have
\begin{eqnarray}
  {\rm AB}+*&\to&{\rm A}+{\rm B},\nonumber\\
  {\rm AB}&\to&*,\\
  2{\rm A}&\to&2*,\nonumber\\
  2{\rm B}&\to&2*,\nonumber
\end{eqnarray}
The kinetic parameters are $\nu_{\rm diss}=\nu_{\rm AB,des}=\nu_{\rm
  A,des}=\nu_{\rm B,des}=10^{13}\,\hbox{cm}^{-1}$, and $E_{\rm
  act,diss}/k_{\rm B}=11500\,$K, $E_{\rm act,AB,des}/k_{\rm
  B}=14500\,$K, $E_{\rm act,A,des}/k_{\rm B}=18000\,$K, and $E_{\rm
  act,B,des}/k_{\rm B}=21000\,$K, We see that the reactions above are
ordered from fast to slow, In a TPD/TPR experiment they would also occur
in that order if there would be no interference,  We use a square grid
for the simulations,

Figure~\ref{fig:abred} shows TPD/TPR spectra of the system and changes
in the coverage, We see that AB dissociation starts at low temperature,
but not all of the AB dissociates even though there are enough vacant
sites, This means that the available sites are not accessible; the A's
and B's prevent the AB molecules from reaching the vacant sites, At
about $T=450\,$K the AB molecules that are still present start to
desorb, This leads to new vacant sites which can be used for
dissociation, At this temperature the dissociation is almost
instantaneous, From the changes in the coverages and the reaction rates
we can see that also not all new vacant sites are used for
dissociation, The mobility of the A's and B's is still too low for AB's
to reach all vacant sites, At $T=550\,$K all AB's are gone and we have
associative desorption of ${\rm A}_2$ and at higher temperature of ${\rm
  B}_2$,
\section{TPD with strong repulsive interactions}

One of the clearest examples of the advantage of DMC simulations over
macroscopic rate equations is formed by simple desorption with strong
lateral interactions, If we take for example a square grid, an initial
situation with all sites occupied, and repulsive interactions between
nearest neighbors, then the following happens, Initially, each adsorbate
has the same tendency to desorb, However, if one adsorbate desorbs then
its former neighbors suddenly feel less repulsion and become adsorbed
more strongly, This means that desorption takes place in two stages,
First about half the adsorbates desorb, because they have many
neighbors, After these adsorbates have desorbed the structure of the
adlayer is that of a checkerboard, with almost all adsorbates having no
neighbors, Because these adsorbates feel little or no repulsion from
other adsorbates they desorb at a later stages,
\begin{figure}[t]
\includegraphics[width=\hsize]{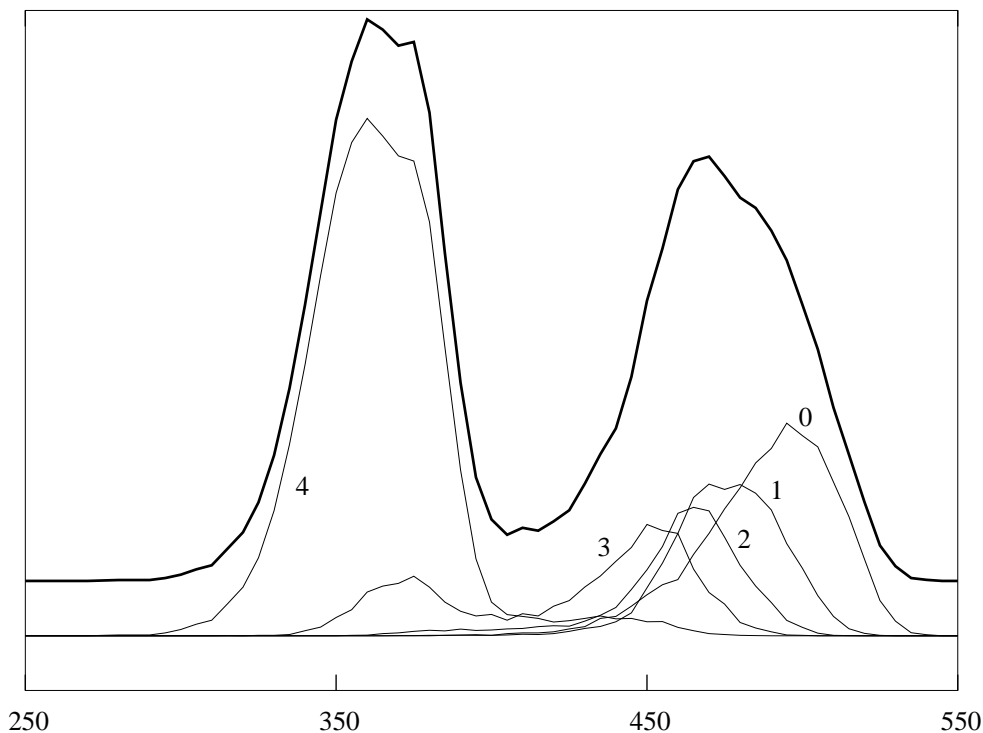}
\caption{Temperature-Programmed Desorption spectrum (desorption rate
  vs, temperature in Kelvin) of adsorbates repelling each other, The fat
  line is the total desorption rate, The thin lines are separate
  contributions of adsorbates desorbing with a number of nearest
  neighbors given by the number next to the curve, Activation energy for
  desorption is $E_{\rm act}/k_{\rm B}=14590\,$K and the preexponential
  factor is $\nu=1.435\cdot 10^{12}\,\hbox{cm}^{-1}$, These numbers were
  taken from CO desorption from Rh(100) at low coverage, The repulsion
  between two adsorbates is $1000\,$K,}
\label{fig:codes}
\end{figure}

Note that the process involves a symmetry breaking, Initially all sites
are equivalent, but after half the adsorbates have desorbed alternate
sites are occupied and vacant, Ordinary macroscopic rate equations are
not able to describe this symmetry breaking, because they assume that
all site are equivalent during the whole process, We can split the
macroscopic rate equations in two; one for the sites with the adsorbates
that desorb first, and one for the sites with adsorbates that desorb
later, However, the equations for both sites are equivalent and the
symmetry breaking only occurs if the initial situation already has a
small difference in the occupations between the sites with the early
desorbers and the sites with the late desorbers, For DMC simulations
such an unrealistic initial condition is not necessary, In DMC
fluctuations in the times when reactions occur cause the symmetry
breaking as they do in real systems, Such fluctuations are not included
in macroscopic rate equations,

The fluctuations determine which adsorbates desorb first, and also
affect the structure of the adlayer when the coverage has been more of
less halved, A perfect checkerboard structure is only found when the
adsorbates also diffuse fast and when the temperature is well below the
order-disorder phase transition temperature, For small system sizes the
diffusion need not be so fast as for large system sizes, Diffusion has
to make sure that no domains are formed as in figure~\ref{fig:latints},
Simulations show that even at relatively small systems with say a
$128\times 128$ grid it is almost impossible to avoid domain formation,

Figure~\ref{fig:codes} shows a TPD spectrum for a system with repulsive
interactions, We show TPD instead of isothermal desorption, because the
latter shows the two stages in the desorption only in how fast the
coverage and the desorption rate decreases, If the coverage is plotted
logarithmically, then we get first a straight line with a large negative
slope, followed by a straight line with a small, in absolute sense,
negative slope, In TPD the two stages are much clearer, because the
desorption rate has two peaks, The figure shows that the second stage
(i,e,, the second peak) has also contributions from adsorbates with one,
two, and even three neighbors, which is due to the fact that the first
stage never forms a perfect checkerboard structure, If the repulsion
becomes very strong, then more than two peaks can be formed,

In real systems the effects of lateral interactions are generally not so
unambiguous, The lateral interactions have to be strong enough so that a
well-defined structure is formed when half the adsorbates have desorbed,
If this is not the case then the lateral interactions only show up by
shifting or broadening a single peak when the initial coverage is
increased, When lateral interactions are strong enough, then they may
also push adsorbates to other adsorption sites, These sites have a lower
adsorption energy, but overall the energy is lowered because it lessens
the repulsion between the adsorbates, The result may even be an adlayer
structure that is incommensurate with the substrate, Finally, lateral
interactions need not just be between nearest neighbors, Interactions
between next- en next-next-nearest neighbors are not uncommon, Longer
range interactions have been excluded,\cite{gex00} but charged
adsorbates might have long-range interactions, which may explain very
broad desorption peaks,
\section{CO electrooxidation on a Pt-Ru electrode}

The reason for doing DMC simulations is that the system is not
homogeneous, This may be because the reactions cause an inhomogeneity in
the adlayer, but it may also be that the substrate is inhomogeneous as
in the case for the system of this section, The precise structure of the
Pt-Ru electrode is not known, We will assume here that the two metals
each form large islands mainly for illustrative purposes,\cite{kop99c}
The idea of having two metals is that each catalyzes only some reactions
well, In the model here the ${\rm CO}_2$ formation proceeds well on Pt
and the ${\rm H}_2{\rm O}$ dissociates on Ru,
\begin{figure}[t]
\includegraphics[width=\hsize]{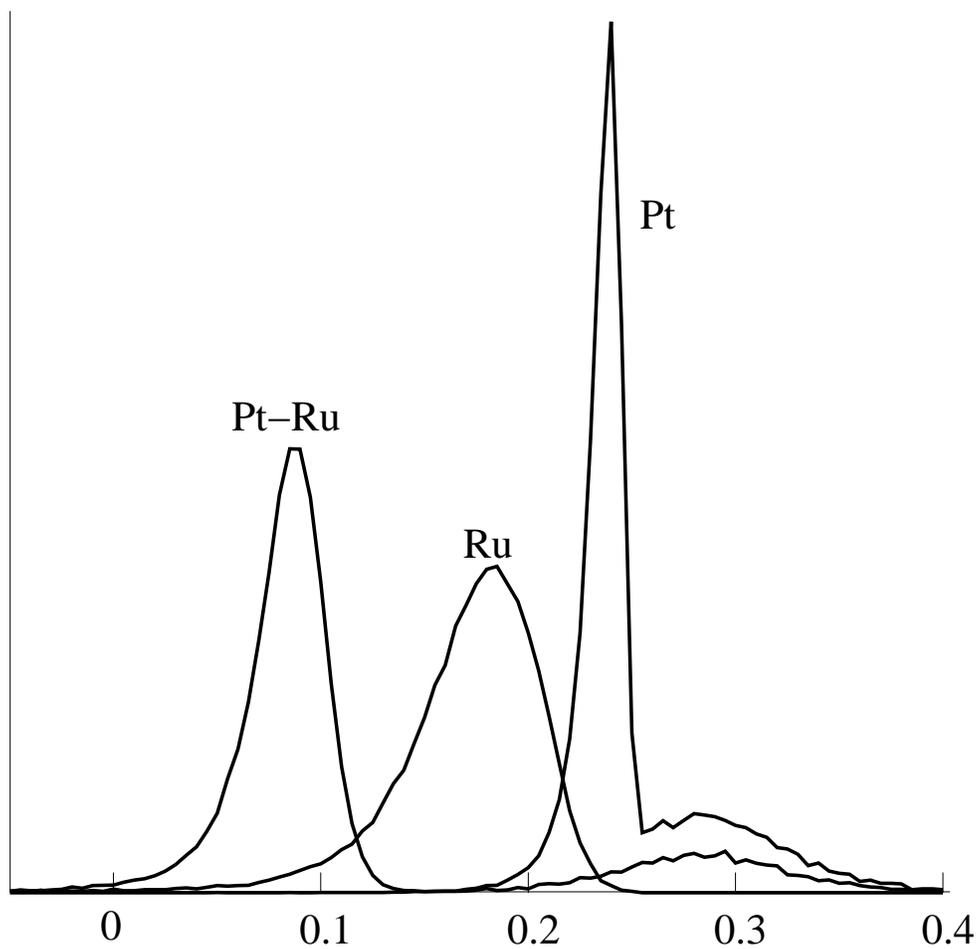}
\caption{Current of CO electrooxidation as a function of electrode
  potential for pure Pt, pure Ru, and a Pt-Ru alloy with randomly
  distributed Pt and Ru atoms, The initial surface is for about 99\%
  covered by CO and the potential is $-0.2\,$V, The potential is
  increased by $0.1\,$V/sec,}
\label{fig:voltam}
\end{figure}
\afterpage{\clearpage}

We model the system with a square grid, We simulate a linear-sweep
voltammetry experiment in which the electrode is initially for 99\%
covered by CO, The following reactions can occur,
\begin{eqnarray}
  {\rm H}_2{\rm O}+*&\to&{\rm OH}+{\rm H}^++e^-,\nonumber\\
  {\rm OH}+{\rm H}^++e^-&\to&{\rm H}_2{\rm O}+*,\\
  {\rm CO}+{\rm OH}&\to&{\rm CO}_2+2*+{\rm H}^++e^-,\nonumber
\end{eqnarray}
or if we just look at the site occupation
\begin{eqnarray}
  *&\to&{\rm OH},\nonumber\\
  {\rm OH}&\to&*,\\
  {\rm CO}+{\rm OH}&\to&2*,\nonumber
\end{eqnarray}
The reactions can take place on both metals, and the latter also on the
border of a Pt and a Ru patch, The CO diffuses rapidly over the surface,
whereas OH remains fixed at the site where it is formed, In the
experiment the current is measured; i,e,, the sum of the rates of the
first and the last reaction minus the second,

Although the reactions can occur on both metals, the rate constants
differ, We can write the rate constants as
\begin{eqnarray}
  W_{\rm ads}^{\rm M}&=&k_{\rm ads}^{\rm M}
    \exp\left[{\alpha_{\rm M}e_0E\over k_{\rm B}T}\right],\nonumber\\
  W_{\rm des}^{\rm M}&=&k_{\rm des}^{\rm M}
    \exp\left[-{(1-\alpha_{\rm M})e_0E\over k_{\rm B}T}\right],\\
  W_{\rm rx}^{{\rm M}_1{\rm M}_2}&=&k_{\rm rx}^{{\rm M}_1{\rm M}_2}
    \exp\left[{\alpha_{{\rm M}_1{\rm M}_2}
     e_0E\over k_{\rm B}T}\right],\nonumber
\end{eqnarray}
The first refers to the dissociative adsorption of water, the second to
the desorption of water, and the last to the formation of ${\rm CO}_2$,
In the last ${\rm M}_1$ is the metal with the CO, and ${\rm M}_2$ the
metal with the OH, The $\alpha$'s are so-called transfer coefficients,
$e_0$ is the elementary charge, and $E$ is the potential of the
electrode,  All transfer coefficient are taken equal to 0.5, $T=300\,$K,
and $k_{\rm ads}^{\rm Pt}=0.2\,{\rm s}^{-1}$, $k_{\rm des}^{\rm
  Pt}=10^4\,{\rm s}^{-1}$, $k_{\rm ads}^{\rm Ru}=40\,{\rm s}^{-1}$,
$k_{\rm des}^{\rm Ru}=5\,{\rm s}^{-1}$, $k_{\rm rx}^{\rm RuRu}=0.1\,{\rm
  s}^{-1}$, $k_{\rm rx}^{\rm RuPt}=0.1\,{\rm s}^{-1}$, $k_{\rm rx}^{\rm
  PtPt}=1\,{\rm s}^{-1}$, and $k_{\rm rx}^{\rm PtRu}=1\,{\rm s}^{-1}$,
The rate constant for CO diffusion is $100\,{\rm s}^{-1}$,

Figure~\ref{fig:voltam} shows the voltammograms for a Pt, a Ru, and a
Pt-Ru electrode, These are obtained by linearly changing the potential
of the electrode from $-0.2\,$ to $0.5\,$V in $7\,$seconds, The
remarkable fact is that the Pt-Ru alloy is more reactive than either of
the pure metals, The reason for this is as follows, The rate limiting
step on Pt is the dissociative adsorption of water, On Ru the rate
limiting step is the formation of ${\rm CO}_2$, We get a synergetic
effect if the oxidation step on the boundary of the Pt and Ru parts of
the alloy is at least as fast as on Pt, In that case OH is formed on Ru
on the boundary, and CO on Pt reacts with this OH, This is most clearly
seen when we do a simulation with large Pt and Ru islands (see
figure~\ref{fig:lrgisl}), There is a depletion of CO on Pt near the
boundary, because there the CO react, There is a higher concentration of
OH on Ru near the boundary because the higher oxidation reactivity there
leads to more vacant sites for water adsorption,
\begin{figure}[t]
\includegraphics[width=\hsize]{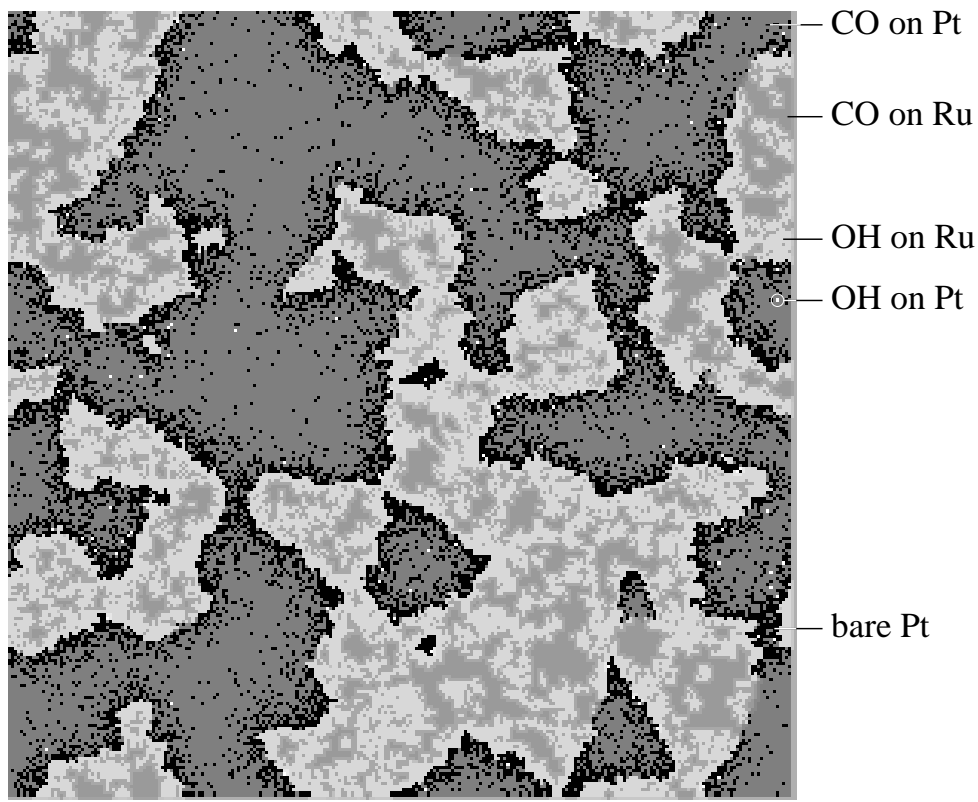}
\caption{Snapshot of CO electrooxidation on a Pt-Ru electrode that
  consists of large islands of Pt and large islands of Ru,}
\label{fig:lrgisl}
\end{figure}
\section{Oscillations of CO oxidation on Pt surfaces}
\label{sec:copt}

One problem for which extensive DMC simulations have been done by
various groups is the problem of CO oscillations on Pt(100) and Pt(110),
A crucial role in these oscillations is played by the reconstruction of
the surface, and the effect of this reconstruction on the adsorption of
oxygen, The explanation of the oscillations is as follows, A bare Pt
surface reconstructs into a structure with a low sticking coefficient
for oxygen, This means that predominantly CO adsorbs on bare Pt,
However, CO lifts the reconstruction, The normal structure has a high
sticking coefficient for oxygen, So after CO has adsorbed in a
sufficient amount to lift the reconstruction oxygen can also adsorb, The
CO and the oxygen react, and form ${\rm CO}_2$, This ${\rm CO}_2$
rapidly desorbs leaving bare Pt which reconstructs again, An important
aspect of this process, and also other oscillatory reactions on
surfaces, is the problem of synchronization, The cycle described above
can easily take place on the whole surface, but oscillations on
different parts on the surface are not necessarily in phase, and the
overall reactivity of a surface is then constant, To get the whole
surface oscillating in phase there has to be a synchronization
mechanism,

The most successful model to describe oscillations on Pt surfaces is the
one by Kortl\"uke, Kuzovkov, and von
Niessen,\cite{kuz99a,kor99b,kuz02,kor98a,kuz98,kor99a} This model has CO
adsorption and desorption, oxygen adsorption, ${\rm CO}_2$ formation, CO
diffusion, and surface reconstruction, The surface is modeled by a
square grid, Each site in the model is either in state $\alpha$ or in
state $\beta$, The $\alpha$ state is the reconstructed state which has a
reduced sticking coefficient for oxygen, The $\beta$ state is the
unreconstructed state with a high sticking coefficient for oxygen, An
$\alpha$ site will convert a neighboring $\beta$ site into an $\alpha$
state if neither sites is occupied by CO, A $\beta$ site will convert a
neighboring $\alpha$ site into $\beta$ if at least one of them is
occupied by CO,

The model shows a large number of phenomena depending on the rate
constants, We will only look at oscillations that occur for reduced rate
constants $y=0.494$, $k=0.1$, and $V=1$,\cite{sal02} The first rate
constant, $y$, is the one for CO adsorption and has the same meaning as
in the ZGB-model (see section~\ref{sec:zgb}), The second, $k$, is the
rate constant for CO desorption, The last, $V$, is the rate constant for
the reconstruction and the lifting of the reconstruction, The rate
constant for oxygen adsorption is as for the ZGB-model $(1-y)/2$ on the
$\beta$ phase, and $s_\alpha (1-y)/2$ on the $\alpha$ phase, We will
look at the Pt(110) surface which has $s_\alpha=\,0.5$, The diffusion
rate constant has been varied,

Figure~\ref{fig:smallD} shows snapshots obtained from some large
simulations in which the diffusion is just about fast enough to lead to
global oscillations provided the initial conditions are
favorable, However, it is also possible to choose the initial conditions
so that the oscillations are not synchronized properly, In that case one
can see the formation of patterns as the right half of the figure shows,
\begin{figure}[t]
\includegraphics[width=\hsize,angle=-90]{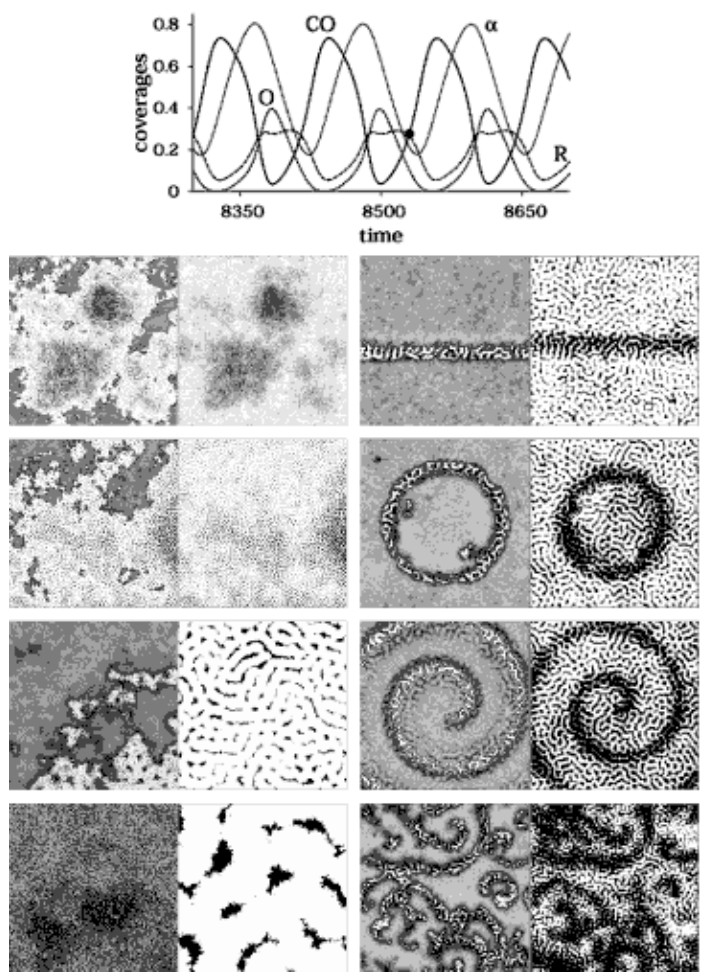}
\caption{Global oscillations and pattern formation with $D=250$, The top
  shows temporal variations of the coverages, the fraction of the
  substrate in the $\alpha$ phase, and the ${\rm CO}_2$ production $R$,
  Each picture has two parts, In the left part we plot the chemical
  species; CO particles are grey and O particles are white, and empty
  sites are black, The right part shows the structure of the surface;
  $\alpha$ phase sites are black, and $\beta$ phase sites are white,
  Sections of the upper-left corner with $L=8192$, $4096$, $1024$, and
  $256$ are shown on the left half of the figure, The sections
  correspond to the dot in the temporal plot at the top, On the right
  half of the figure we have a wave front, a target, a spiral, and
  turbulence ($L=2048$), which can be obtained with different initial
  conditions,}
\label{fig:smallD} 
\end{figure}

Synchronization is obtained when the diffusion rate is fast enough, The
minimal value is related to the so-called Turing-like structures that
are formed in the substrate, These structure can best be seen in the
lower two pictures on the left and all pictures on the right of
figure~\ref{fig:smallD}, If diffusion is so fast that within one
oscillatory period CO can move from one phase ($\alpha$ or $\beta$) to a
neighboring island of the other phase, then the oscillation are well
synchronized, If the diffusion rate is slower, then we get pattern
formation, Note that the system has two length scales, The
characteristic length scale of the adlayer is much larger than the
characteristic length scale of the Turing-like structures as can be seen
in the right half of the figure,
%
%
%
%\bibliography{,/dmcrefs}

%
%
%
\end{document}